\documentclass[12pt]{article}
\usepackage{amssymb,latexsym,amsmath}
\usepackage{epsfig}
\usepackage{graphicx}
\usepackage{amsfonts}
\usepackage{mathrsfs}
\usepackage[dvips]{color}

\newcommand{\R}{\mathbb{R}}

\newcommand{\fa}{\mathfrak{a}}

\newcommand{\fu}{\mathfrak{u}}

\newcommand{\fn}{{\mathfrak{n}}}

\newcommand{\fz}{\mathfrak{z}}

\newcommand{\bM}{\mathbf{M}}

\newcommand{\cE}{\mathcal{E}}
\newcommand{\cH}{\mathcal{H}}

\newcommand{\cO}{\mathcal{O}}
\newcommand{\cP}{\mathcal{P}}

\newcommand{\cT}{\mathcal{T}}
\newcommand{\cU}{\mathcal{U}}

\newcommand{\cW}{\mathcal{W}}
\newcommand{\cX}{\mathcal{X}}
\newcommand{\be}{\begin{equation}}
\newcommand{\ee}{\end{equation}}
\newcommand{\bea}{\begin{eqnarray}}
\newcommand{\eea}{\end{eqnarray}}
\newcommand{\nn}{\nonumber}
\newcommand{\kt}{\rangle}
\newcommand{\br}{\langle}

\newcommand{\ed}{\end{document}}

\newcommand{\bi}{\begin{itemize}}
\newcommand{\ei}{\end{itemize}}

\newcommand{\bce}{\begin{center}}
\newcommand{\ece}{\end{center}}

\newcommand{\sE}{\mathscr{E}}
\newcommand{\sF}{\mathscr{F}}

\newcommand{\sH}{\mathscr{H}}

\newcommand{\sT}{\mathscr{T}}
\newcommand{\RE}{{\rm Re}}

\begin{document}

\title{Spectral Singularities in the TE and TM modes of a $\cP\cT$-Symmetric Slab System: Optimal conditions for realizing a CPA-Laser}

\author{Ali~Mostafazadeh\thanks{E-mail address: amostafazadeh@ku.edu.tr}~~and 
Mustafa Sar{\i}saman\thanks{E-mail address: msarisaman@ku.edu.tr}\\ \\
Departments of Mathematics and Physics, Ko\c{c} University, \\ 34450 Sar{\i}yer,
Istanbul, Turkey}

\date{ }
\maketitle

\begin{abstract}

Among the interesting outcomes of the study of the physical applications of spectral singularities in $\cP\cT$-symmetric optical systems is the discovery of CPA-lasers. These are devices that act both as a threshold laser and a coherent perfect absorber (CPA) for the same values of their physical parameters. Unlike a homogeneous slab that is made to act as a CPA, a slab CPA-laser would absorb the incident waves coming from the left and right of the device provided that they have appropriate intensity and phase contrasts. We provide a comprehensive study of one of the simplest experimentally accessible examples of a CPA-laser, namely a $\cP\cT$-symmetric optical slab system consisting of a balanced pair of adjacent or separated gain and loss components. In particular, we give a closed form expression describing the spectral singularities of the system which correspond to its CPA-laser configurations. We determine the intensity and phase contrasts for the TE and TM waves that are emitted (absorbed) whenever the slab acts as a laser (CPA). We also investigate the behavior of the time-averaged energy density $\br u\kt$ and Poynting vector $\br\vec S\kt$ for these waves. This is necessary for determining the optimal values of the physical parameters of the system that make it act as a CPA-laser. These turn out to correspond to situations where the separation distance $s$ between the gain and loss layers is an odd multiple of a characteristic length scale $s_0$. A curious by-product of our study is that, except for the cases where $s$ is an even integer multiple of $s_0$, there is a critical angle of polarization beyond which the energy of the waves emitted from the lossy layer can be larger than the energy of those emitted from the gain layer.
\vspace{2mm}

\vspace{2mm}

\end{abstract}

\section{Introduction}

A basic fact about lasers is that they begin functioning once the gain coefficient $g$ of the laser material exceeds a critical value known as the threshold gain $g_\star$. The relation $g=g_\star$ is therefore called the `laser threshold condition' \cite{silfvast}. A few years ago \cite{pra-2011a} it was noticed that this condition coincided with the requirement that the continuous spectrum of the corresponding optical potential includes certain points known to mathematicians as spectral singularities \cite{naimark,ss-math}. Spectral singularities entered into physics literature as mathematical obstructions \cite{samsonov,jpa-2006,jpa-2009} to a Hermitization procedure developed to construct unitary quantum systems using certain non-Hermitian Hamiltonian operators \cite{jpa-2004,review}. Spectral singularities turn out to have an interesting physical meaning \cite{prl-2009}; they correspond to a special class of scattering states with a real and positive energy that behave exactly like resonances. This observation has led to a detailed study of the physical aspects of spectral singularities \cite{pra-2009,ss-phys-1,samsonov2,ss-phys-2,pla-2011,prsa-2012,jpa-2012,ss-phys-3,ss-phys-4,ss-phys-5,aalipour,HR} and their nonlinear generalizations \cite{prl-2013,liu,reddy,sap-2014}. Ref.~\cite{p123} gives a brief survey of developments in the subject.

An interesting application of spectral singularities is in the description of the phenomenon of coherent perfect absorption (CPA) which is also called antilasing \cite{antilaser1,longhi-PT,antilaser2,antilaser3,antilaser4}. It turns out that there are special circumstances where an optical potential absorbs certain incident coherent waves. Given an optical potential $v$ supporting a spectral singularity, the time-reversed optical system determined by the complex-conjugate of $v$ displays CPA. An interesting situation is when a spectral singularity accompanies its time-reversal \cite{jpa-2012}. This happens for $\cP\cT$-symmetric scattering potentials \cite{longhi-PT} and corresponds to a peculiar optical device that functions as a laser emitting coherent waves except when it is subject to certain incident coherent waves in which case it acts as an absorber. It is, therefore, called a CPA-laser. Today, CPA-lasers are theoretical constructs awaiting experimental realization.

The discovery of CPA-lasers is one of the most notable by-products of the recent interest in the manifestations and applications of $\cP\cT$-symmetric potentials in optics \cite{rdm,muss,markis,ruter}. The role of $\cP\cT$-symmetry in the context of CPA-lasers is similar to its role in the study of unidirectional indivisibility \cite{lin}. $\cP\cT$-symmetry implies that the condition for the emergence of a spectral singularity coincides with that of its time-reversal \cite{longhi-PT}. This makes $\cP\cT$-symmetric CPA-lasers the primary examples of CPA-lasers  \cite{jpa-2012}. The same holds in the study of unidirectional invisibility. Because under $\cP\cT$ the equations governing this effect are mapped to an equivalent set of equations, $\cP\cT$-symmetric unidirectionally invisible potentials have a much simpler structure \cite{pra-2013a}.

A basic problem in making a CPA-laser to act as a CPA is that the waves incident onto the device will not be absorbed unless they have correct amplitude and phase properties. This does not cause a major difficulty for a (non-lasing) CPA made out of a homogeneous slab of lossy material such as those considered in Refs.~\cite{antilaser3,pra-2015a}, for the following reasons.
    \begin{enumerate}
    \item The optical potential for a homogeneous slab is $\cP$-invariant, where $\cP$ is the reflection about the plane parallel to the slab that passes through its center.
    \item The equations governing the spectral singularities are $\cP$-invariant \cite{prl-2009,pra-2011a,prl-2013}.
    \end{enumerate}
These imply that the incoming waves from the left and right of the slab are absorbed provided that they have identical amplitude and phase. In contrast, the optical potential associated with a CPA-laser (with planar symmetry) is never $\cP$-invariant and the amplitude and phase properties of the absorbed waves depend on the details of the corresponding optical potential.

\begin{figure}
    \begin{center}
    \includegraphics[scale=.55]{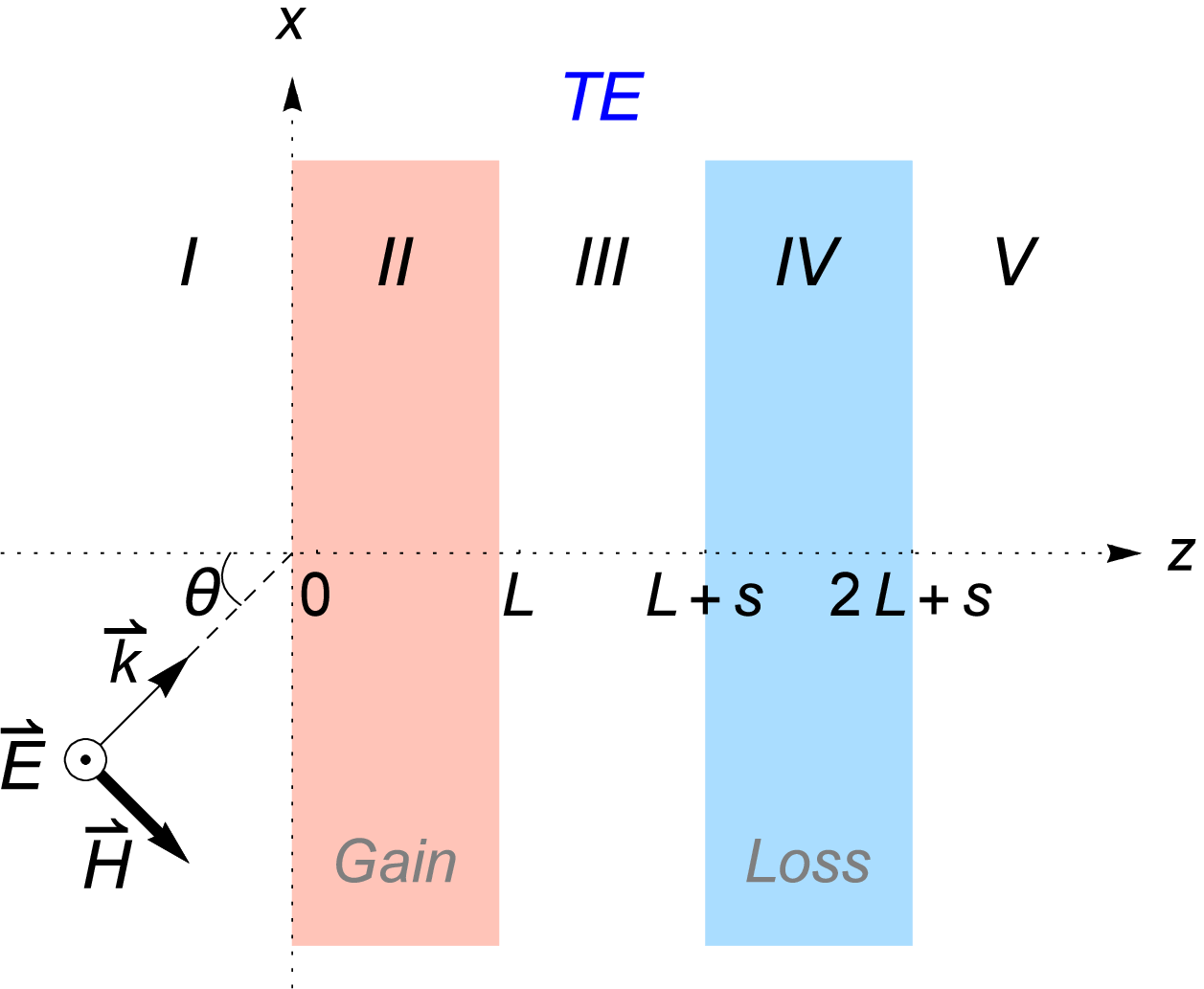}~~~~~~~~
    \includegraphics[scale=.55]{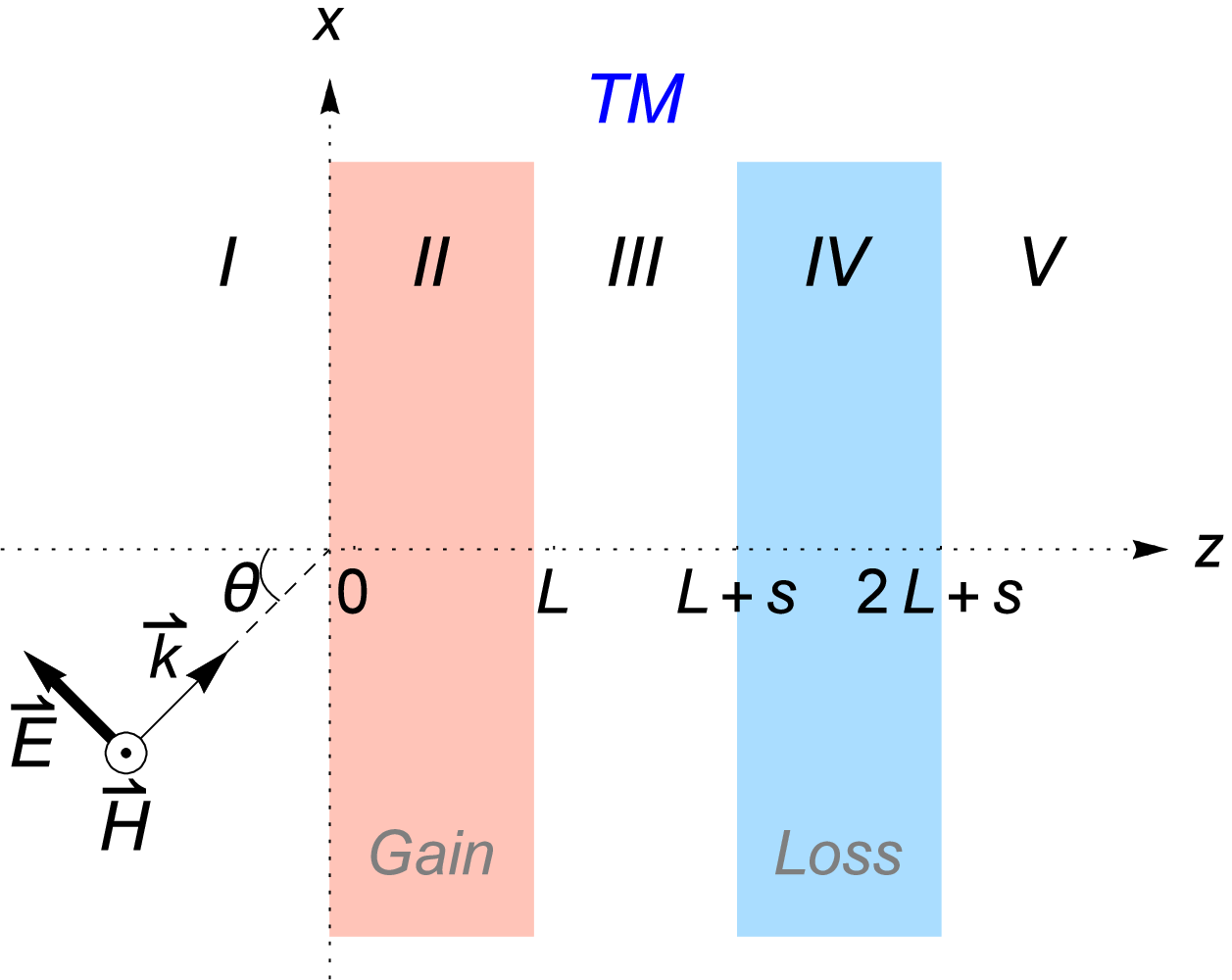}
    \caption{(Color online)  TE (on the left) and TM (on the right) modes of a slab system consisting of a pair of gain and loss layers of thickness $L$ placed a distance $s$ apart in vacuum. The symbols $I$, $I\!I$, $I\! I\! I$, $I\!V$, and $V$ respectively label the regions of the space corresponding to $z<0$, $0<z<L$, $L<z<L+s$, $L+s<z<2L+s$, and $z>2L+s$.}
    \label{fig1}
    \end{center}
    \end{figure}

Recently, we have investigated the behavior of spectral singularities in the oblique TE (transverse electric) and TM (transverse magnetic) modes of an infinite planar slab of homogeneous gain (or lossy) medium and determined the laser threshold and CPA conditions for these modes \cite{pra-2015a}. This revealed a number of unusual phenomena related to the behavior of the energy density and Poynting vector for the TM modes with an incidence angle larger than the Brewster's angle. In the present article we use a similar approach to study a $\cP\cT$-symmetric planar slab system that consists of a pair of balanced gain and loss layers of thickness $L$ separated by a distance $s\geq 0$, as depicted in Fig.~\ref{fig1}.
Unlike the homogeneous slab of Ref.~\cite{pra-2015a}, this system is capable of serving as a CPA-laser. By studying the spectral singularities in its TE and TM modes, we obtain all possible configurations of this system that support CPA-laser action. In order to determine the practically most desirable choices among these, we examine the behavior of the energy density and Poynting vector for the spectrally singular TE and TM waves. We also give a complete solution for the problem of finding the intensity and phase contrasts that make the system absorb the incident waves. This provides valuable information for a possible experimental realization of a slab CPA-laser.

The organization of the article is as follows. In Sec.~\ref{S2}, we determine the explicit form of the TE and TM waves for the general case where the layers consist of arbitrary homogeneous optical material. In Sec.~\ref{S3} we compute the transfer matrix for the system and derive an analytic expression for the condition of the presence of spectral singularities. In Sec.~\ref{S4} we confine our attention to the $\cP\cT$-symmetric case where the system consists of layers with balanced gain and loss. In Sec.~\ref{S5} we investigate the effects of dispersion. In Sec.~\ref{S6} we obtain explicit expressions for the spectrally singular TE and TM waves and examine the behavior of their energy density and the Poynting vector. In Sec.~\ref{S8} we give the amplitude and phase conditions necessary for our $\cP\cT$-symmetric CPA-laser to act as a perfect absorber, and in Sec.~\ref{S9} we present our concluding remarks.

\section{TE and TM Modes of a Parallel Pair of Slabs}
\label{S2}

Consider the system depicted in Fig.~\ref{fig1}. Suppose that the regions $I\!I$ and $I\!V$ are respectively filled with gain and loss material having constant complex refractive indices $\fn_1$ and $\fn_2$. Let $\vec\cE$ and $\vec\cH$ denote the electric and magnetic fields interacting with this system, and
    \be
    \fz(z):=\left\{\begin{array}{ccc}
    \fn_1^2 & {\rm for} & z\in I\!I,\\
    \fn_2^2 & {\rm for} & z\in I\!V,\\
    1 & &{\rm otherwise}.
    \end{array}\right.
    \label{e1}
    \ee
Then Maxwell's equations for the time-harmonic electromagnetic fields, $\vec{\cE}(\vec r,t)=e^{-i\omega t}\vec{E}(\vec r)$ and $\vec{\cH}(\vec r,t)=e^{-i\omega t}\vec{H}(\vec r)$, take the form
    \begin{align}
    &\left[\nabla^{2} +k^2\fz(z)\right] \vec{E}(\vec{r}) = 0, &&
    \vec{H}(\vec{r}) = -\frac{i}{k Z_{0}} \vec{\nabla} \times \vec{E}(\vec{r}),
    \label{equation4}\\
    &\left[\nabla^{2} +k^2\fz(z)\right] \vec{H}(\vec{r}) = 0,&&
    \vec{E}(\vec{r}) = \frac{i Z_{0}}{k \fz(z)} \vec{\nabla} \times \vec{H}(\vec{r}),
    \label{equation5}
    \end{align}
where $\vec r:=(x,y,z)$, $k:=\omega/c$ is the wavenumber, $c:=1/\sqrt{\mu_{0}\varepsilon_{0}}$ is the the speed of light in vacuum, and $Z_{0}:=\sqrt{\mu_{0}/\varepsilon_{0}}$, $\varepsilon_0$, and $\mu_0$ are respectively the impedance, permittivity, and permeability of the vacuum.

The TE and TM waves correspond to the solutions of (\ref{equation4}) and (\ref{equation5}) for which $\vec E(\vec{r})$ and $\vec H(\vec{r})$ are respectively parallel to the surface of the slabs. We use a coordinate system in which they are aligned along the $y$-axis. Suppose that for $z<0$, $\vec E(\vec{r})$ (respectively $\vec H(\vec{r})$) coincides with a plane wave with wavevector $\vec k$ in the $x$-$z$ plane, i.e.,
    \begin{align}
    &\vec k=k_x \hat e_x+ k_z \hat e_z, && k_x:=k\sin\theta, &&k_z:=k\cos\theta,
    \end{align}
where $\hat e_x,\hat e_y,$ and $\hat e_z,$ are respectively the unit vectors along the $x$-, $y$- and $z$-axes, and $\theta$ is the incidence angle (See Fig.~\ref{fig1}.) Then the electric field for the TE waves and the magnetic field for the TM waves are respectively given by
    \begin{align}
    &\vec E(\vec{r})=\sE(z)e^{ik_{x}x}\hat e_y,
    && \vec H(\vec{r})=\sH (z)e^{ik_{x}x}\hat e_y,
    \label{ez1}
    \end{align}
where $\sE$ and $\sH$ are solutions of the Schr\"odinger equation
    \be
    -\psi''(z)+v(z)\psi(z)=k^2\psi(z),~~~~~~~~~~z\notin\{ 0,L,L+s,2L+s\},
    \label{sch-eq}
    \ee
for the potential
    \[v(z):=k^2[1+\sin^2\theta-\fz(z)].\]
Because $v(z)$ is a piecewise constant potential, we can easily solve (\ref{sch-eq}) to obtain
    \be
    \psi(z):=\left\{\begin{array}{ccc}
    a_1\,e^{ik_z z} + b_1\,e^{-ik_z z} & {\rm for} & z\in I,\\
    a_2\,e^{i{\tilde k}_1z} + b_2\,e^{-i{\tilde k}_1z} & {\rm for} & z\in I\!I,\\
    a_3\,e^{ik_z z} + b_3\,e^{-ik_z z} & {\rm for} & z\in I\!I\!I,\\
    a_4\,e^{i{\tilde k}_2z} + b_4\,e^{-i{\tilde k}_2z} & {\rm for} & z\in I\!V,\\
    a_5\, e^{ik_z z} + b_5\, e^{-ik_z z} & {\rm for} & z\in V,
    \end{array}\right.
    \label{E-theta}
    \ee
where $a_i$ and $b_i$, with $i=1,2,3,4,5$, are complex coefficients, and
    \begin{align}
    &{\tilde k}_{j}:=k\sqrt{\fn_{j}^2-\sin^2\theta}=k_{z}\tilde\fn_{j},
    &&\tilde\fn_{j}:=\sec\theta\, \sqrt{\fn_{j}^2 -\sin^2\theta}.
    \label{tilde-parm}
    \end{align}

Substituting (\ref{ez1}) in the second equation in (\ref{equation4}) and (\ref{equation5}), we can find the magnetic field for the TE waves and the electric field for the TM waves inside and outside the slabs. We then impose the appropriate boundary conditions for the problem to relate the coefficients $a_i$ and $b_i$. These amount to the requirement that the tangential components of $\vec E$ and $\vec H$ must be continuous functions of $z$ at $z=0$, $z=L$, $z=L+s$ and $z=2L+s$. Table~\ref{table01} gives explicit expressions for the components of the electric and magnetic fields, and Table~\ref{table02} lists the corresponding boundary conditions.%
    \begin{table}[!htbp]
    \begin{center}
	{
    \begin{tabular}{|c|c|}
    \hline
    TE-Fields & TM-Fields \\
    \hline & \\[-6pt]
    $\begin{aligned}
    & E_{x}=E_{z}=H_{y}=0\\[3pt]
    & E_{y}=\sE(z)\,e^{ik_{x}x}\\[3pt]
    &H_{x} =-\frac{\sF(z)}{Z_0}\,\sT(x,z)\\[3pt]
    &H_{z} =\frac{\sin\theta\, e^{ik_{x}x}\sE(z)}{Z_0}\\[3pt]
    \end{aligned}$ &
    $\begin{aligned}
    & E_{y}=H_{x}=H_{z}=0\\[2pt]
    & E_{x} =\frac{Z_{0}\,\sF(z)}{\fz(z)}\,\sT(x,z)\\[3pt]
    & E_{z} =- \frac{Z_{0}\sin\theta\, e^{ik_{x}x}\sH(z)}{\fz(z)}\\
    & H_{y} =\sH(z)\,e^{ik_{x}x}\\[-8pt]
    &
    \end{aligned}$\\
    \hline
    \end{tabular}}
    \vspace{6pt}
    \caption{Components of the TE and TM fields in cartesian coordinates. Here $\sE(z)$ is given by the right-hand side of (\ref{E-theta}), and $\sF(z)$ and $\sT(x,z)$ are respectively defined by (\ref{F-theta}) and (\ref{T1-2}).}
    \label{table01}
    \end{center}
    \end{table}%
    \begin{table}[!htbp]
    \begin{center}
	{
    \begin{tabular}{|c|c|}
    \hline
    &\\[-10pt]
    $z=0$ &
    $\begin{aligned}
    & a_1 + b_1 = a_2+ b_2, && b_1 - a_1 = \fu_1 (b_2- a_2)\\[3pt]
    \end{aligned}$\\
    \hline
    &\\[-8pt]
    $z=L$ & $\begin{aligned}
    & a_2 e^{i{\tilde k}_1L} + b_2 e^{-i{\tilde k}_1L}= a_3 e^{ik_z L} + b_3 e^{-ik_z L} \\[3pt]
    & \fu_1 (a_2 e^{i{\tilde k}_1L} - b_2 e^{-i{\tilde k}_1L})=
    a_3 e^{i k_z L} - b_3 e^{-i k_z L}
    \end{aligned}$\\[-8pt]
    &\\
    \hline
    &\\[-8pt]
    $z=L+s$ & $\begin{aligned}
    & a_3 e^{i k_z (L+s)} + b_3 e^{-i k_z (L+s)}= a_4 e^{i\tilde{k}_2 (L+s)} + b_4 e^{-i\tilde{k}_2(L+s)} \\[3pt]
    & a_3 e^{ik_z(L+s)} - b_3 e^{-ik_z(L+s)}=
    \fu_2 (a_4 e^{i\tilde{k}_2(L+s)} - b_4 e^{-i\tilde{k}_2(L+s)})
    \end{aligned}$\\[-8pt]
    &\\
    \hline
    &\\[-8pt]
    $z=2L+s$ & $\begin{aligned}
    & a_4 e^{i{\tilde k}_2(2L+s)} + b_4 e^{-i{\tilde k}_2(2L+s)}= a_5 e^{ik_{z}(2L+s)} + b_5 e^{-ik_{z}(2L+s)} \\[3pt]
    & \fu_2 (a_4 e^{i{\tilde k}_2(2L+s)} - b_4 e^{-i{\tilde k}_2(2L+s)})=
    a_5 e^{ik_{z}(2L+s)} - b_5 e^{-ik_{z}(2L+s)}
    \end{aligned}$\\[-8pt]
    &\\
    \hline
    \end{tabular}}
    \vspace{6pt}
    \caption{Boundary conditions for the TE and TM waves.}
    \label{table02}
    \end{center}
    \end{table}
They involve the following quantities.
    \bea
    \sF(z)&:=&\left\{\begin{array}{ccc}
    a_1\,e^{ik_{z}z} - b_1\,e^{-ik_{z}z} & {\rm for} & z\in I,\\
    a_2\,e^{i{\tilde k}_1z} - b_2\,e^{-i{\tilde k}_1z} & {\rm for} & z\in I\!I,\\
    a_3\,e^{ik_{z}z} - b_3\,e^{-ik_{z}z} & {\rm for} & z\in I\!I\!I,\\
    a_4\,e^{i{\tilde k}_2z} - b_4\,e^{-i{\tilde k}_2z} & {\rm for} & z\in I\!V,\\
    a_5\, e^{ik_{z}z} - b_5\, e^{-ik_{z}z} & {\rm for} & z\in V,
    \end{array}\right.
    \label{F-theta}\\[6pt]
    \sT(x,z)&:=&\left\{\begin{array}{cc}
    \sqrt{\fn_{1}^2-\sin^2\theta}\,e^{ik_{x}x}  & {\rm for} ~z\in I\!I,\\
    \sqrt{\fn_{2}^2-\sin^2\theta}\,e^{ik_{x}x}  & {\rm for} ~z\in I\!V,\\
    \cos\theta\, e^{ik_{x}x}  & {\rm otherwise},
    \end{array}\right.
    \label{T1-2}\\[6pt]
    \fu_{j}&:=&\left\{\begin{array}{cc}
    \tilde \fn_{j}=\sec\theta\,\sqrt{\fn_{j}^2 -\sin^2\theta}~~~~~~& \mbox{for TE waves},\\
    \displaystyle\frac{\tilde\fn_{j}}{\fn_{j}^2}=\fn_{j}^{-2}\sec\theta\,\sqrt{\fn_{j}^2 -\sin^2\theta} & \mbox{for TM waves}.
    \end{array}\right.
    \label{u=}
    \eea

\section{Transfer Matrix and Spectral Singularities}
\label{S3}

Transfer matrix formalism provides an effective method of computing the scattering properties of multilayer systems. For the system we consider, the transfer matrix of the slabs placed in regions $I\!I$ and $I\!V$ and the transfer matrix of the whole system are respectively the $2\times 2$ matrices $\bM_1$, $\bM_2$, and $\bM=[M_{ij}]$ satisfying
    \begin{align}
    &\left[\begin{array}{c}
    a_3\\ b_3\end{array}\right]=\bM_1 \left[\begin{array}{c}
    a_1\\ b_1\end{array}\right],
    &&\left[\begin{array}{c}
    a_5\\ b_5\end{array}\right]=\bM_2 \left[\begin{array}{c}
    a_3\\ b_3\end{array}\right],
    &&\left[\begin{array}{c}
    a_5\\ b_5\end{array}\right]=\bM \left[\begin{array}{c}
    a_1\\ b_1\end{array}\right].
    \nn
    \end{align}
These in particular imply the well-known composition relation,
    \be
    \bM=\bM_2\bM_1.
    \label{comp}
    \ee

Ref.~\cite{pra-2015a} gives an explicit expression for $\bM_1$. We can easily compute $\bM_2$ using this expression and the transformation property of the transfer matrices under translations, $z\stackrel{T_a}{\longrightarrow}z-a$. In terms of the entries $M_{ij}$ of a generic transfer matrix, this  takes the form \cite{pra-2014b}:
    \begin{align}
    &M_{11}\stackrel{T_a}{\longrightarrow}M_{11},
    &&M_{12}\stackrel{T_a}{\longrightarrow}e^{-2iak_z}M_{12},
    &&M_{21}\stackrel{T_a}{\longrightarrow}e^{2iak_z}M_{21},
    &&M_{22}\stackrel{T_a}{\longrightarrow}M_{22}.\nn
    \end{align}
Having $\bM_1$ and $\bM_2$ computed, we can determine $\bM$ using (\ref{comp}). Here we only give the expression for $M_{22}$.
    \bea
    M_{22} &=&\cos\fa_1\cos\fa_2\Big[1 - i\,\mathfrak{u}^{+}_1 \tan\fa_1- i\,\mathfrak{u}^{+}_2  \tan\fa_2+ \nn\\
    &&(\fu_1^-\fu_2^-e^{2ik_z s}-\fu_1^+\fu_2^+)\tan\fa_1 \tan\fa_2 \Big] e^{2ik_zL},
    \label{M22=x}
    \eea
where we have introduced
    \begin{align}
    &\fa_j:=k_zL\tilde\fn_j,
    &&\fu_j^\pm:=\frac{1}{2}\left(\fu_{j} \pm \fu^{-1}_{j}\right).\nn
    \end{align}

As noted in Refs.~\cite{jpa-2009,prl-2009}, spectral singularities correspond to the real values of the wavenumber $k$ for which $M_{22}=0$. In light of (\ref{M22=x}), we can express this equation in the form
    \be
    e^{2i\fa_2} = \frac{(\fu_1^2-1)(\fu_2^2-1)e^{2ik_zs}(e^{2i\fa_1}-1)+(\fu_2+1)^2[(\fu_1+1)^2-(\fu_1-1)^2
    e^{2i\fa_1}]}{(\fu_1^2-1)(\fu_2^2-1)e^{2ik_zs}(e^{2i\fa_1}-1)+
    (\fu_2-1)^2[(\fu_1+1)^2-(\fu_1-1)^2e^{2i\fa_1}]}.
    \label{spectsingularitywithvacuum}
    \ee
For a bilayer slab, $s=0$, and this relation takes the following simpler form
    \be
    e^{2i\fa_2}= \left(\frac{\fu_2+1}{\fu_2-1}\right)\left[
    \frac{e^{2i\fa_1}(\fu_1-1)(\fu_2-\fu_1)+
    (\fu_1+1)(\fu_1+\fu_2)}{e^{2i\fa_1}(\fu_1-1)(\fu_1+\fu_2)+(\fu_1+1)(\fu_2-\fu_1)}\right].
    \label{spectsingularity}
    \ee
For $\fn_1=1$ (similarly $\fn_2 =1$), this relation reduces to the condition for the presence of a spectral singularity in the TE and TM modes of a homogeneous slab \cite{pra-2015a}. For normally incident waves, where $\theta=0$, it reproduces the results of Ref.~\cite{jpa-2012}.

\section{$\cP\cT$-Symmetric Configurations}
\label{S4}

Consider the situation that the gain and loss components of our system balance one another, i.e., $\fn_2^*=\fn_1=:\fn$. Then it is $\cP\cT$-symmetric, and we have
    \begin{align}
    &\tilde\fn_2^*=\tilde\fn_1=:\tilde\fn,
    &&\fu_2^*=\fu_1=:\fu,
    &&\fa_2^*=\fa_1=:\fa.
    \label{eq251}
    \end{align}
First we examine the case of a $\cP\cT$-symmetric bilayer slab with adjacent gain and loss layers, i.e., set $s=0$. Then in view of (\ref{u=}) and (\ref{eq251}), Eq.~(\ref{spectsingularity}) takes the form
    \be
    e^{2i\fa^\ast} =      \left(\frac{\tilde{\fn}^{\ast}+\fn^{\ast\ell}}{\tilde{\fn}^{\ast}-\fn^{\ast\ell}}\right)
    \left[
    \frac{e^{2i\fa}(\tilde{\fn}^\ast \fn^{\ell}-\tilde{\fn}\,\fn^{\ast\ell})(\tilde{\fn}-\fn^{\ell})
    +(\tilde{\fn}\,\fn^{\ast\ell}+\tilde{\fn}^\ast \fn^{\ell})(\fn^{\ell}+\tilde{\fn})}{
    e^{2i\fa}
    (\tilde{\fn}^\ast\fn^{\ell}+\tilde{\fn}\,\fn^{\ast\ell})(\tilde{\fn}-\fn^{\ell})
    +(\tilde{\fn}^\ast \fn^{\ell}-\tilde{\fn}\,\fn^{\ast\ell})(\fn^{\ell}+\tilde{\fn})}\right],
    \label{spectsingularity2}
    \ee
where
    \[\ell:=\left\{\begin{array}{ccc}
    0 & {\rm for~TE~waves},\\
    2 & {\rm for~TM~waves}.\end{array}\right.\]

In order to clarify the physical meaning of (\ref{spectsingularity2}), we expand both sides of this relation in powers of the imaginary part $\kappa$ of $\fn$ and ignore quadratic and higher order terms. This yields a reliable approximation provided that
    \be
    |\kappa|\ll \eta-1<\eta,
    \label{condi-pert}
    \ee
where $\eta$ stands for the real part of $\fn$, so that
    \be
    \fn = \eta + i \kappa.
    \label{eta-kappa=}
    \ee
Similarly, we use $\tilde\eta$ and $\tilde\kappa$ to denote the real and imaginary parts of $\tilde\fn$, respectively. To leading order in $\kappa$, we have
    \begin{align}
    &\tilde{\eta}\approx\sec\theta\sqrt{\eta^2 -\sin^2\theta}, && \tilde{\kappa}\approx\frac{\sec\theta\,\eta\,\kappa}{\sqrt{\eta^2-\sin^2\theta}}.
    \label{approx-001}
    \end{align}
Substituting $\tilde\fn=\tilde\eta+i\tilde\kappa$ in (\ref{spectsingularity2}), making use of (\ref{approx-001}), and keeping the leading order terms in $\kappa$ we can reduce (\ref{spectsingularity2}) to the following pair of real equations.
    \begin{align}
    g&\approx \frac{\sqrt{\eta^2-\sin^2\theta}}{\eta L}\ln\left|\frac{2\eta^{\ell}\tilde{\eta}
    (\eta^{2\ell}+\tilde{\eta}^2)}{\sigma_\ell\,\tilde\kappa(\eta^{2\ell}-\tilde{\eta}^2 )}\right|,
    \label{gainTE}\\
    \lambda &\approx\frac{8L\sqrt{\eta^2-\sin^2\theta}}{2m+1}.
    \label{lamdaTE}
    \end{align}
Here $g$ is the gain coefficient given by
    \be
    g:=-2k\kappa=-\frac{4\pi\kappa}{\lambda},
    \label{gain-def}
    \ee
$\lambda:=2\pi/k$ is the wavelength, $m=1,2,3,\cdots$ is a mode number, and
    \[\sigma_\ell:=
    \left\{\begin{array}{ccc}
    1&{\rm for}&\ell=0,\\
    2\sin^2\theta-\eta^2&{\rm for}&\ell=2.
    \end{array}\right.\]

Equations (\ref{gainTE}) and (\ref{lamdaTE}) provide the laser threshold and phase conditions in the TE and TM modes of our $\cP\cT$-symmetric bilayer slab provided that it is made out of typical optical material satisfying (\ref{condi-pert}).

\begin{figure}
	\begin{center}
    \includegraphics[scale=0.6]{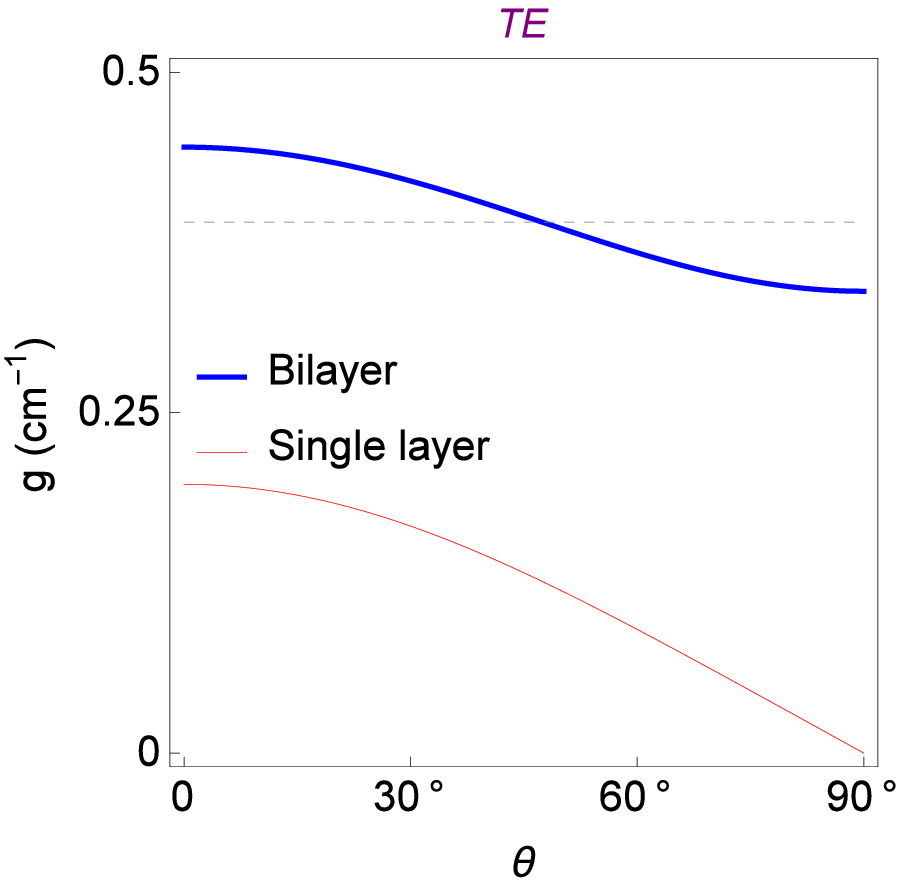}~~~~~~~
    \includegraphics[scale=0.6]{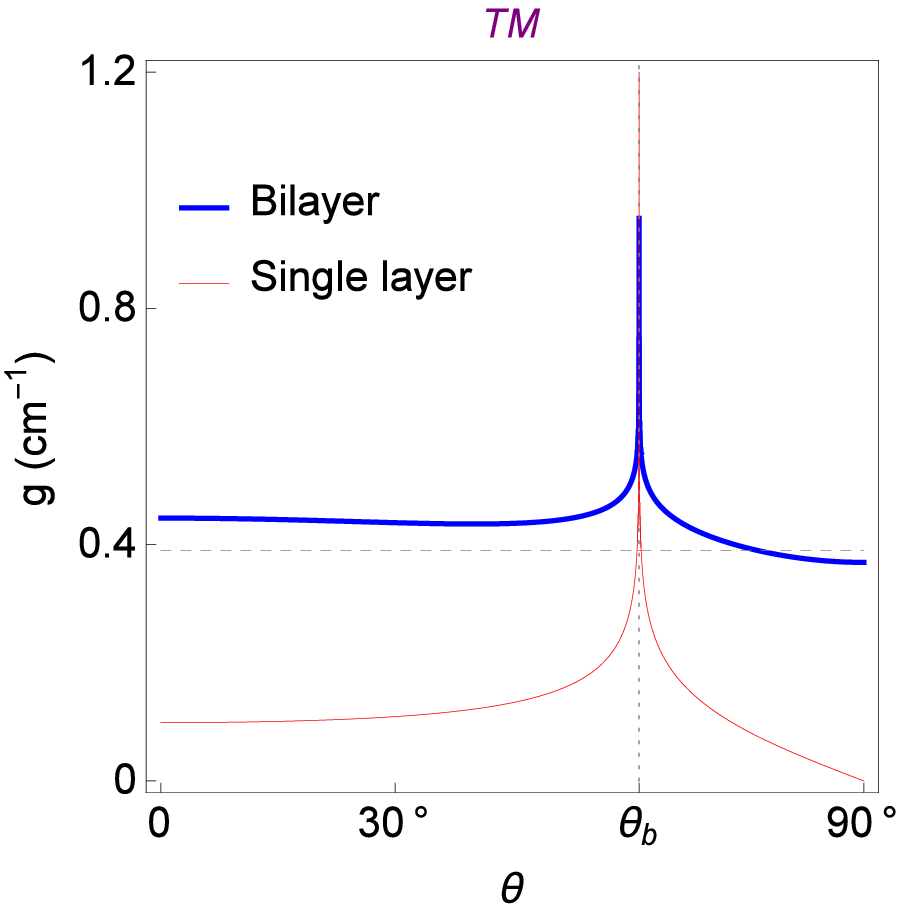}
	\caption{(Color online) Plots of the threshold gain coefficient $g$ as a function of the incidence angle $\theta$ for the TE and TM modes of a homogenous active slab of thickness $25~{\rm cm}$ (thin red curves) and a $\cP\cT$-symmetric bilayer slab of layer thickness $25~{\rm cm}$ (thick blue curves) made out of Nd:YAG crystals with $\eta = 1.8217$. $\theta_b$ labels the Brewster's angle and has the value $61.24^\circ$. The dotted horizontal line marks the experimental upper bound on the gain coefficient.}
    \label{g0TE}
    \end{center}
    \end{figure}

Figure \ref{g0TE} shows the graphs of the threshold gain coefficient $g$ as a function of $\theta$ for a homogeneous slab of gain material and a $\cP\cT$-symmetric bilayer slab made out of Nd:YAG crystals. The main difference is that the threshold gain coefficient for the latter does not tend to zero at the grazing angle ($\theta=90^{\circ}$).

Comparing the phase condition (\ref{lamdaTE}) for the $\cP\cT$-symmetric bilayer with that of a homogeneous slab of thickness $2L$, we find that the value of $\lambda$ for the latter is given by (\ref{lamdaTE}) if we change $m$ to $m-1/2$ (see \cite{pra-2015a}). Given that typically $m$ takes very large values, the wavelength $\lambda$ is essentially the same for the two slabs.

Next, we consider the general case, where the gain and loss components of our system are placed at a distance $s$ apart. A similar analysis of (\ref{spectsingularitywithvacuum}) leads to the following expressions for the threshold gain coefficient and wavelength.
    \begin{align}
    &g \approx \frac{\sqrt{\eta^2-\sin^2\theta}}{2\eta L}\ln\left[\frac{1-\mathcal{A}_{\ell} - \sqrt{1-2\mathcal{A}_{\ell}}}{\mathcal{A}_{\ell}-\mathcal{B}_{\ell}}\right],
    \label{g=s}
    \\[6pt]
    &\lambda\approx\frac{4L\sqrt{\eta^2-\sin^2\theta}}{m + \mathcal{C}_{\ell}},
    \label{lambda=s}
    \end{align}
    where
        \be
        \mathcal{A}_{\ell}:=
        \frac{\sin^4(k_z s) +\tau^+_\ell\sin^2(2k_z s)}{2[\sin^2(k_z s) + 4\tau^-_\ell]^2},~~~~~
        \mathcal{B}_{\ell}:= \frac{(\tilde{\eta}^2 -\eta^{2\ell})\tilde{\kappa}\sigma_{\ell}\sin(2k_z s)}{(\tilde{\eta}^2 +\eta^{2\ell})^2[\sin^2(k_z s) + 4\tau^-_\ell]},~~~~~\tau^\pm_\ell:=\frac{\tilde{\eta}^2 \eta^{2\ell}}{(\tilde{\eta}^2 \pm \eta^{2\ell})^2},\nn
    \ee
    \be
    \mathcal{C}_{\ell}:=
    \frac{1}{\pi}\arccos\left\{\left[
    \frac{(1-\mathcal{B}_{\ell}-\sqrt{1-2\mathcal{A}_{\ell}})\sin^2(k_z s)}{
    2\sqrt{(\mathcal{A}_{\ell}-\mathcal{B}_{\ell})
    (1-\mathcal{A}_{\ell}-\sqrt{1-2\mathcal{A}_{\ell}})}\,[\sin^2(k_z s) + 4\tau_\ell^-]}\right]\right\}.\nn
    \ee
We have checked that in the limit $s\rightarrow 0$, Eqs.~(\ref{g=s}) and (\ref{lambda=s}) do actually tend to Eqs.~(\ref{gainTE}) and (\ref{lamdaTE}).

\begin{figure}
	\begin{center}
     \includegraphics[scale=0.65]{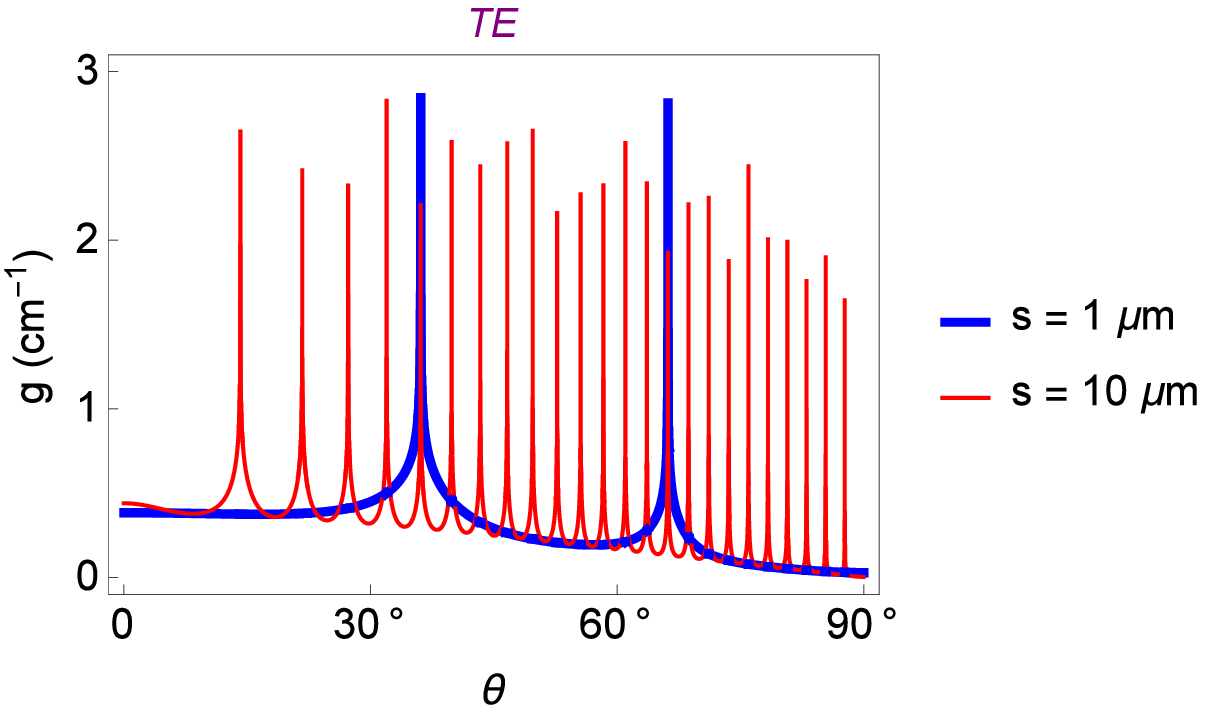}
     ~~~\includegraphics[scale=0.65]{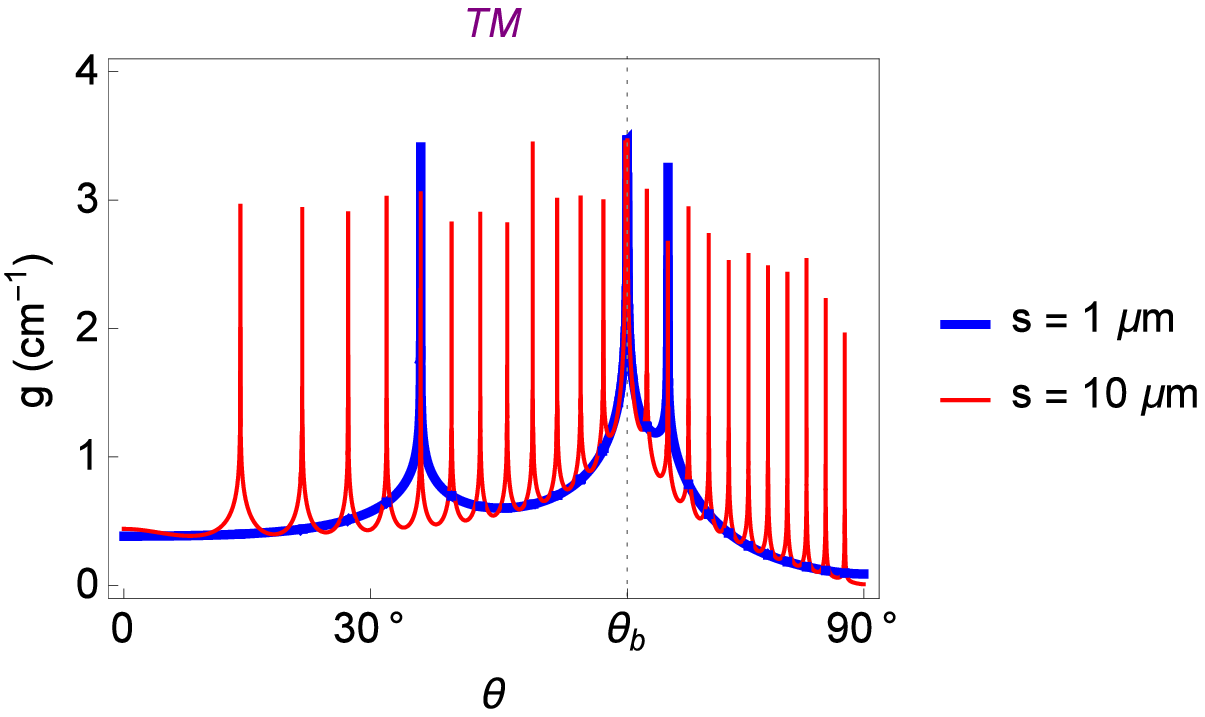}
	\caption{(Color online) Plots of gain coefficient $g$ as a function of the incidence angle $\theta$ for the TE and TM modes of a $\cP\cT$-symmetric two-slab system made out of Nd:YAG crystals with $\eta = 1.8217$, $L=25~{\rm mm}$, and separation distances  $s=1~\mu{\rm m}$ (thick blue curves) and $s=10~\mu{\rm m}$ (thin red curves). For larger values of $s$ there are more peaks, but the general behavior of $g$ does not change.}
    \label{g0TE2}
    \end{center}
    \end{figure}

\begin{figure}
    \begin{center}
    \includegraphics[scale=.6]{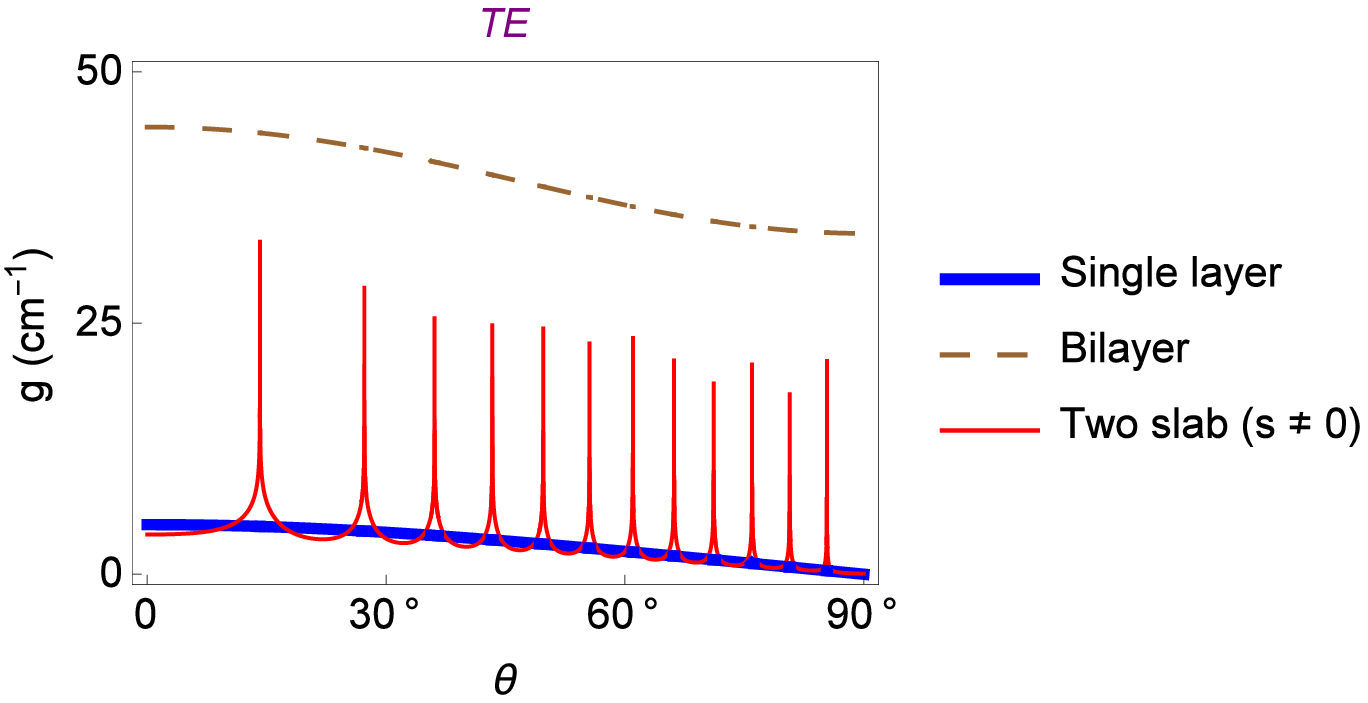}
    \includegraphics[scale=.6]{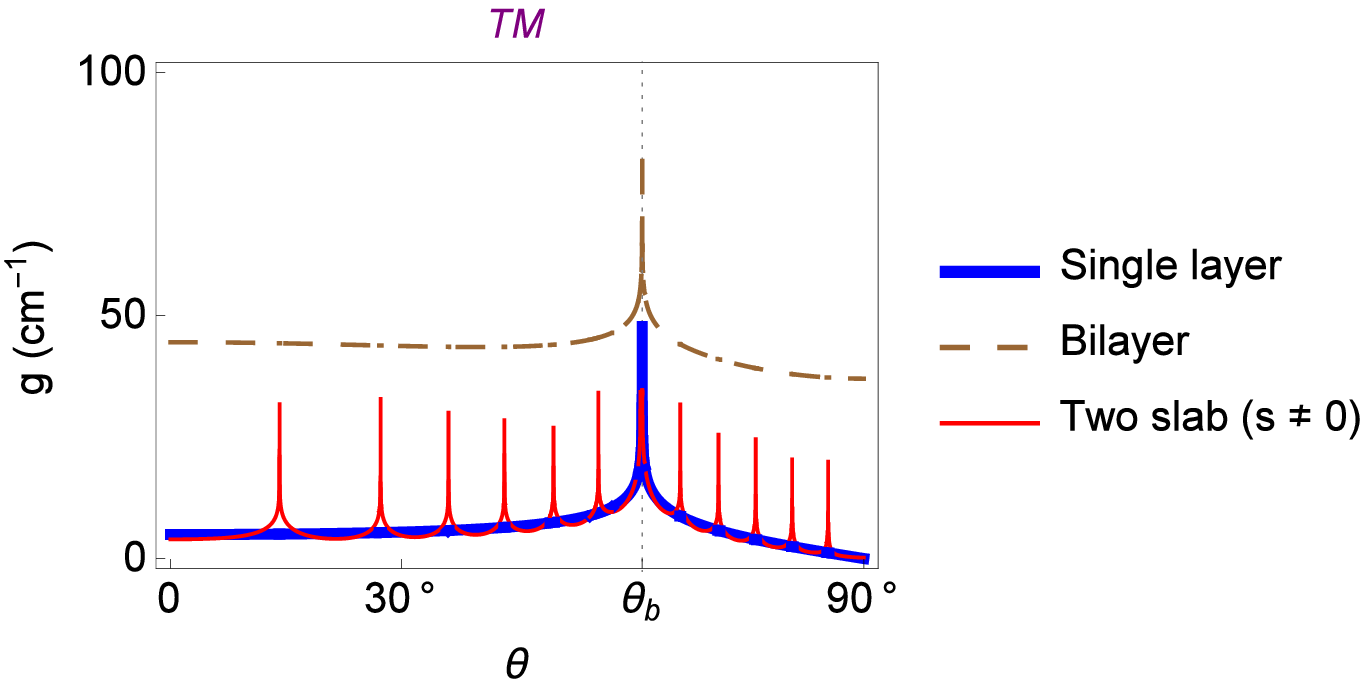}
	\caption{(Color online) Plots of the threshold gain coefficient $g$ as a function of the incidence angle $\theta$ for single layer gain slab (thick blue curves), a $\cP\cT$-symmetric bilayer slab (dashed curves), and a $\cP\cT$-symmetric two-slab system (thin red curves) all made of Nd:YAG crystals. We have taken the thickness of the single layer slab to be $5~ {\rm mm}$, while for the bilayer slab and the two-slab system we have set $L=2.5~{\rm mm}$. For the latter we have taken $s=5~\mu{\rm m}$.}
    \label{g0theta2}
    \end{center}
    \end{figure}

Figs.~\ref{g0TE2} and \ref{g0theta2} show the graphs of the threshold gain coefficient $g$ as a function of $\theta$ for the TE and TM modes of a $\cP\cT$-symmetric two-slab system for different separation distances $s$. The peaks are consequences of internal reflections in the gap between the two slabs. Their number is an increasing function of $s$. For $\theta\to 90^\circ$, the presence of the gap changes the behavior of $g$ drastically. Unlike for the case $s=0$ and similarly to the single layer slab considered in Ref.~\cite{pra-2015a}, here $g$ tends to zero as $\theta\to 90^\circ$.

\begin{figure}
	\begin{center}
    \includegraphics[scale=.65]{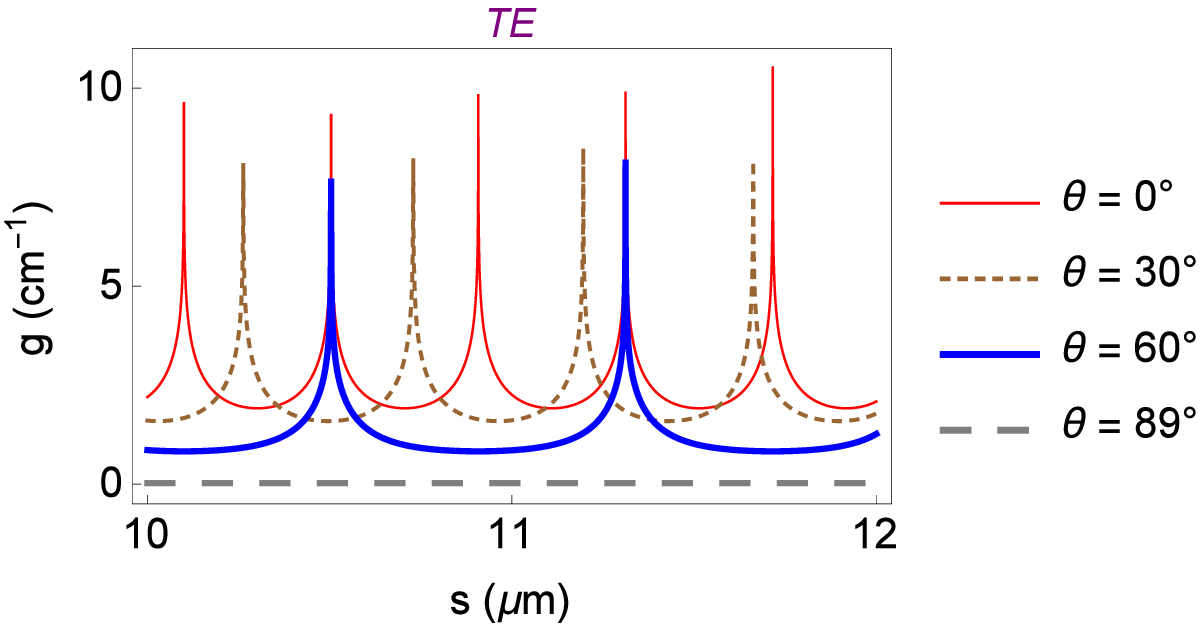}
    \includegraphics[scale=.65]{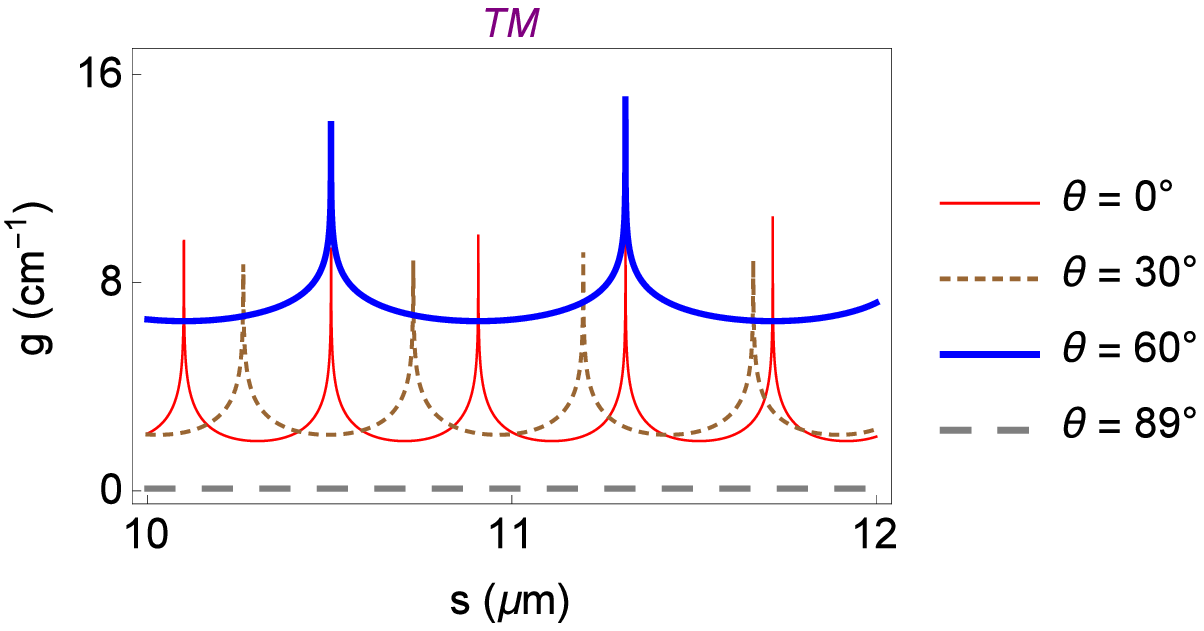}
 	\caption{(Color online) Plots of the threshold gain coefficient $g$ as a function of the separation distance $s$ for the TE and TM modes of a $\cP\cT$-symmetric two-slab system made out of Nd:YAG crystals with $\eta = 1.8217$ and $L=5~\textrm{mm}$ for various incidence angles $\theta$. The number of peaks is a decreasing function of $\theta$.}
    \label{g0theta2n}
    \end{center}
    \end{figure}

Fig.~\ref{g0theta2n} shows the plots of the threshold gain coefficient $g$ as a function of the separation distance $s$ for different values of the incidence angle.

\section{Inclusion of the Effects of Dispersion}
\label{S5}

In the preceding section we have ignored the fact that the index of refraction depends on the wavenumber $k$. In this section we consider the influence of the $k$-dependence of $\fn$ on the spectral singularities in the TE and TM modes of our $\cP\cT$-symmetric two-slab system.

Suppose that the active material filling the left-hand (gain) layer is obtained by doping a host medium of refraction index $n_0$ and its refractive index satisfies the dispersion relation
    \be
    \fn^2= n_0^2-
    \frac{\hat\omega_p^2}{\hat\omega^2-1+i\hat\gamma\,\hat\omega},
    \label{epsilon}
    \ee
where $\hat\omega:=\omega/\omega_0$, $\hat\gamma:=\gamma/\omega_0$, $\hat\omega_p:=\omega_p/\omega_0$, $\omega_0$ is the resonance frequency, $\gamma$ is the damping coefficient, and $\omega_p$ is the plasma frequency. We can express $\hat\omega_p^2$ in terms of the imaginary part $\kappa_0$ of $\fn$ at the resonance wavelength $\lambda_0:=2\pi c/\omega_0$ according to $\hat\omega_p^2=2n_0\hat\gamma\kappa_0+\cO(\kappa_0^2)$, where $\cO(\kappa_0^2)$ stands for the quadratic and higher order terms in $\kappa_0$, \cite{pra-2011a}.
Substituting this equation in (\ref{epsilon}), using (\ref{eta-kappa=}), and neglecting quadratic and higher order terms in $\kappa_0$, we obtain \cite{pla-2011}
     \begin{align}
    &\eta\approx n_0+\frac{\kappa_0\hat\gamma(1-\hat\omega^2)}{(1-\hat\omega^2)^2+
    \hat\gamma^2\hat\omega^2},
    &&\kappa\approx\frac{\kappa_0\hat\gamma^2\hat\omega}{(1-\hat\omega^2)^2+
    \hat\gamma^2\hat\omega^2}.
    \label{eqz301}
    \end{align}

\begin{figure}
    \begin{center}
    \includegraphics[scale=.50]{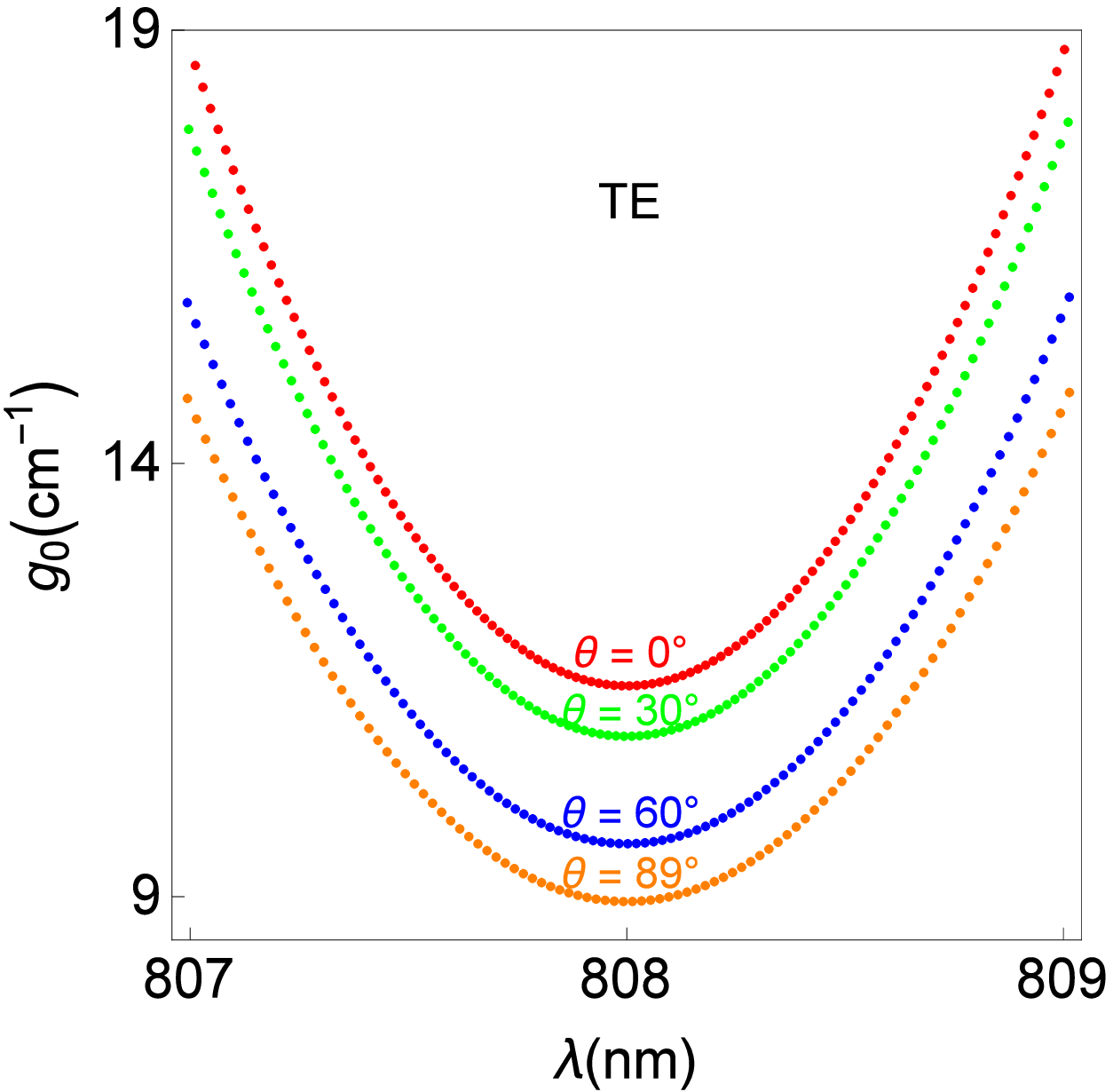}~~~~~~~~~~
    \includegraphics[scale=.50]{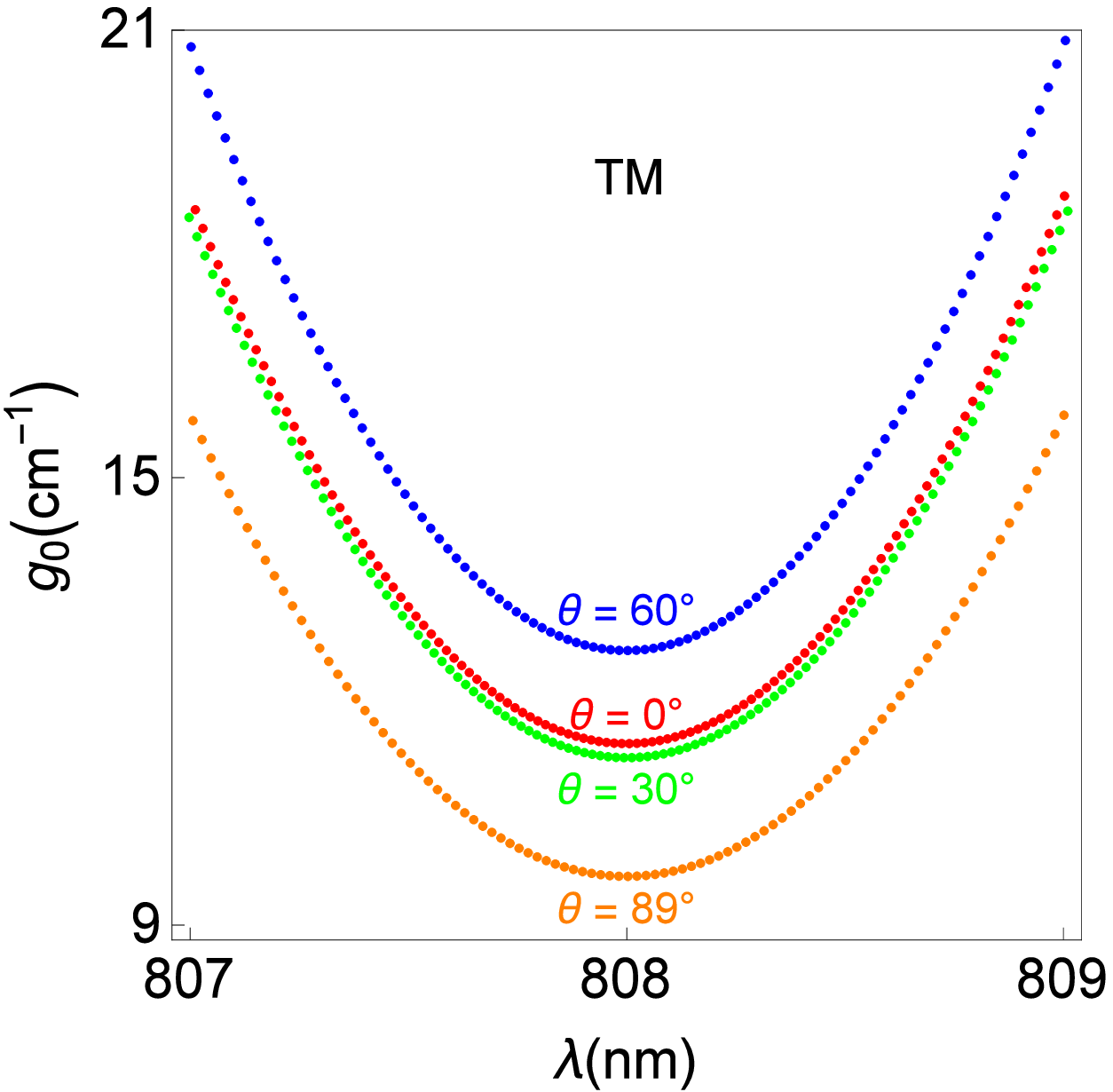}
    \caption{(Color online) The threshold gain $g_0$ and the wavelength $\lambda$ for the spectral singularities in the TE and TM modes of a $\cP\cT$-symmetric bilayer slab of layer thickness $L=1~ \textrm{cm}$ made out of the gain medium (\ref{specific}) for different incidence angles $\theta$. The minimum value of $g_0$ is obtained for the spectral singularities at the resonance wavelength. The threshold gain coefficient takes smaller values for larger incidence angles.}
    \label{wavelengthg0TE}
    \end{center}
    \end{figure}

\begin{figure}
    \begin{center}
    \includegraphics[scale=.55]{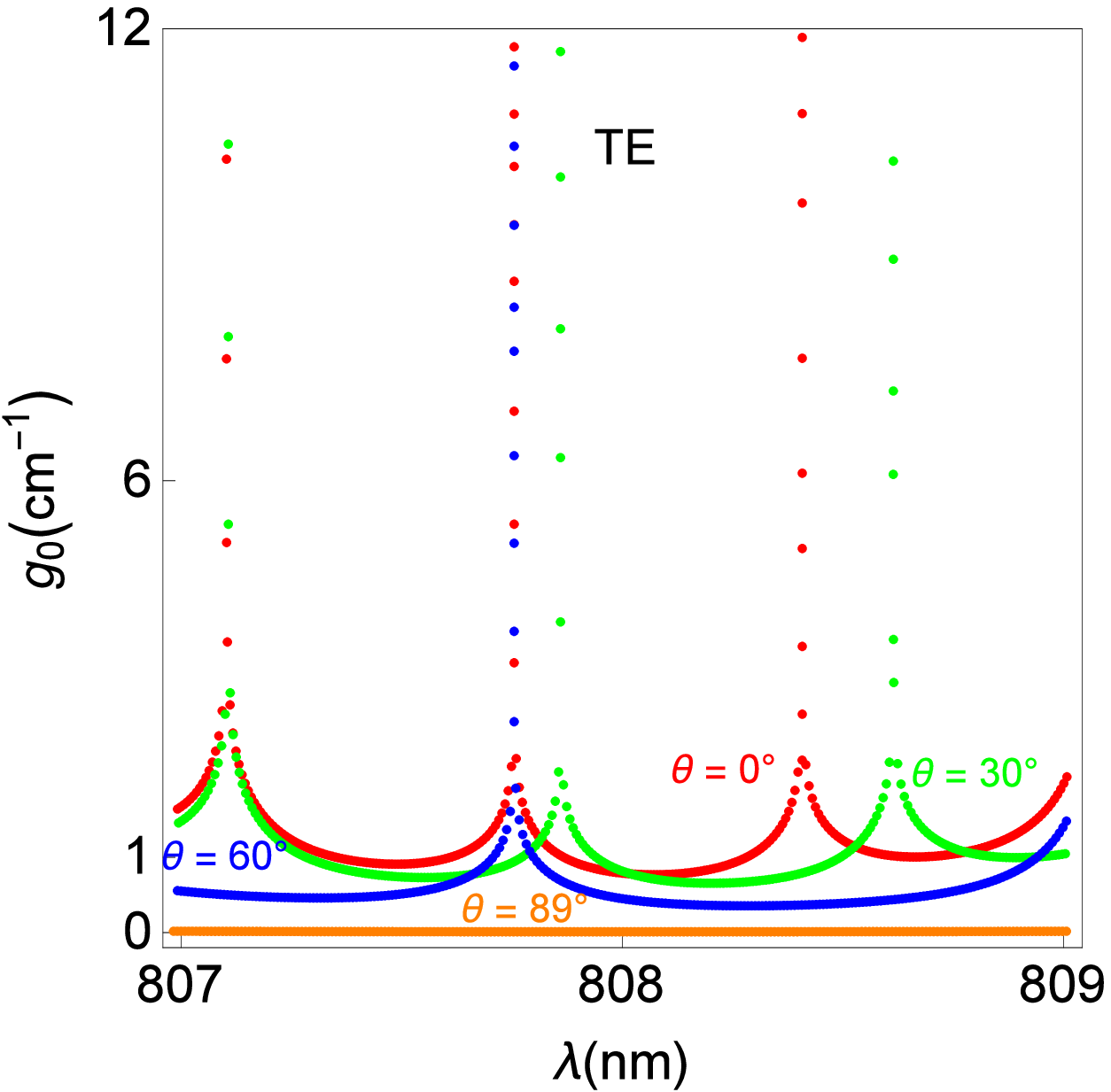}~~~~~~
    \includegraphics[scale=.55]{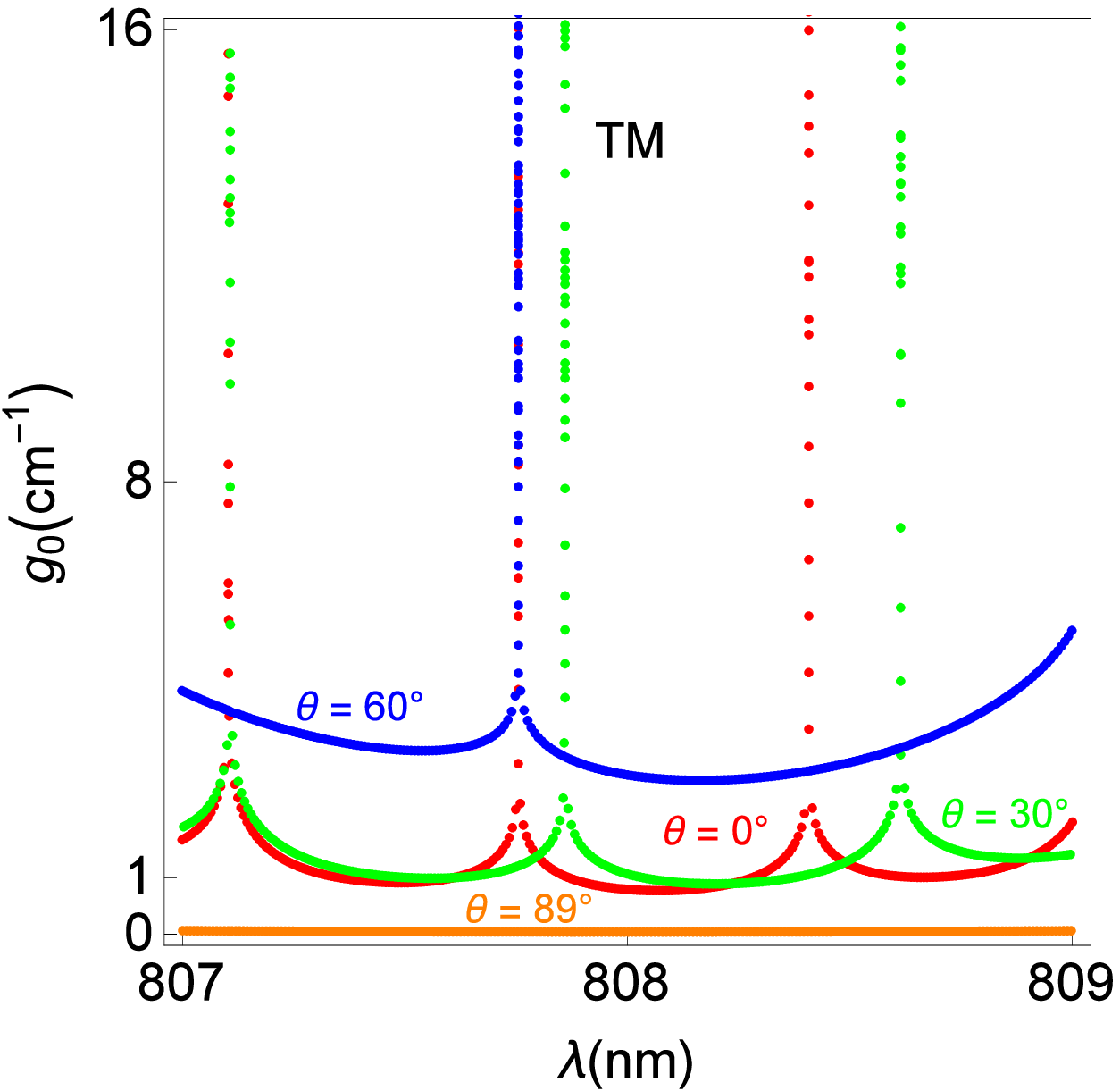}
    \caption{(Color online) The threshold gain $g_0$ and the wavelength $\lambda$ for the spectral singularities in the TE and TM modes of a $\cP\cT$-symmetric two-slab system with layer thickness $L=2.5~ \textrm{cm}$ and separation distance $s=0.5~ \textrm{mm}$ made out of the gain medium (\ref{specific}) for different incidence angles $\theta$. There are certain wavelengths corresponding to extremely large values of $g_0$ where lasing is impossible. The number of these increases with the separation distance. As the incidence angle increases, $g_0$ decreases.}
    \label{wavelengthg0TEspace}
    \end{center}
    \end{figure}

At a resonance wavelength, (\ref{gain-def}) reads $\kappa_0=-\lambda_{0}g_0/4\pi$. Inserting this relation in (\ref{eqz301}) and making use of (\ref{eta-kappa=}), (\ref{spectsingularitywithvacuum}) and (\ref{spectsingularity}), we can determine the $\lambda$ and $g_{0}$ values for the spectral singularities. Figures~\ref{wavelengthg0TE} and \ref{wavelengthg0TEspace} respectively show the location of the spectral singularities in the $\lambda$-$g_0$ plane for a $\cP\cT$-symmetric bilayer with layer thickness $L=1\,{\rm cm}$ and a $\cP\cT$-symmetric two-slab system with layer thickness $L=2.5\,{\rm cm}$ and the separation distance $s=0.5\,{\rm mm}$. Both of these are made of Nd:YAG crystals with the following specifications \cite{silfvast}:
    \begin{align}
    &n_0=1.8217, &&\lambda_1=808\,{\rm nm}, &&
    \hat\gamma=0.003094.
    \label{specific}
    \end{align}

\section{Spectrally Singular TE and TM Waves}
\label{S6}

Because spectral singularities are associated with the singularities of the reflection and transmission amplitudes \cite{prl-2009}, they correspond to solutions of the wave equation with purely outgoing boundary conditions. For the system we consider, these are characterized by
    \be
    a_1=b_5=0.
    \label{zq1}
    \ee
In this section we examine the TE and TM modes of our $\cP\cT$-symmetric slab system that fulfill
this condition. Following Ref.~\cite{pra-2015a} we refer to them as `singular waves' or `singular modes'.

We begin our study of these modes by enforcing Eqs.~(\ref{spectsingularitywithvacuum}) and (\ref{zq1}), and the boundary conditions given in Table~\ref{table02}. This gives
    \be
    a_5=V_s(\tilde{\fn}_1,\tilde{\fn}_2, L)\,e^{-ik_z(2L+s)}~b_1,
    \label{zq2}
    \ee
where for each $z\in\R$,
    \begin{align}
    V_{s}(\tilde{\fn}_1, \tilde{\fn}_2, z) &:= U_1 (\tilde{\fn}_1, z) e^{i k_z s} +
    \left(\frac{\fu_2 - 1}{\fu_2 + 1}\right)U_2 (\tilde{\fn}_1, z) e^{-ik_z s},
    \label{Vs-def}\\
    U_{1}(\tilde{\fn}, z) &:=\frac{1}{2}\left[U_{+}(\tilde{\fn}, z)+\fu\,U_{-}(\tilde{\fn}, z)\right],\\
    U_{2}(\tilde{\fn}, z) &:=\frac{1}{2} (\fu+1)\,U_{+}(\tilde{\fn}, z),\\
    U_{\pm}(\tilde{\fn}, z) &:= \frac{1}{2\fu} \left[(\fu-1)\,e^{ik_z z \tilde{\fn}}\pm (\fu +1)\,e^{-ik_z z \tilde{\fn}}\right].
    \label{Upm-def}
    \end{align}

Next we substitute (\ref{zq1}) in the formulas given in Table~\ref{table01} to obtain the explicit form of the singular TE and TM waves. Table~\ref{table03} gives the result of this calculation in terms of the functions:
    \begin{align}
    &F^{s}_\pm(\tilde{\fn}_1,\tilde{\fn}_2,z):=\left\{\begin{array}{ccc}
    \pm e^{-ik_z z} &{\rm for} & z\in I,\\[6pt]
    \fu_1^{(1\mp 1)/2}
    U_\pm(\tilde{\fn}_1, z) &{\rm for} & z\in I\!I,\\[6pt]
    V_\pm(\tilde{\fn}_1, z-L) &{\rm for} & z\in I\!I\!I,\\[6pt]
    \fu_2^{(1\mp 1)/2}
    U_\pm(\tilde{\fn}_2, 2L+s-z)\,V_{s}(\tilde{\fn}_1, \tilde{\fn}_2, L) &{\rm for} & z\in I\!V,\\[6pt]
    V_{s}(\tilde{\fn}_1, \tilde{\fn}_2, L) \,e^{ik_z [z-2L-s]} &{\rm for} & z\in V,\end{array}\right.
    \label{FFG-all}
    \end{align}
where
    \begin{align}
    V_{\pm}(\tilde{\fn}, z) &:= U_1 (\tilde{\fn}, L) e^{ik_z z} \pm U_2 (\tilde{\fn}, L) e^{-i k_z z}.
    \label{V-pm=}
    \end{align}
Note that Region $I\!I\!I$ is absent for $s=0$.
    \begin{table}
    \begin{center}
	{
    \begin{tabular}{|c|c|}
    \hline
    Spectrally Singular TE-Fields & Spectrally Singular TM-Fields \\
    \hline & \\[-6pt]
    $\begin{aligned}
    & E_{x}=E_{z}=H_{y}=0\\[3pt]
    & E_{y}=b_1 e^{ik_x x}F^{s}_+( \tilde{\fn}_1,\tilde{\fn}_2,z)\\[0pt]
    &H_{x} =-\frac{b_1\cos\theta}{Z_0}\, e^{ik_x x} F^{s}_-( \tilde{\fn}_1,\tilde{\fn}_2,z)\\[0pt]
    &H_{z} =\frac{b_1\sin\theta}{Z_0}\, e^{ik_x x} F^{s}_+( \tilde{\fn}_1,\tilde{\fn}_2,z)\\[3pt]
    \end{aligned}$ &
    $\begin{aligned}
    & E_{y}=H_{x}=H_{z}=0\\[0pt]
    & E_{x} =b_1 Z_0\cos\theta\,  e^{ik_x x} F^{s}_-( \tilde{\fn}_1,\tilde{\fn}_2,z)\\[0pt]
    & E_{z} =- b_1 Z_0\sin\theta \, e^{ik_x x}\frac{F^{s}_+(\tilde{\fn}_1,\tilde{\fn}_2,z)}{\fz(z)}\\[0pt]
    & H_{y} =b_1 e^{ik_x x} F^{s}_+(\tilde{\fn}_1,\tilde{\fn}_2,z)\\[-8pt]
    &
    \end{aligned}$\\
    \hline
    \end{tabular}}
    \vspace{6pt}
    \caption{Components of the spectrally singular TE and TM fields in cartesian coordinates. Here $\tilde\fn_j$ and $F^{s}_\pm(\tilde{\fn}_1,\tilde{\fn}_2,z)$ are respectively defined by (\ref{tilde-parm}) and (\ref{FFG-all}).}
    \label{table03}
    \end{center}
    \end{table}%

In order to acquire a better understanding of the behavior of singular waves, we examine the behavior of the corresponding time-averaged energy density and Poynting vector, $\br u\kt $ and $\langle\vec{S}\rangle$. Because the explicit expression for these quantities are rather complicated, in what follows we provide a graphical demonstration of their consequences and give their derivation in the appendix.

For convenience we respectively express the numerical values of $\br u\kt$ and $|\langle\vec{S}\rangle|$ in units of
    \begin{align}
    &\br u_I\kt:=\frac{|b_1|^2}{2}\times \left\{\begin{array}{cc}
    \epsilon_0 &{\rm for~TE~waves},\\
    \mu_0 &{\rm for~TM~waves},\end{array}\right.
    && |\br \vec S_I\kt|:=\frac{|b_1|^2}{2}\times\left\{\begin{array}{cc}
    Z_0^{-1} &{\rm for~TE~waves},\\
    Z_0 &{\rm for~TM~waves}.\end{array}\right.
    \label{units}
    \end{align}
These are the values of $\br u\kt$ and $|\langle\vec{S}\rangle|$ in Region~$I$ of Fig.~\ref{fig1}, i.e., for $z<0$.

It turns out that the presence of a gap between the gain and loss components of our system has drastic effects on the behavior of $\br u\kt$ and $\langle\vec{S}\rangle $. For this reason we examine the cases $s=0$ and $s\neq 0$ separately.

\begin{figure}
    \begin{center}
    \includegraphics[scale=.55]{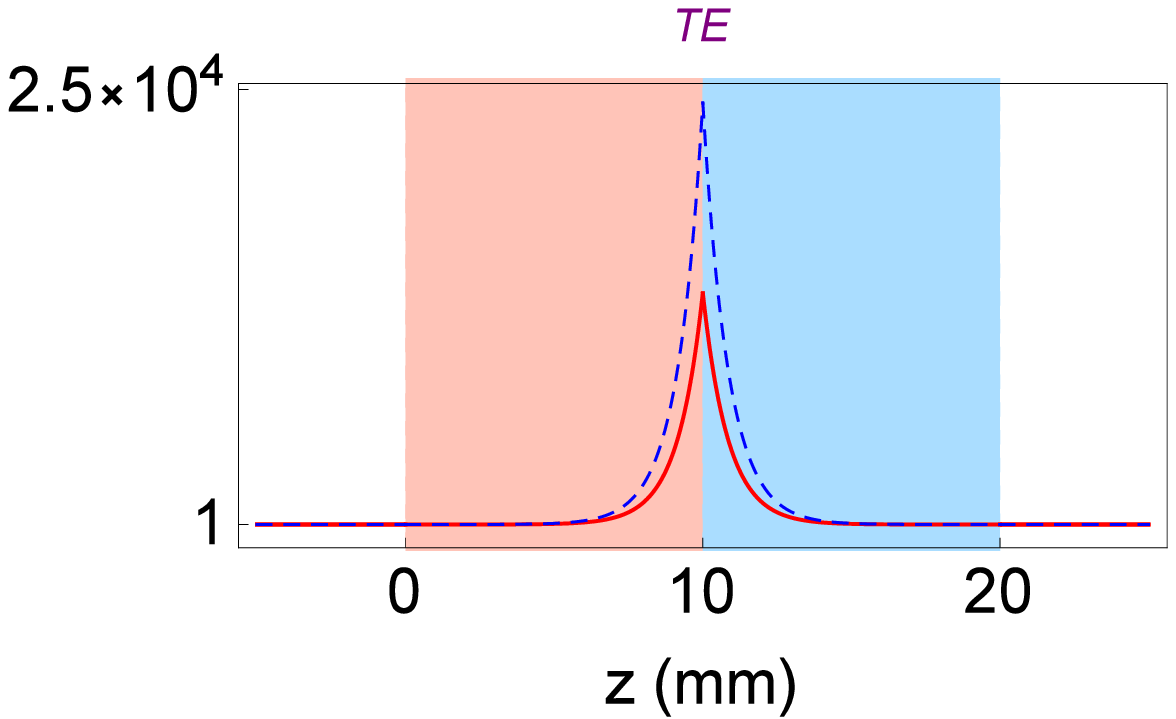}~~~~~~
    \includegraphics[scale=.55]{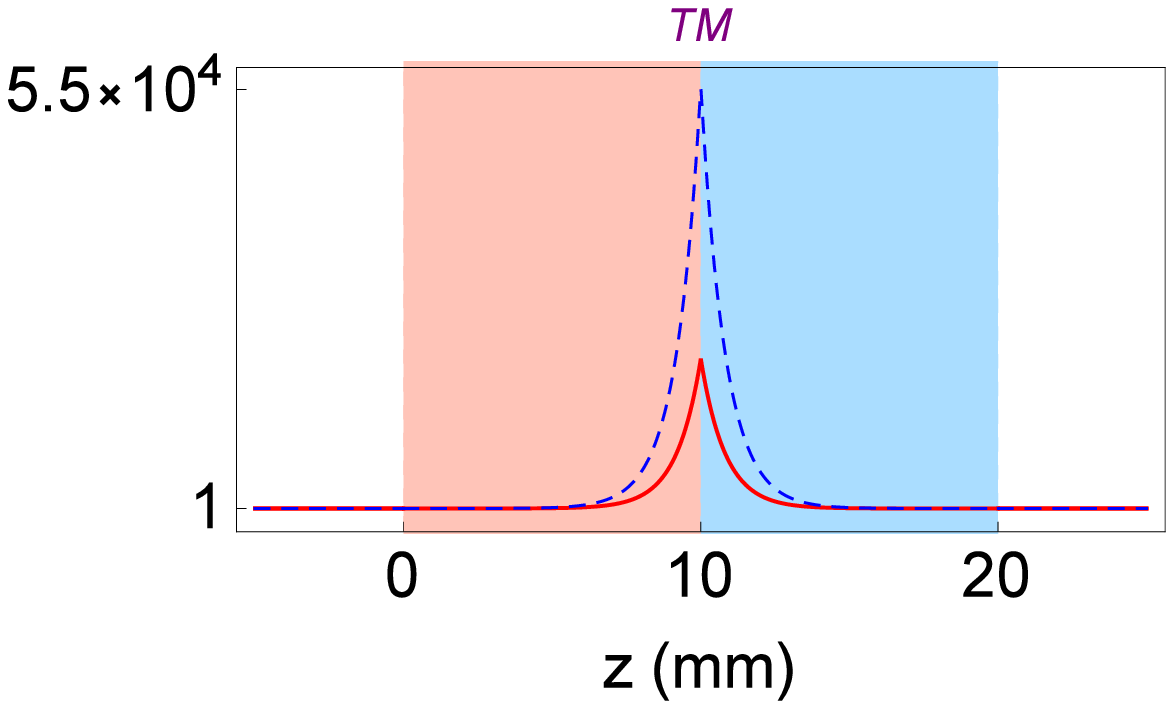}\\
    \includegraphics[scale=.55]{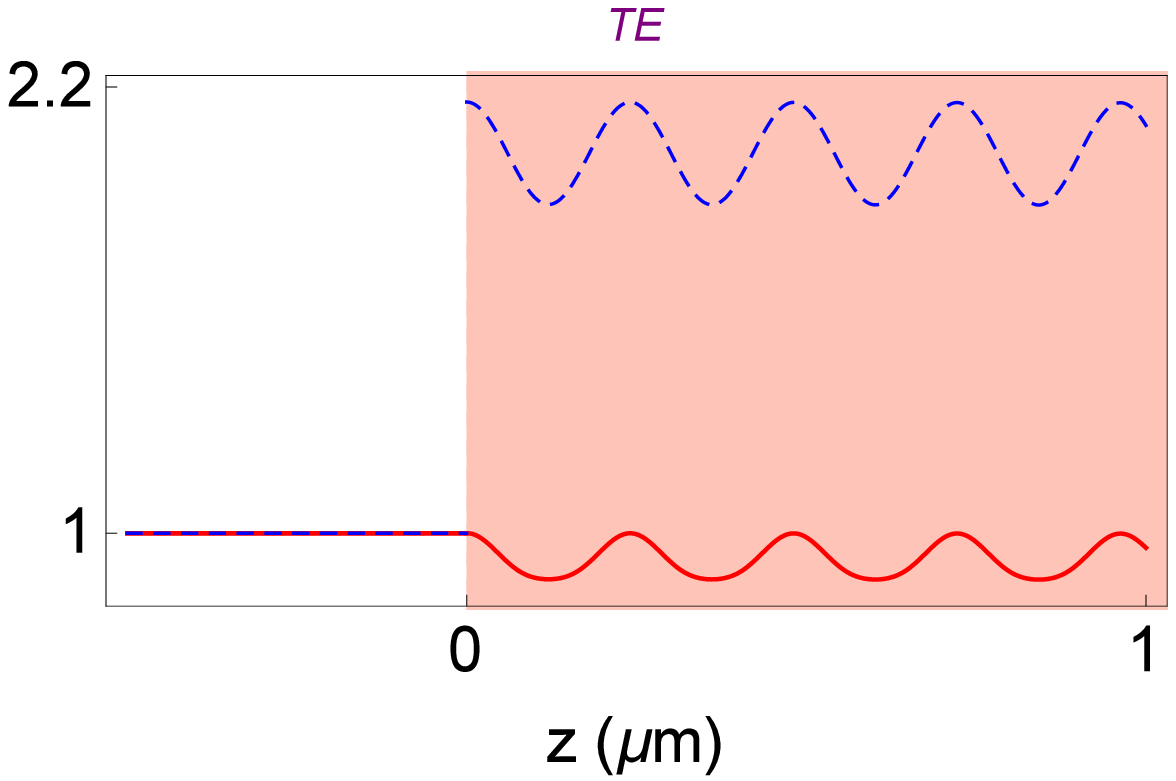}~~~~~~
    \includegraphics[scale=.55]{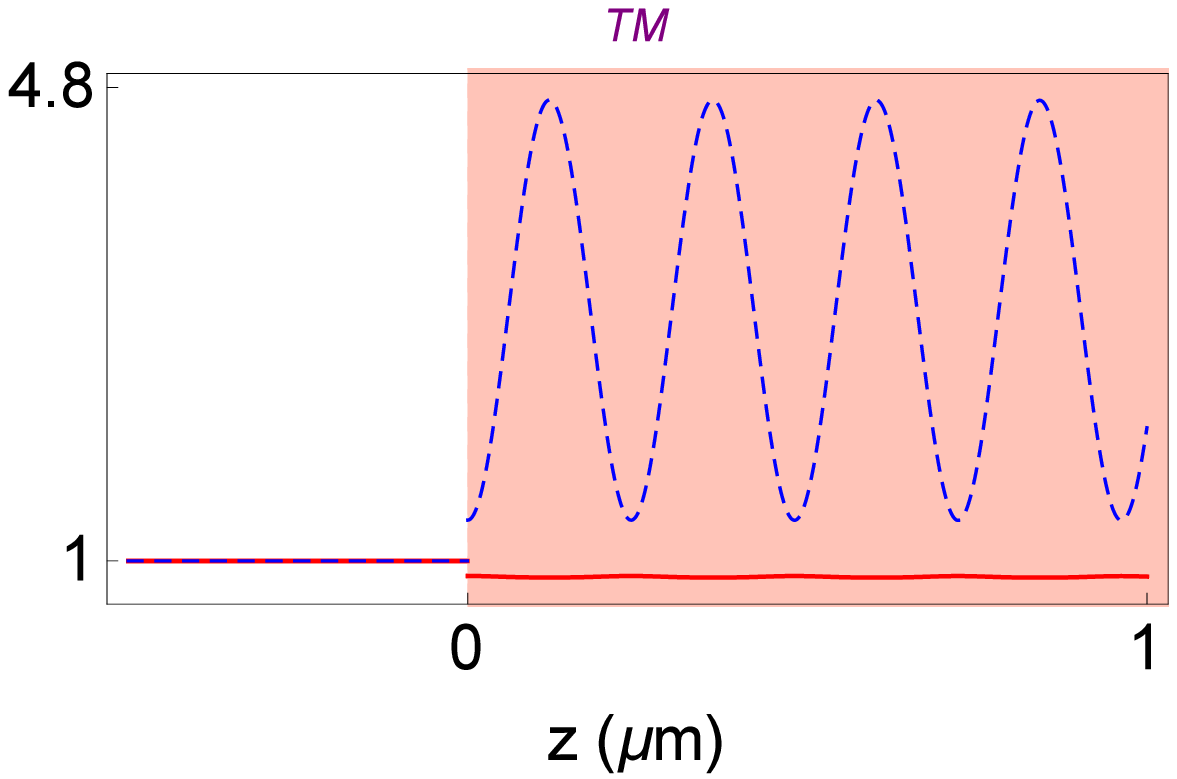}\\
    \includegraphics[scale=.55]{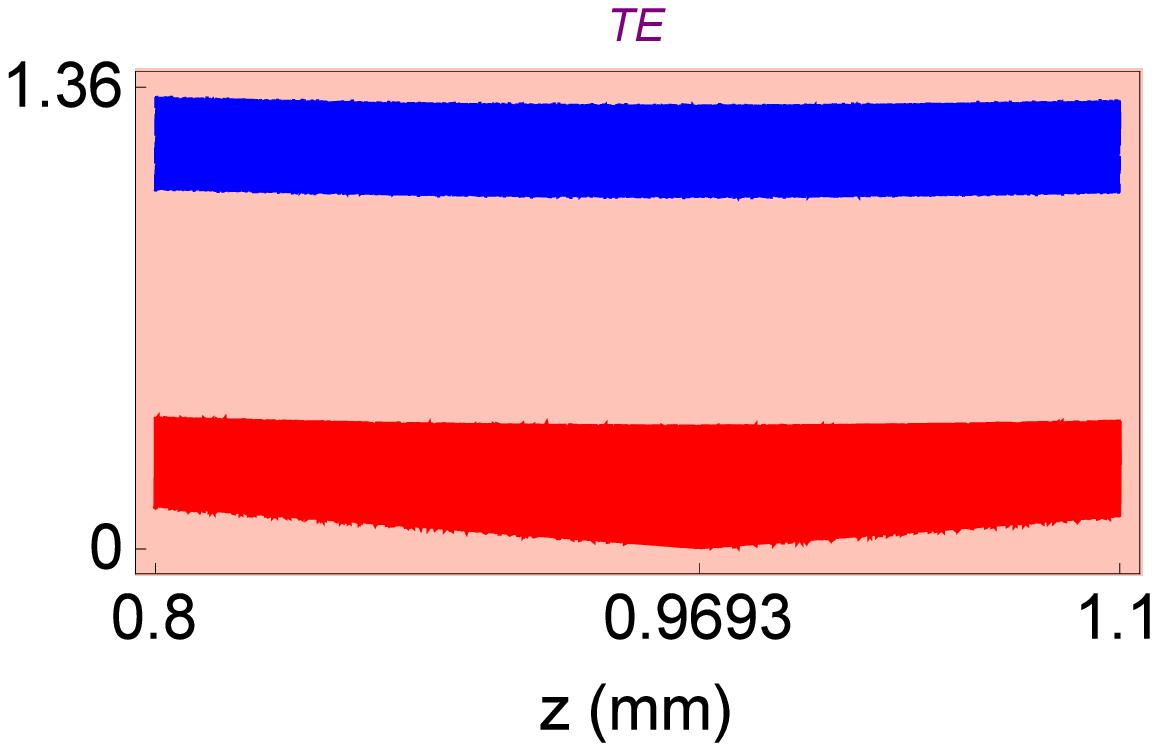}~~~~~~
    \includegraphics[scale=.55]{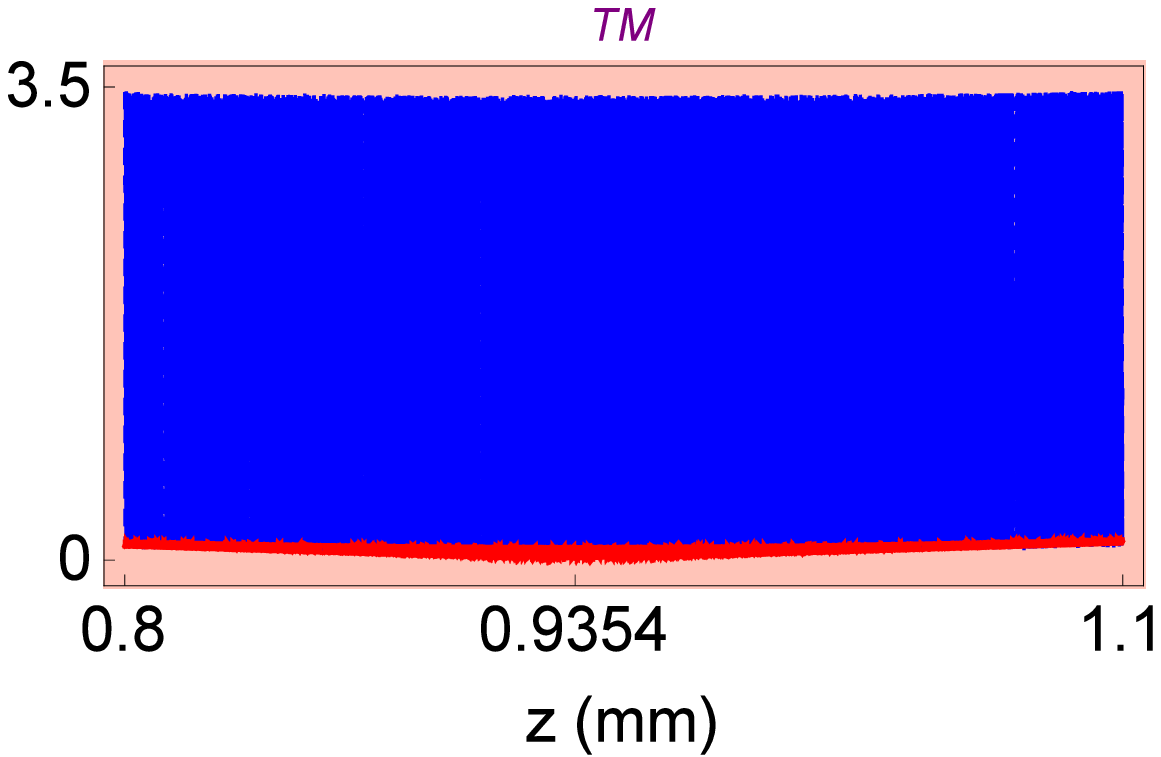}\\
    \includegraphics[scale=.55]{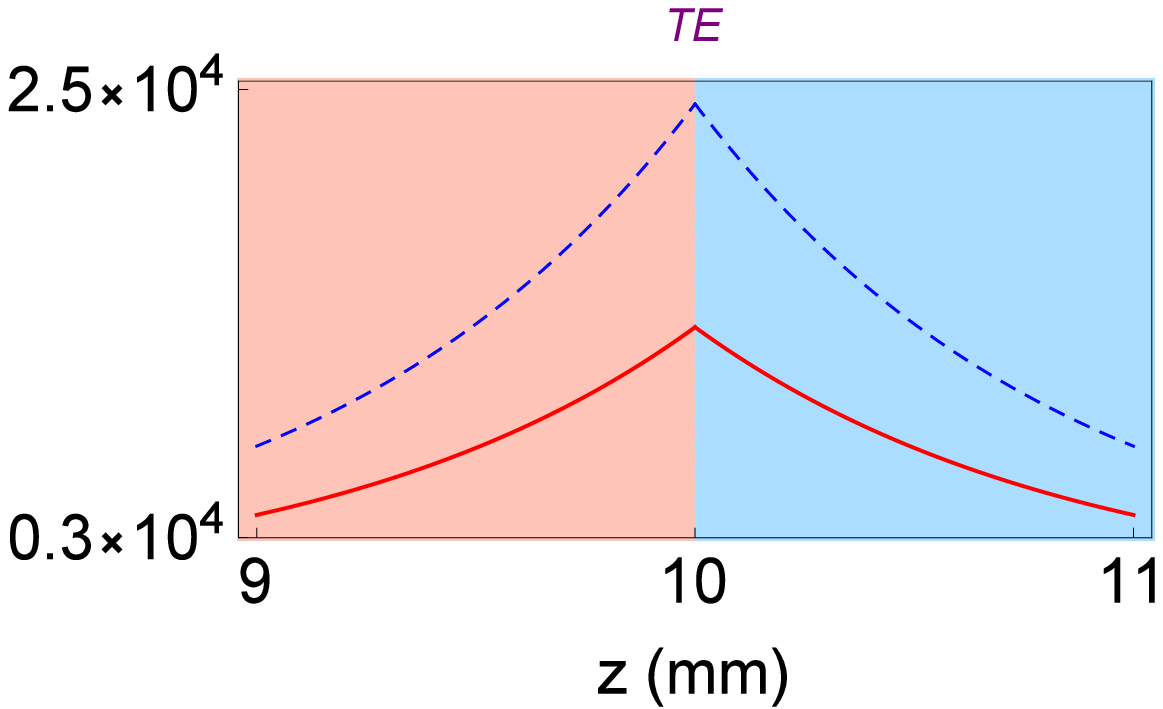}~~~~~~
    \includegraphics[scale=.55]{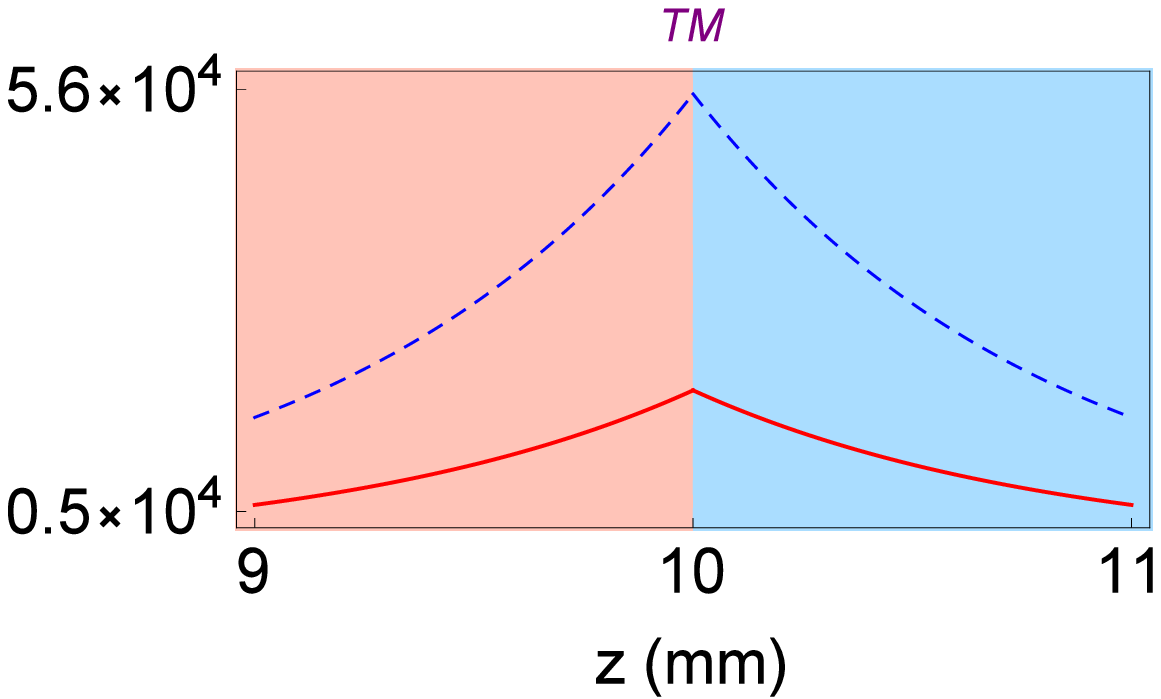}\\
    \includegraphics[scale=.55]{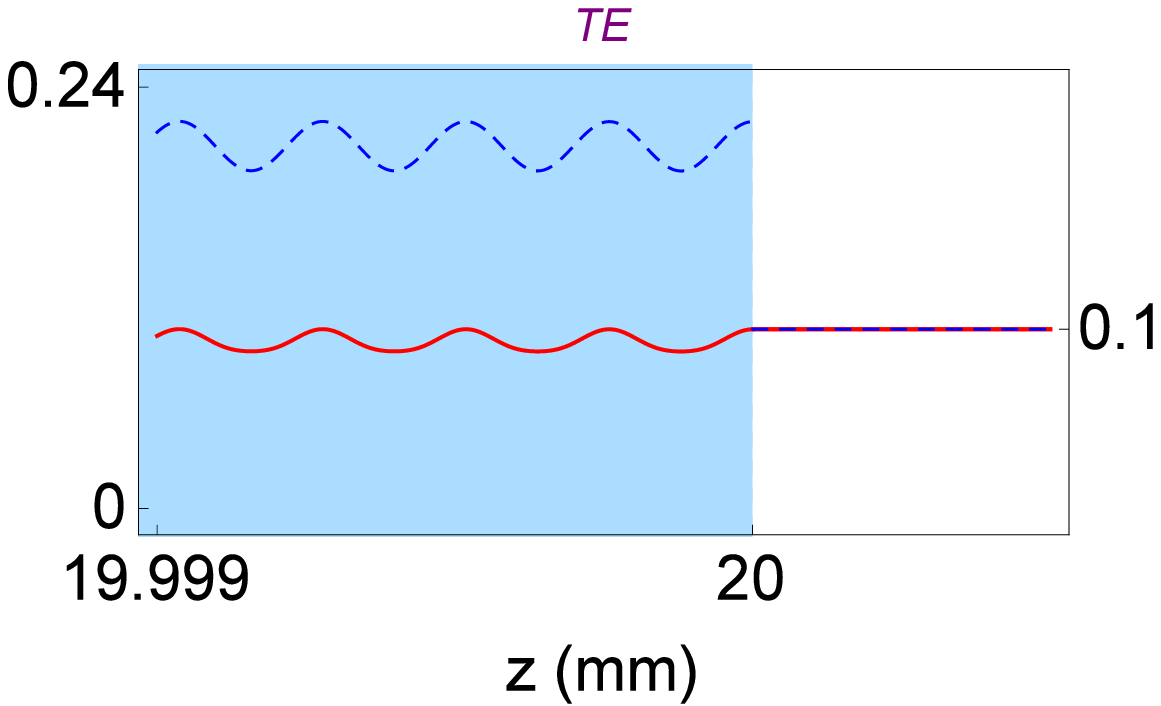}~~~~~~
    \includegraphics[scale=.55]{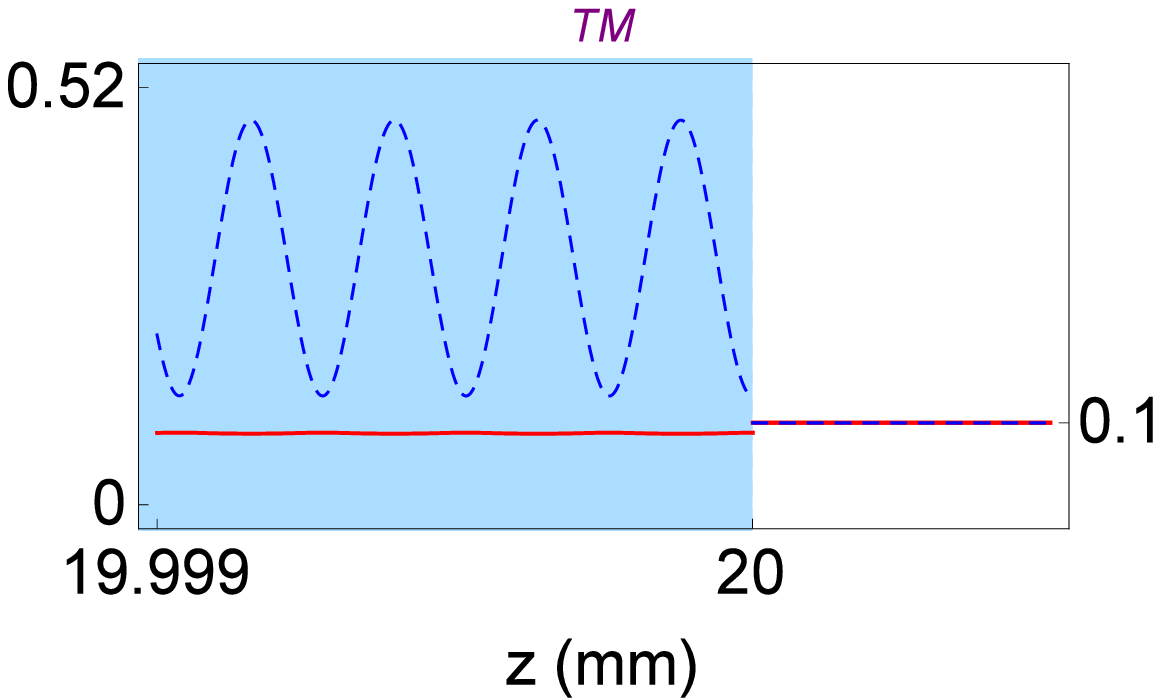}
    \caption{(Color online) Graphs of $\br u\kt$ in units of $\br u_I\kt$ (dashed navy curves) and $|\langle\vec{S}\rangle|$ in units of $|\langle\vec{S}_I\rangle|$ (solid red curves) for singular TE and TM modes of the $\cP\cT$-symmetric Nd:YAG bilayer determined by  (\ref{SS-30=1}) and (\ref{SS-30=}). The pink and blue regions correspond to the layers with gain and loss, respectively.}
    \label{fig8a}
    \end{center}
    \end{figure}

\begin{figure}
    \begin{center}
    \includegraphics[scale=.55]{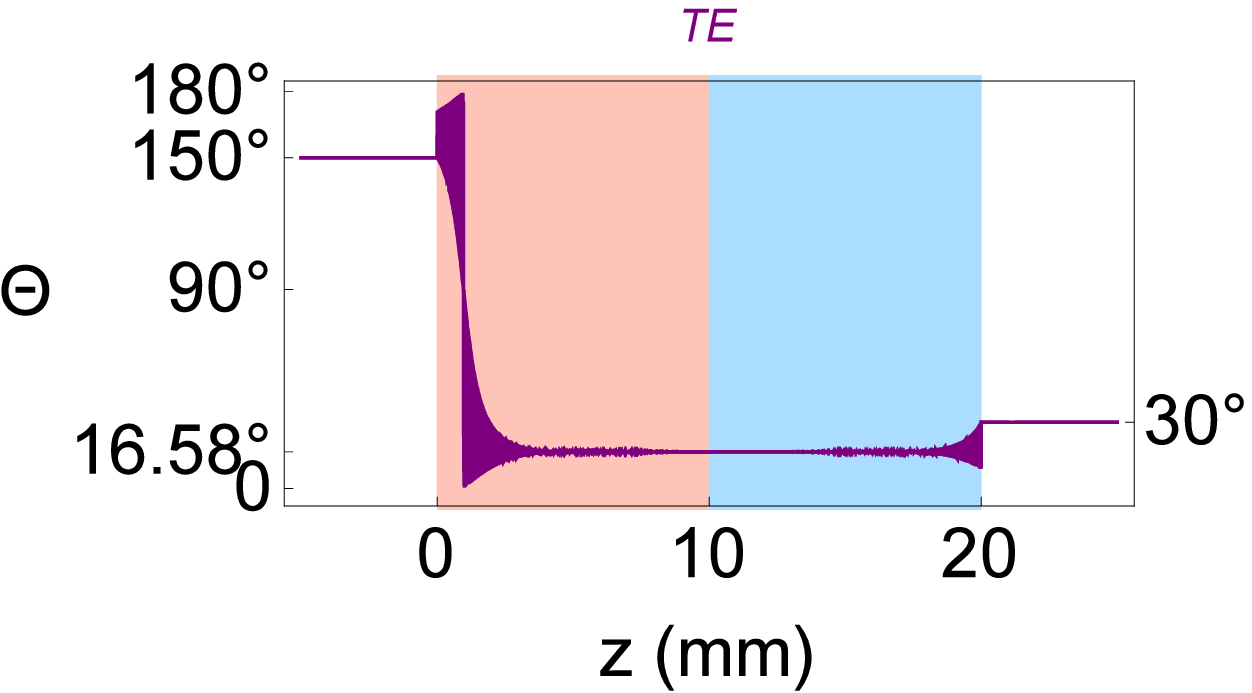}~~~~~~
    \includegraphics[scale=.55]{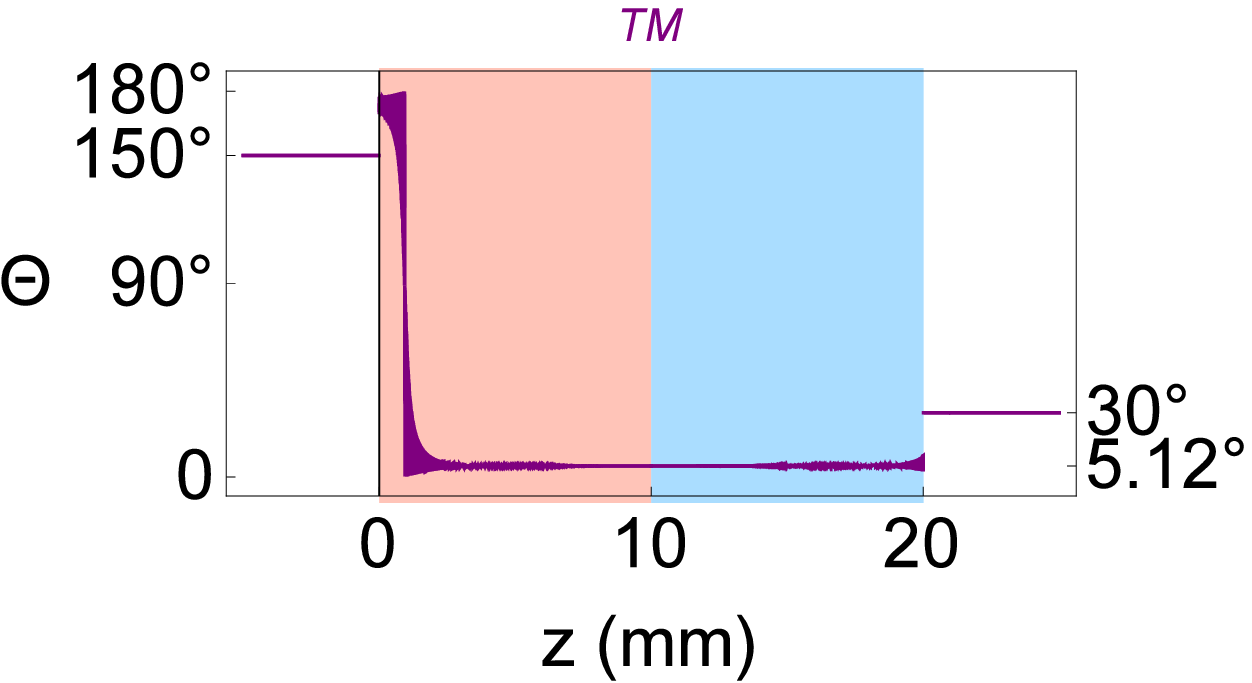}\\
    \includegraphics[scale=.55]{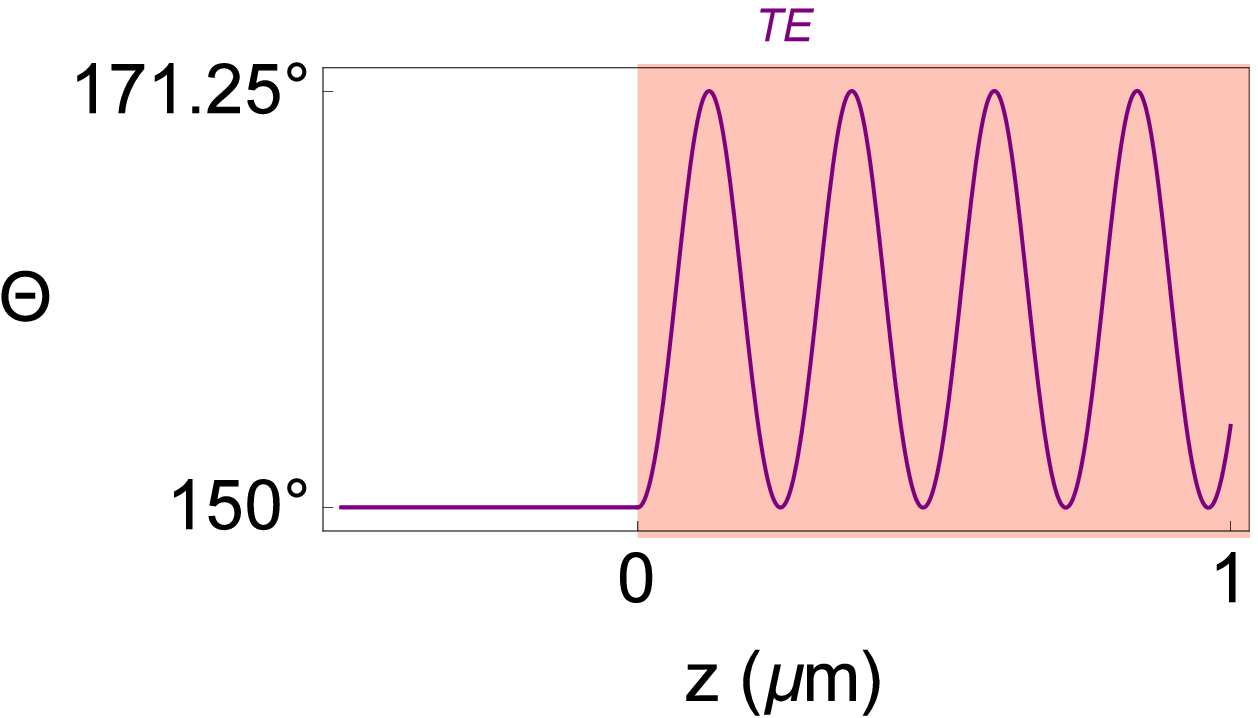}~~~~~~
    \includegraphics[scale=.55]{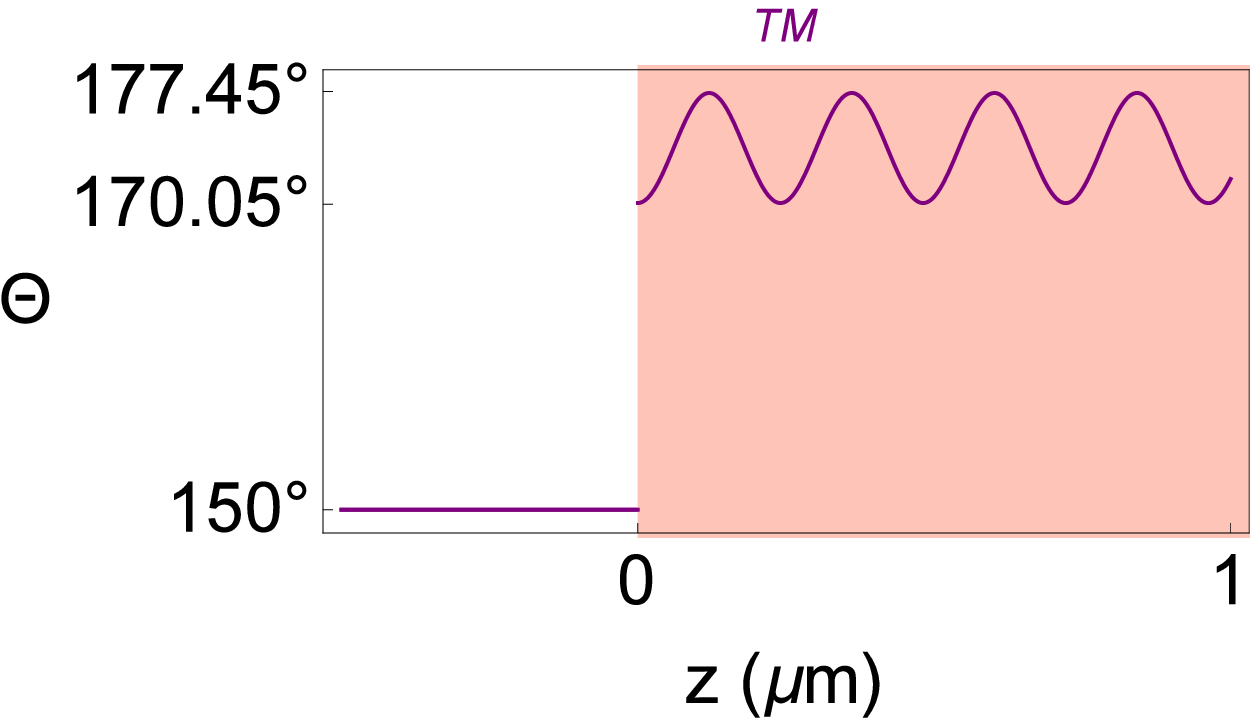}\\
    \includegraphics[scale=.55]{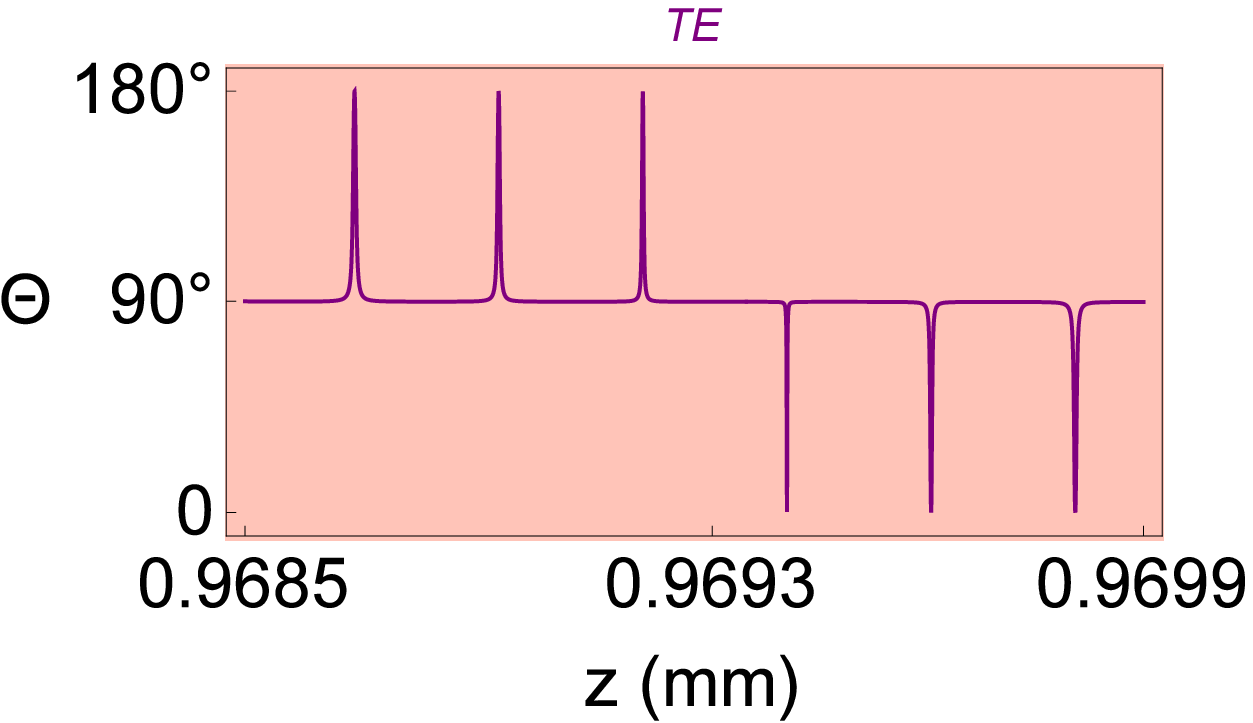}~~~~~~
    \includegraphics[scale=.55]{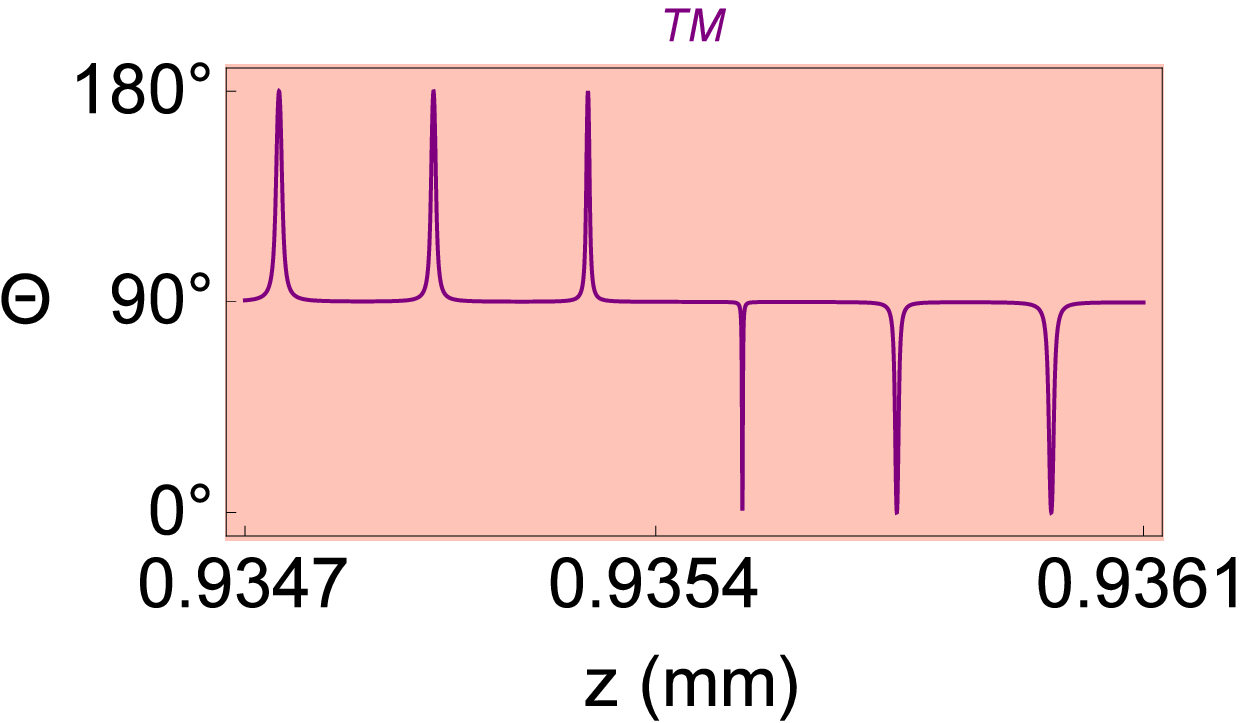}\\
    \includegraphics[scale=.55]{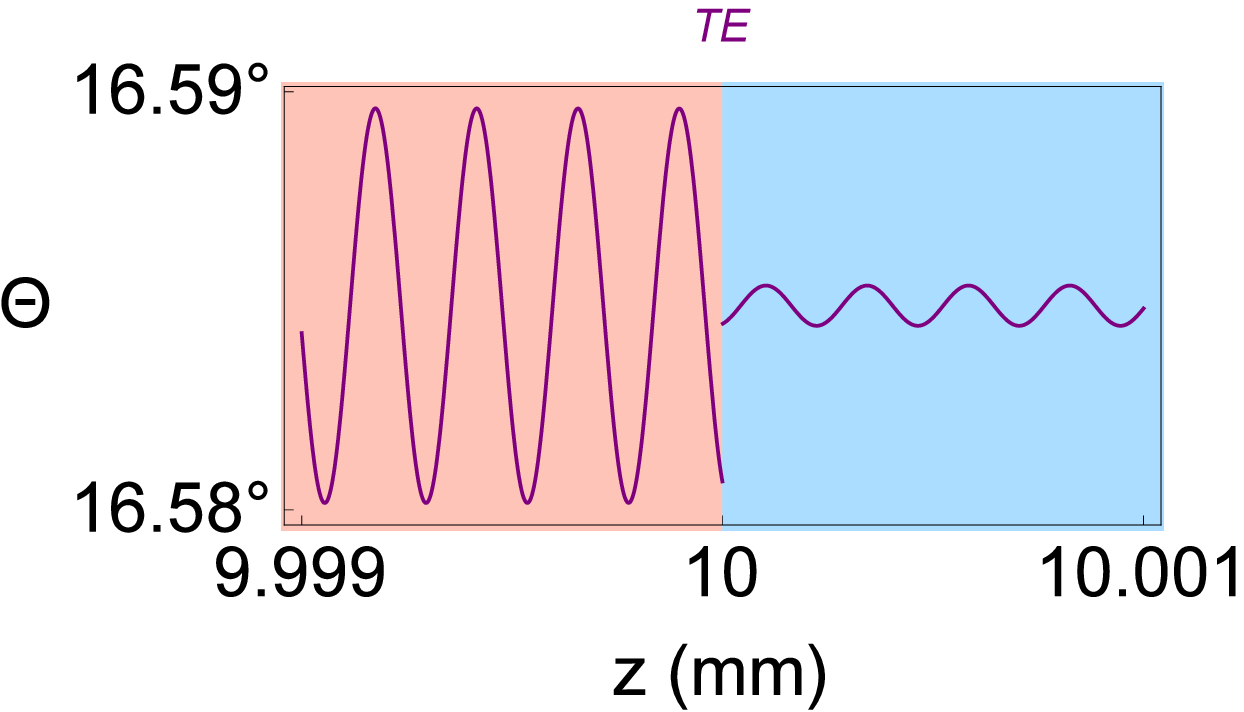}~~~~~~
    \includegraphics[scale=.55]{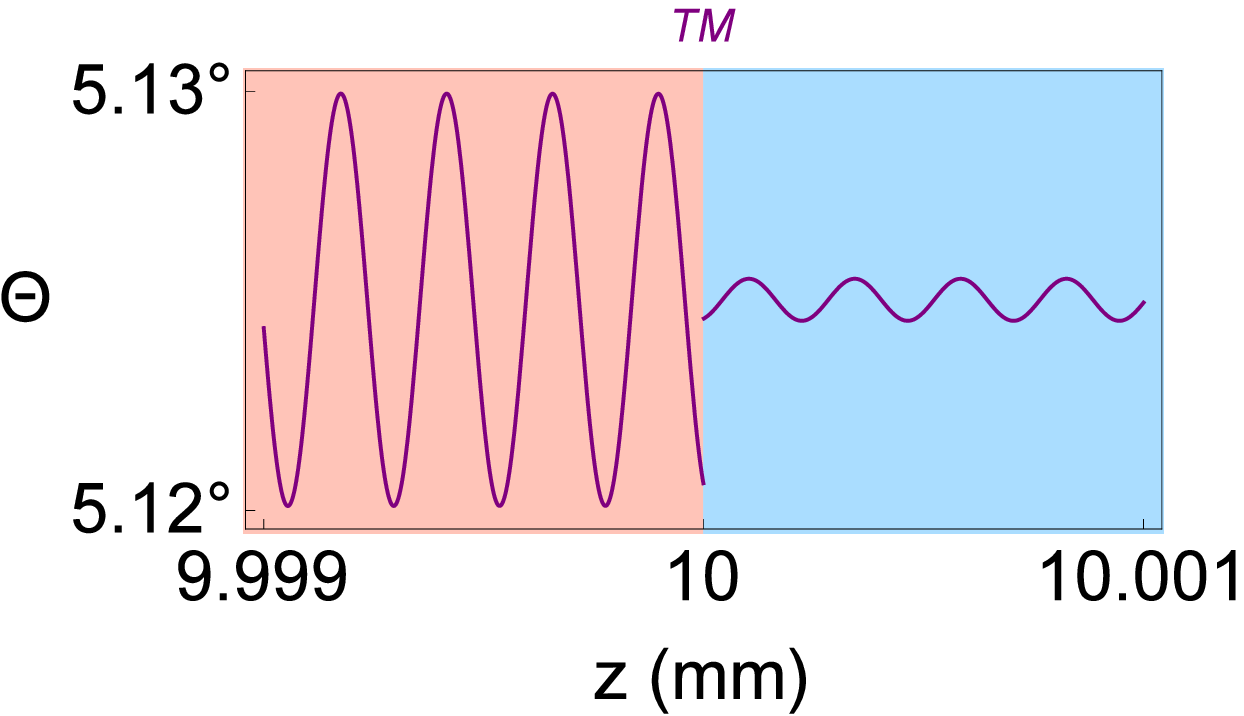}\\
    \includegraphics[scale=.55]{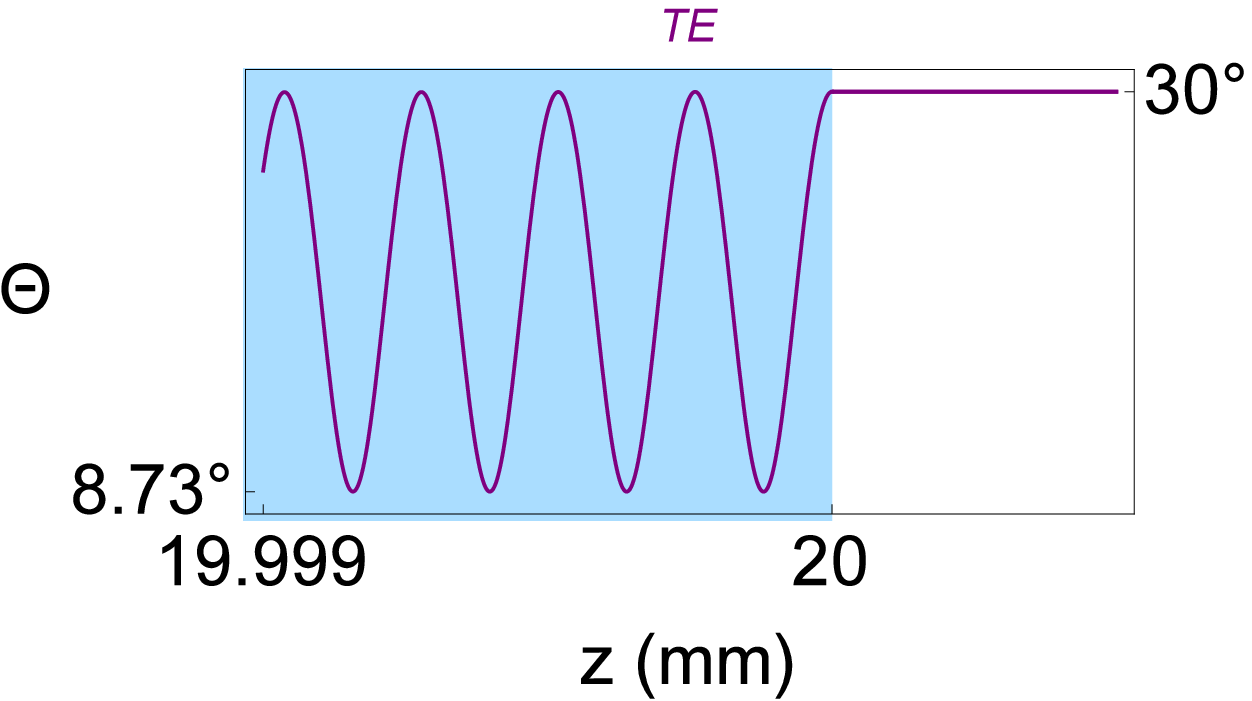}~~~~
    \includegraphics[scale=.55]{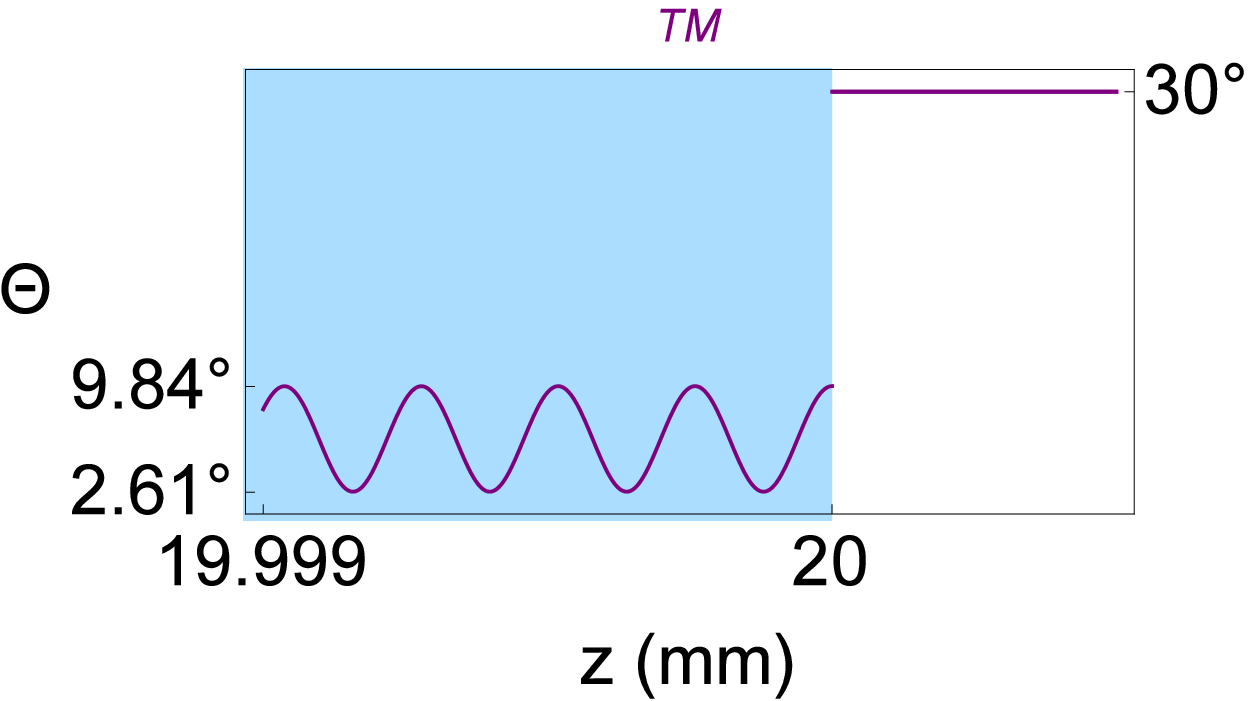}
    \caption{(Color online) Graphs of the angle $\Theta$ between the Poynting vector $\langle\vec{S}\rangle$ and positive $z$-axis for the same singular TE and TM modes as in Fig.~\ref{fig8a}. $\langle\vec{S}\rangle$ points away from the critical plane $z=0.9693~{\rm mm}$ for the TE mode and  $z=0.9354~{\rm mm}$ for the TM mode.}
    \label{fig8b}
    \end{center}
    \end{figure}

Figures~\ref{fig8a} and \ref{fig8b} show the plots of $\br u\kt $, $|\langle\vec{S}\rangle|$, and the angle between $\langle\vec{S}\rangle$ and the positive $z$-axis, i.e.,
    \be
    \Theta:= \arctan\left(\frac{\langle\vec{S}\rangle\cdot\hat e_x}{\langle\vec{S}\rangle\cdot\hat e_z}\right),
    \label{big-theta}
    \ee
for singular TE and TM modes of a $\cP\cT$-symmetric Nd:YAG bilayer slab ($s=0$) obtained for the following values of the physical parameters.
    \begin{align}
    & L=1~{\rm cm}, && \eta=1.8217, && \theta=30^\circ,
    \label{SS-30=1}\\
    &\lambda^{(E/M)}=807.993~{\rm nm}, && g^{(E)}=10.851\,{\rm cm}^{-1},
    && g^{(M)}=11.244\,{\rm cm}^{-1}.
    \label{SS-30=}
    \end{align}%
As seen from these figures, $\br u\kt$ and $|\br\vec S\kt|$ oscillate throughout the gain and loss regions. As one increases $z$ starting from zero, their envelope decreases slightly until they attain their minimum value at about $z=z_c:=0.9693~{\rm mm}$ for the TE mode and $z=z_c:=0.9354~{\rm mm}$ for the TM mode. For the values of $z$ in the vicinity of $z_c$ the direction of $\br\vec S\kt$ undergoes rapid changes, with $\Theta$ oscillating between $90^\circ$ and $180^\circ$ for $z<z_c$ and between $0^\circ$ and $90^\circ$ for $z>z_c$. This shows that $\br\vec S\kt$ always points away from the plane $z=z_c$, \cite{foot1}. For $z\in(z_c,L)$ the envelops of $\br u\kt$ and $|\br\vec S\kt|$ increase steadily. In the lossy region, i.e., $z\in(L,2L)$, they are monotonically decreasing functions of $z$. Furthermore, $\br u\kt$ and $|\br\vec S\kt|$ take much smaller values for $z>2L$ compared with $z<0$. Therefore, as expected, the emitted laser light from the outer boundary of the lossy layer has a smaller energy and power than the one emitted from that of the gain layer.

\begin{figure}
    \begin{center}
    \includegraphics[scale=.4]{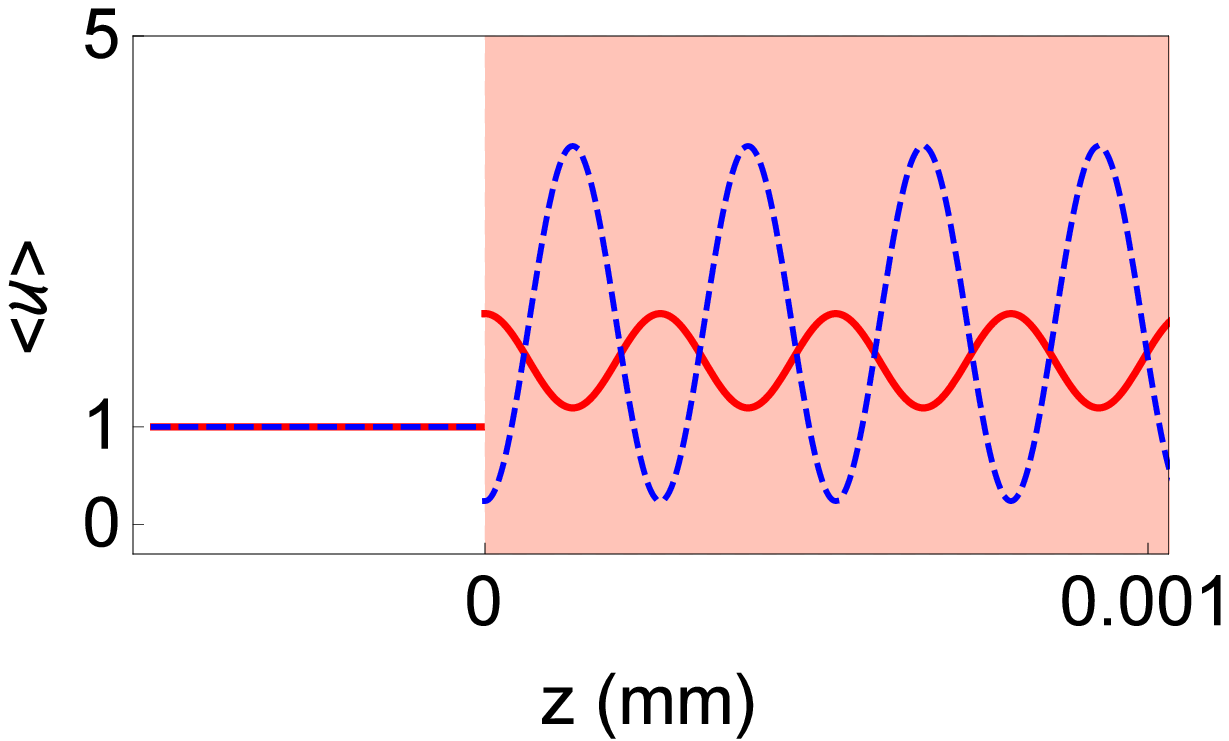}~~
    \includegraphics[scale=.4]{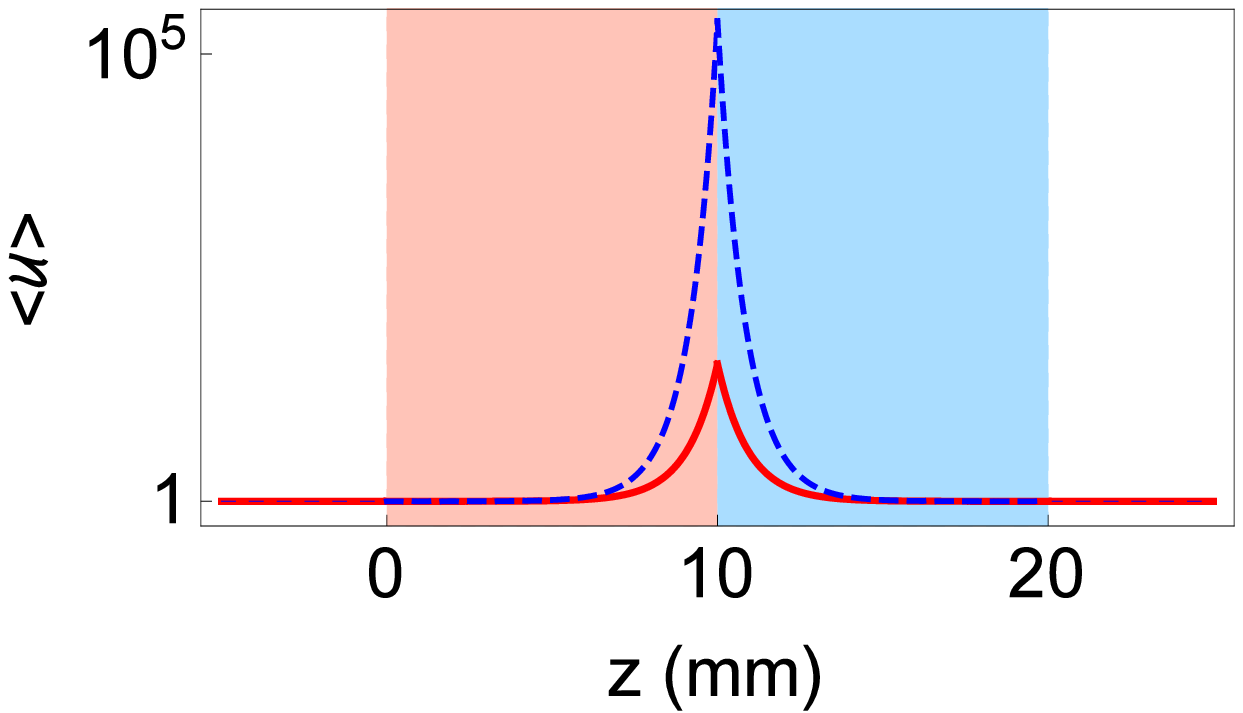}~~
    \includegraphics[scale=.4]{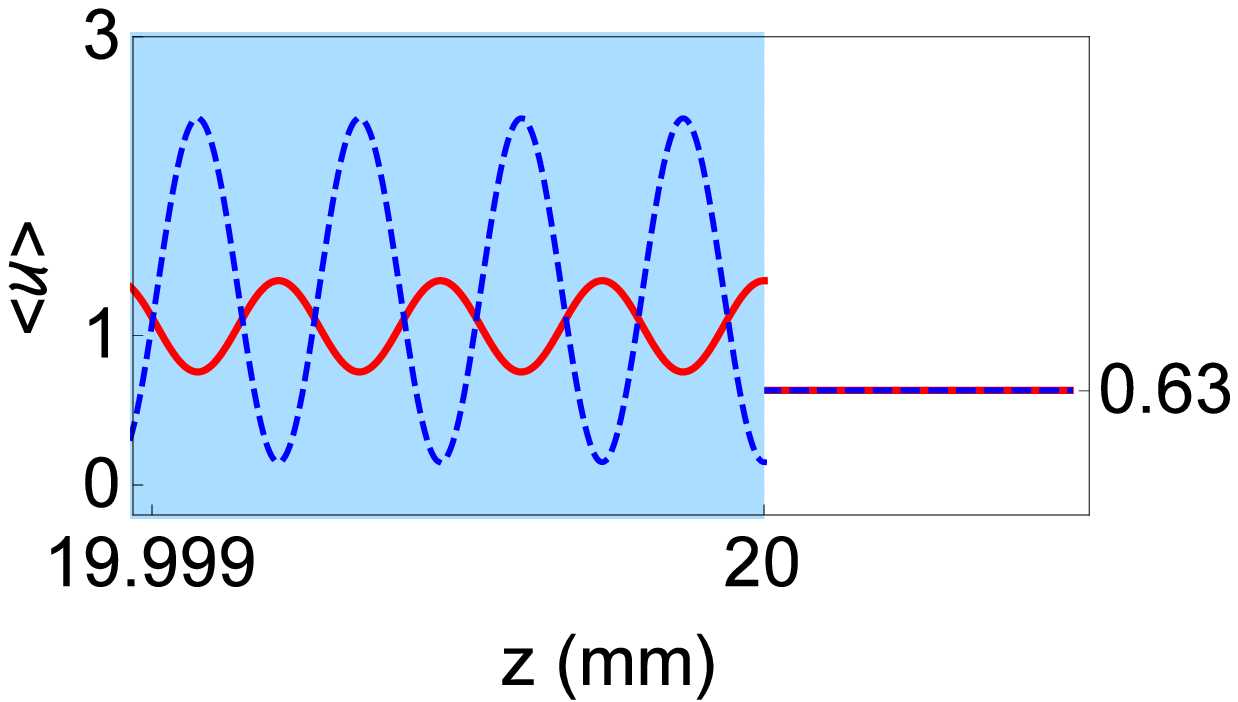}
    \caption{(Color online) Graphs of $\langle u \rangle$ (in units of $\langle u_I\rangle$) as a function of $z$ for the singular TE (solid red curve) and TM (blue dashed curve) modes of a $\cP\cT$-symmetric Nd:YAG bilayer slab with $\theta=80^\circ>\theta_b$. The relevant physical parameters are given by (\ref{SS-30=2}). The pink and blue regions correspond to the layers with gain and loss, respectively.}
    \label{fig10}
    \end{center}
    \end{figure}

We have also examined the situation where $\theta$ exceeds the Brewster's angle $\theta_b$. For a homogeneous gain slab, the condition $\theta>\theta_b$ implies that the time-averaged energy density $\br u\kt$ of the singular TM modes takes smaller values inside the slab than outside it \cite{pra-2015a}. This surprising effect is also present in the $\cP\cT$-symmetric bilayer we are considering. For the singular TM modes with $\theta>\theta_b$, the value of $\br u\kt$ drops sharply as one enters either the gain or lossy layer from the outer boundary of the bilayer. But it then begins oscillating with a much larger amplitude than the magnitude of the difference of values of $\br u\kt$ inside and outside the bilayer. This is depicted in Fig.~\ref{fig10} where we offer a graphical comparison of $\br u\kt$ for the singular TE and TM modes obtained for the following values of the physical parameters.
    \be
    \begin{aligned}
    & L=1~{\rm cm},~~~~~\eta=1.8217,~~~~~\theta=80^\circ,\\
    &\lambda^{(E)}=808.009~{\rm nm},~~~~~g^{(E)}=9.025\,{\rm cm}^{-1},\\
    &\lambda^{(M)}=807.999~{\rm nm},~~~~~g^{(M)}=9.903\,{\rm cm}^{-1}.
    \end{aligned}
    \label{SS-30=2}
    \ee

Next, we summarize the consequences of the presence of a gap between the gain and loss layers, i.e., $s>0$:
    \begin{enumerate}
    \item Again there is a critical plane $z=z_c$ located inside the gain layer where $\br u\kt$ and $|\br\vec S\kt|$ attain their minimum value. Moreover, $\br\vec S\kt$ points away from this plane.
    \item The behavior of the $\br u\kt$ and $|\br\vec S\kt|$ is highly sensitive to the ratio of $s$ to
            \[s_0:=\frac{\pi}{2k_z}=\frac{\lambda}{4\cos\theta}.\]
        If $s/s_0$ is an even integer, the presence of the gap does not lead to any significant changes in the behavior of $\br u\kt$ and $\br\vec S\kt$, and the situation resembles that of the case $s=0$. In contrast, if $s/s_0$ is an odd integer, $\br u\kt$ and $|\br\vec S\kt|$ take much smaller values within the slabs and the gap in between, while they take larger values in the lossy layer and to its right. In other words, the emitted waves from the outer boundary of the lossy layer has larger energy density and power as compared to the case $s=0$.
    \end{enumerate}
The fact that for even values of $s/s_0$ the intensity of the waves inside the system can take extremely large values is due to the constructive interference of these waves. For this reason we use the terms `constructive', `destructive', and `generic configurations' to refer to the cases with even, odd, and non-integer values of $s/s_0$, respectively.

The destructive configurations are more desirable, because for these configurations the waves interacting with the content of the slabs have much smaller intensities. Therefore the nonlinearities arising from this interaction are suppressed and the linear treatment of the problem that we offer is more reliable.

\begin{figure}
    \begin{center}
    \includegraphics[scale=.5]{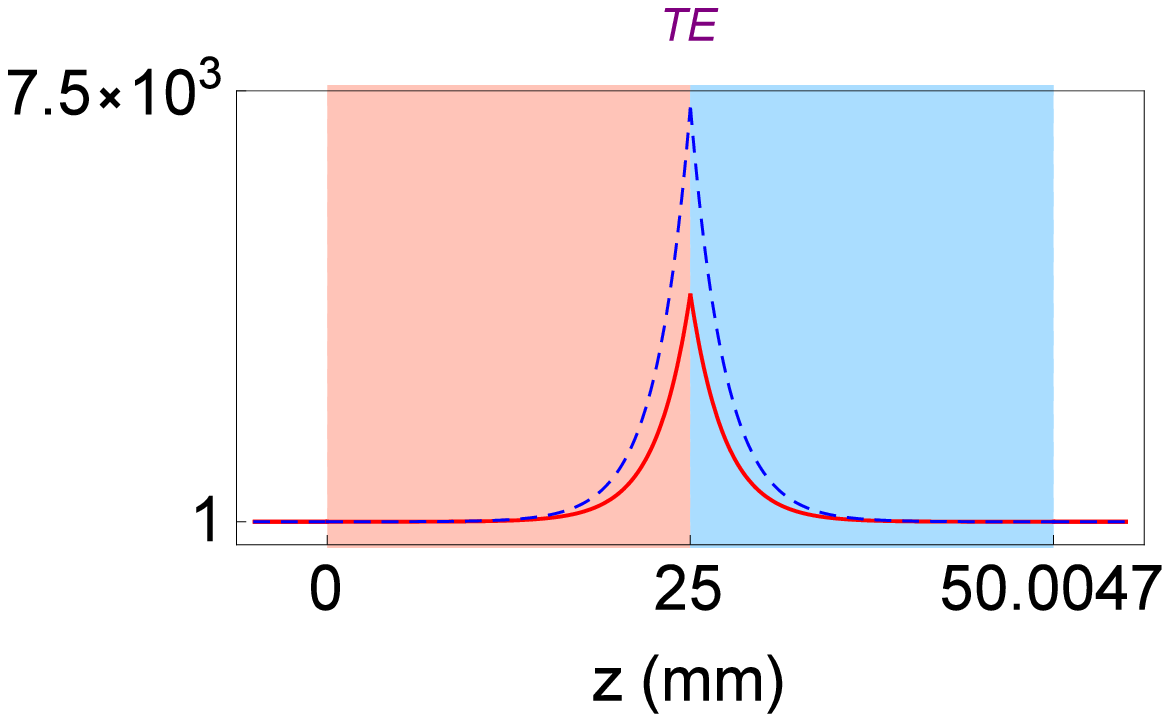}~~~~~~
    \includegraphics[scale=.5]{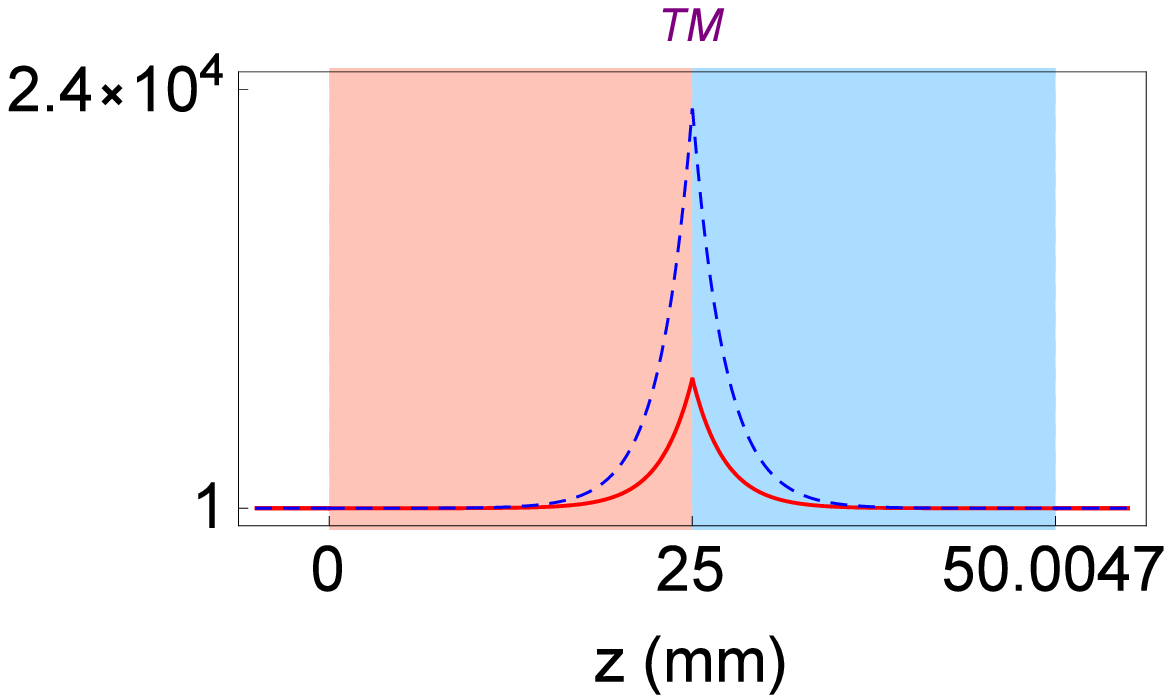}\\
    \includegraphics[scale=.5]{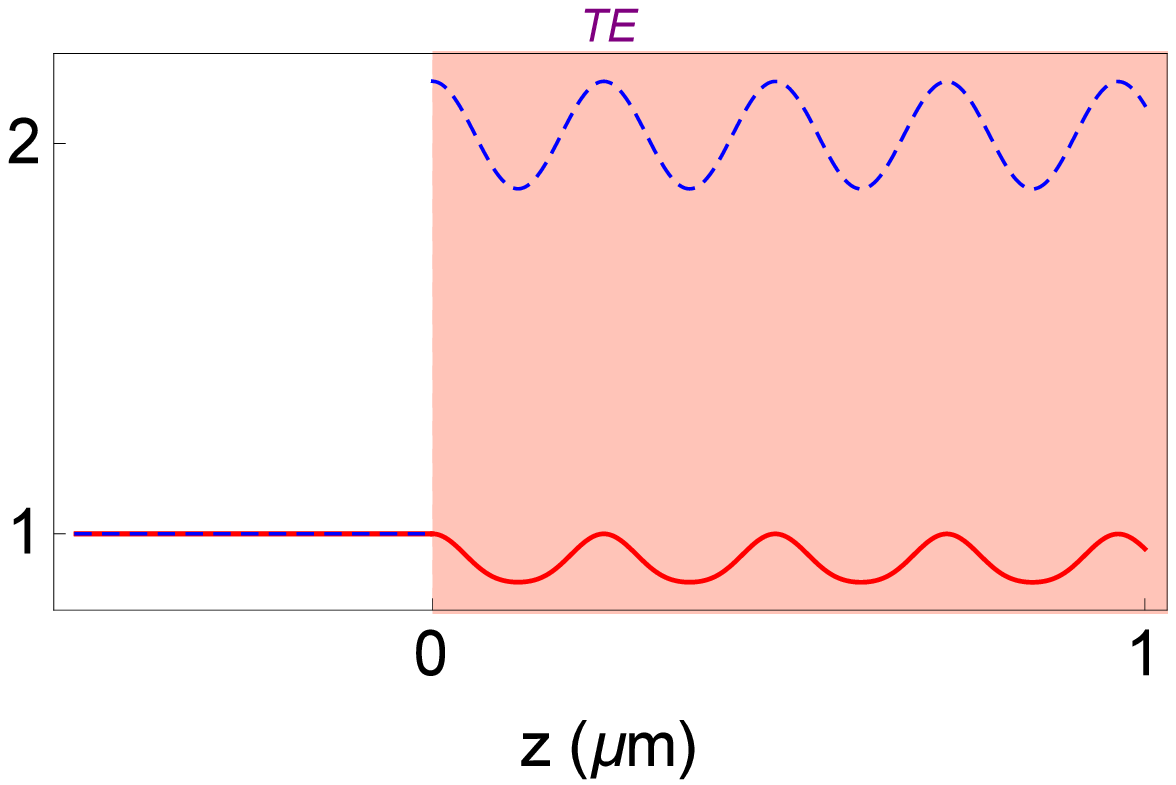}~~~~~~
    \includegraphics[scale=.5]{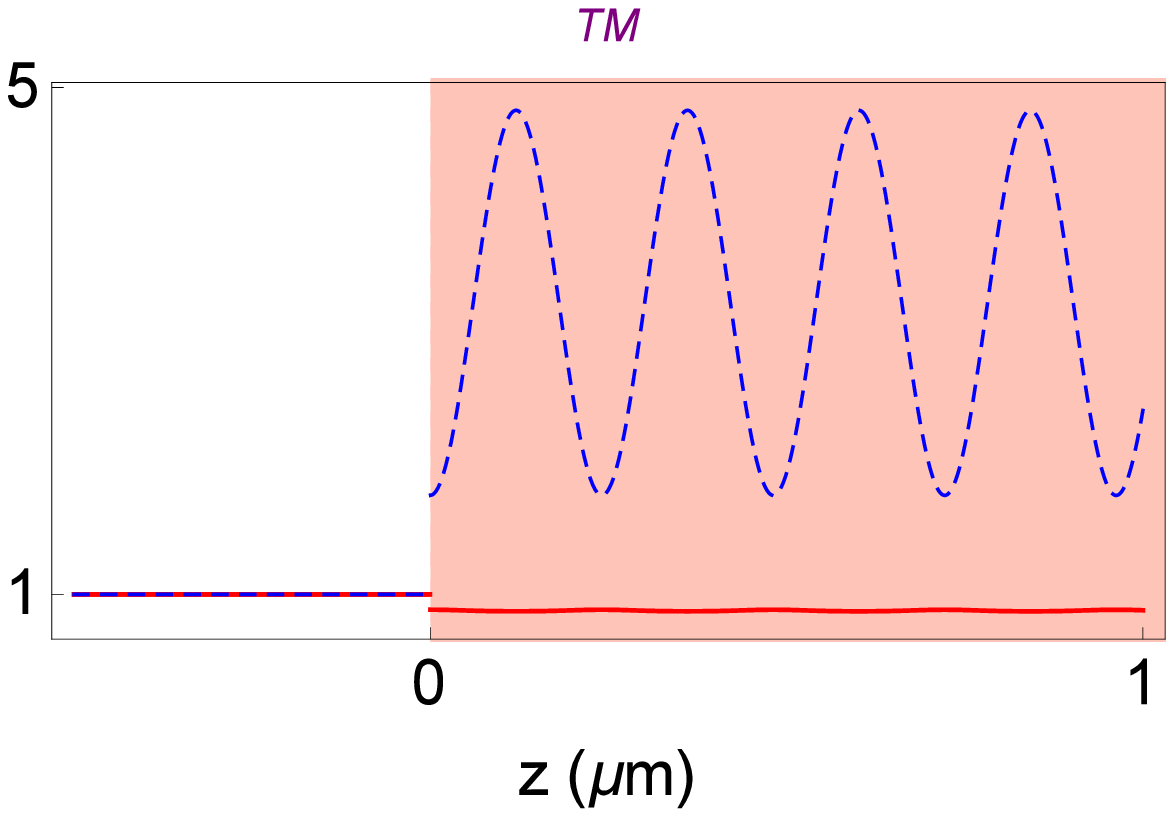}\\
    \includegraphics[scale=.5]{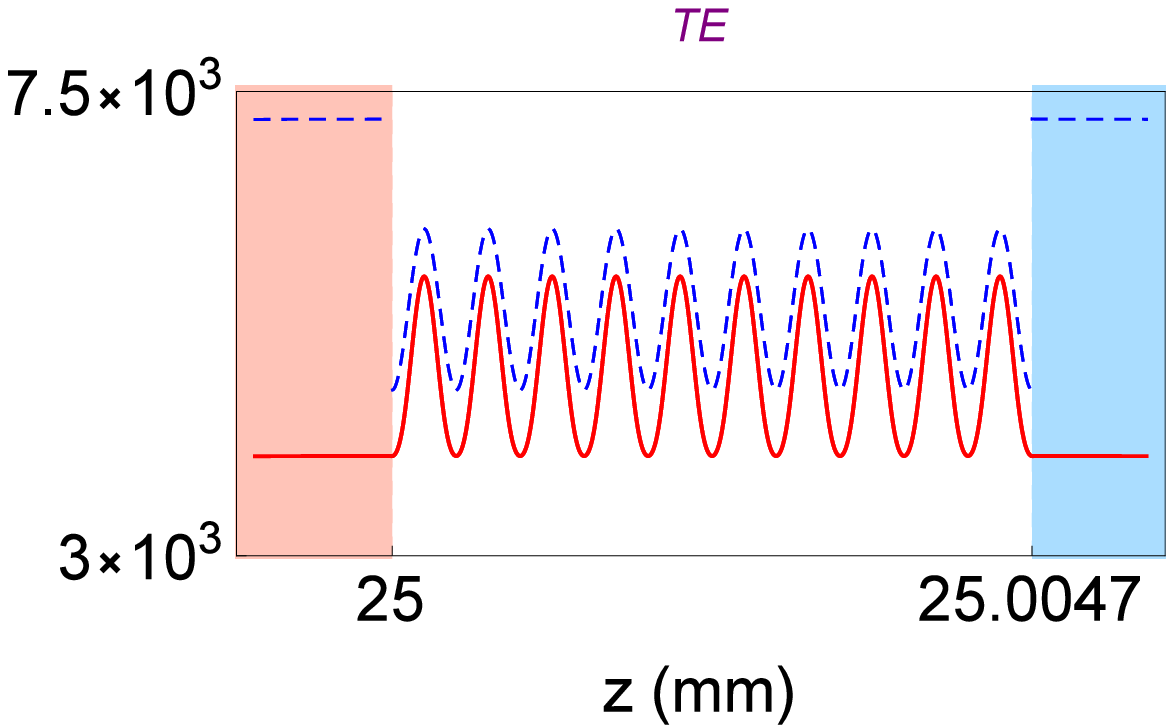}~~~~~~
    \includegraphics[scale=.5]{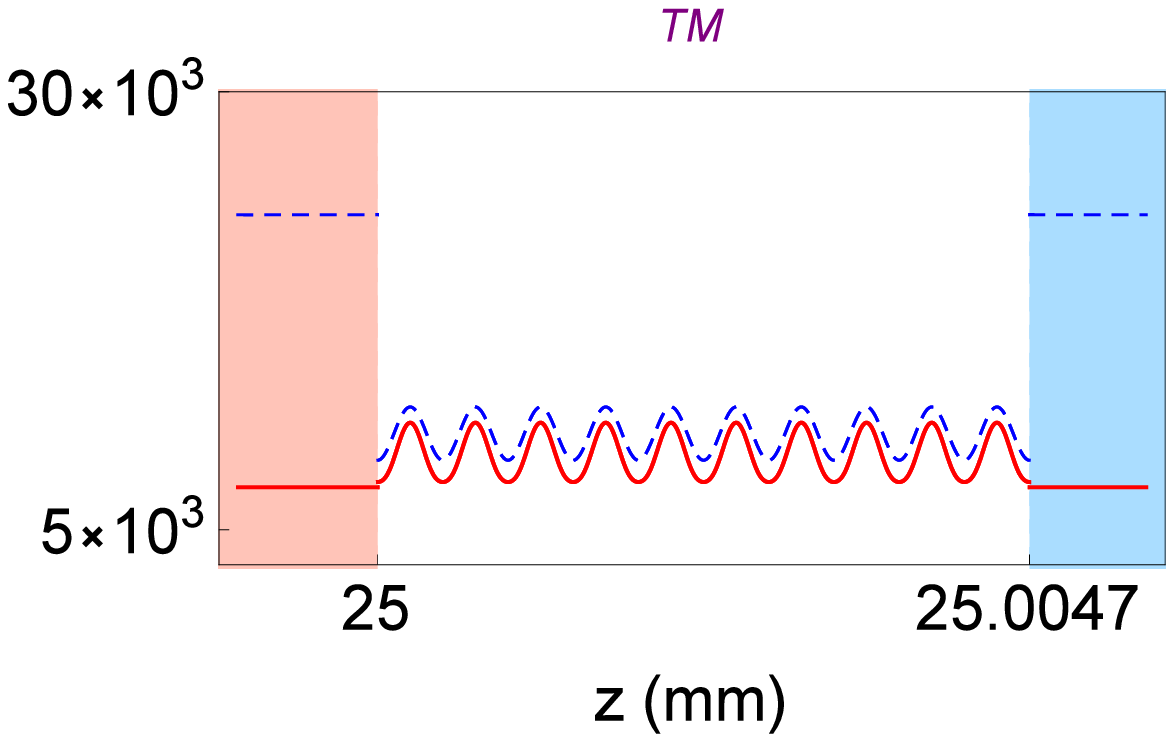}\\
    \includegraphics[scale=.5]{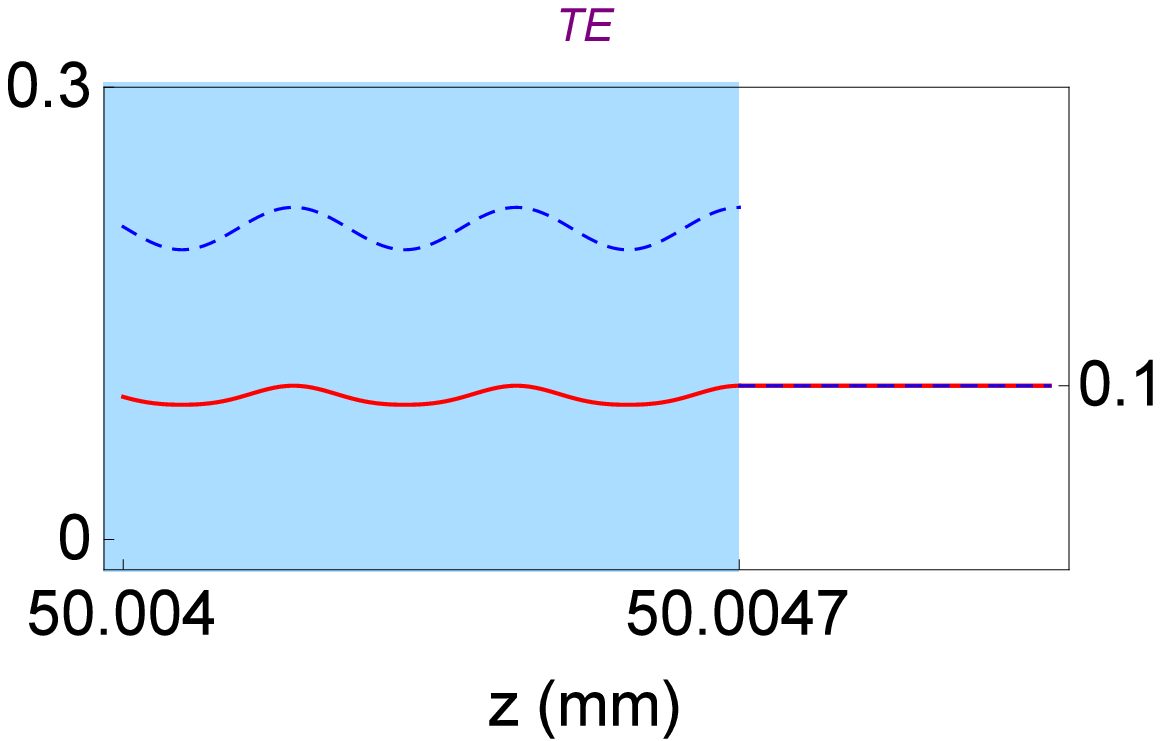}~~~~~~
    \includegraphics[scale=.5]{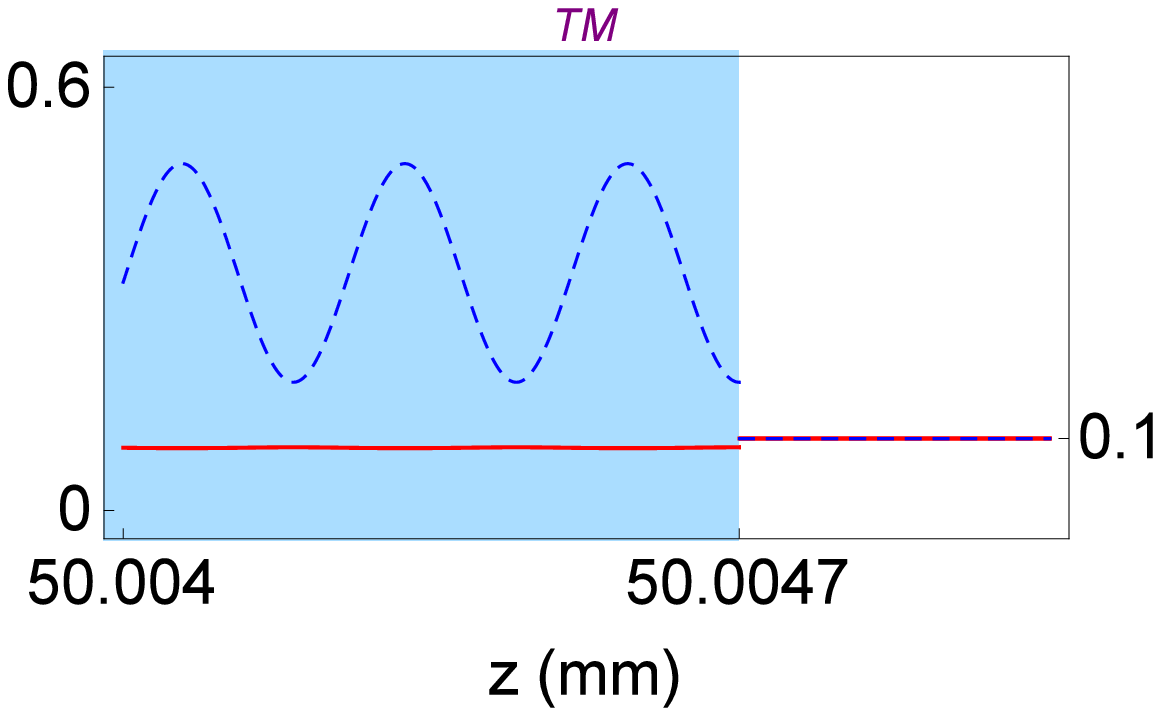}
    \caption{(Color online) Graphs of $\br u\kt$ in units of $\br u_I\kt$ (dashed navy curves) and $|\langle\vec{S}\rangle|$ in units of $|\langle\vec{S}_I\rangle|$ (solid red curves)  for the singular TE and TM modes of the constructive configuration of a $\cP\cT$-symmetric Nd:YAG two-slab system given by (\ref{sp-gen}) and (\ref{SEM-30=constructivespace}).}
    \label{fig11a}
    \end{center}
    \end{figure}

\begin{figure}
    \begin{center}
    \includegraphics[scale=.55]{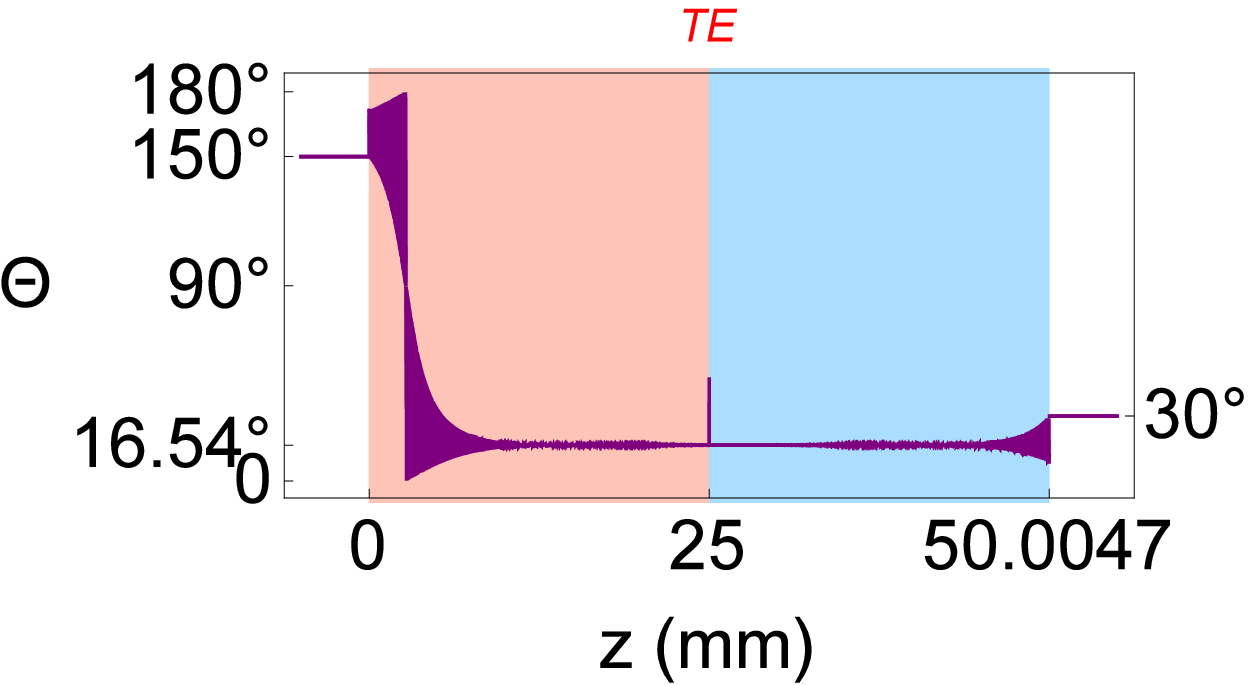}~~~~~~
    \includegraphics[scale=.55]{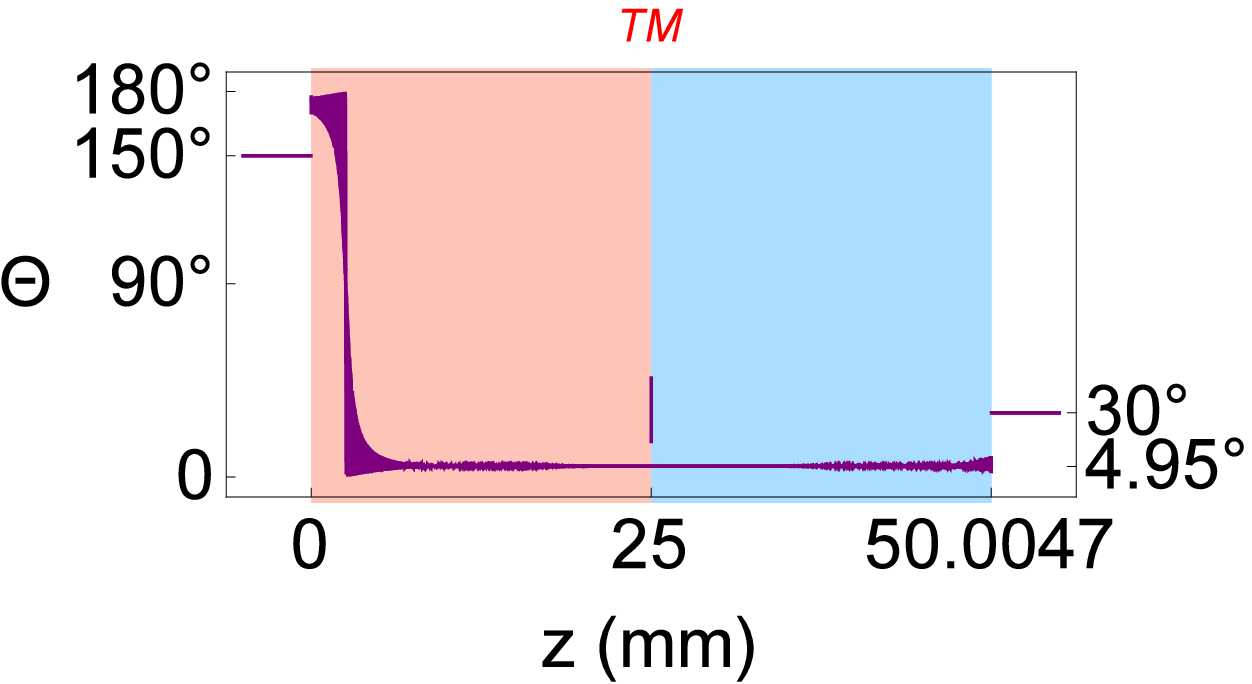}
    \caption{(Color online) Graphs of the angle $\Theta$ between the Poynting vector $\langle\vec{S}\rangle$ and positive $z$-axis for the TE and TM modes considered in Fig.~\ref{fig11a}. $\langle\vec{S}\rangle$ points away from the critical plane $z=2.701~{\rm mm}$ for the TE mode and $z=2.541~{\rm mm}$ for the TM.}
    \label{fig11b}
    \end{center}
    \end{figure}

\begin{figure}
    \begin{center}
    \includegraphics[scale=.55]{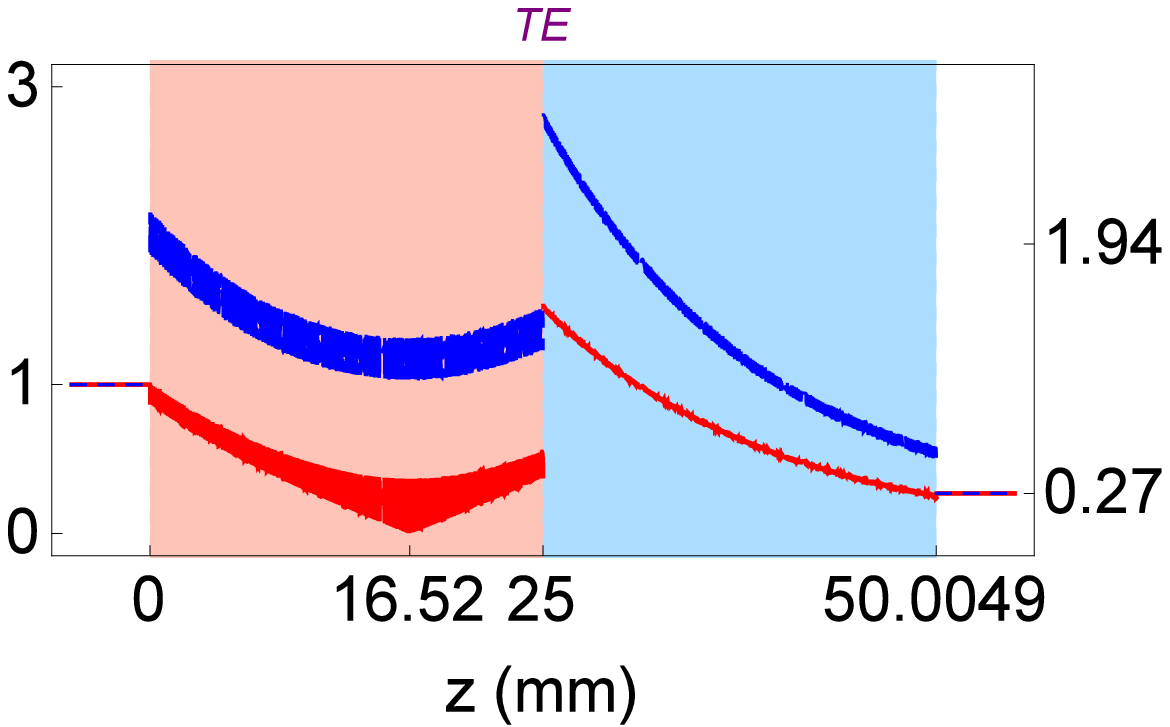}~~~~~~
    \includegraphics[scale=.55]{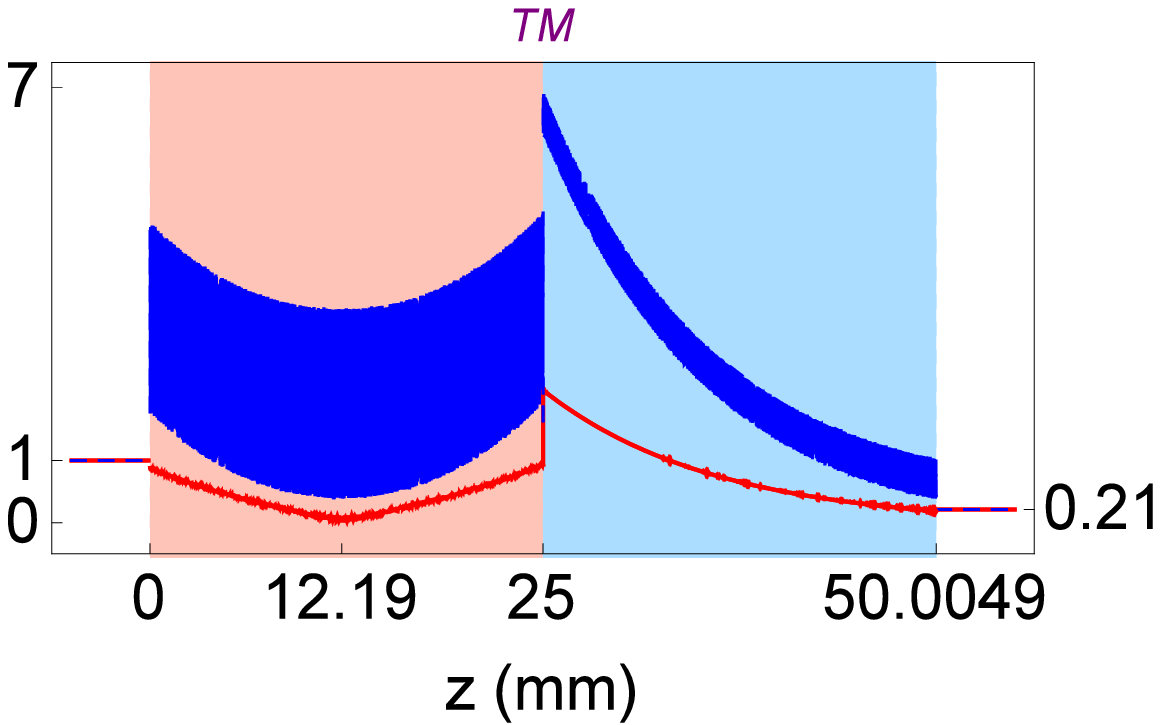}\\
    \includegraphics[scale=.55]{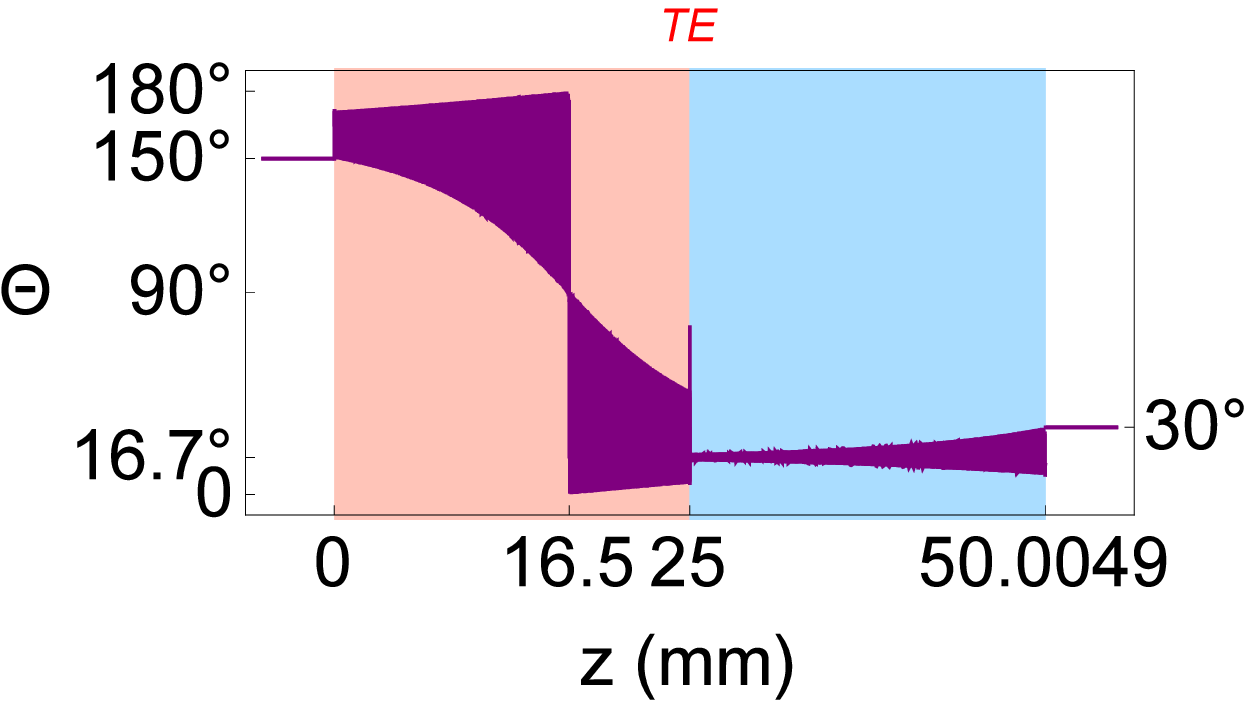}~~~~~~
    \includegraphics[scale=.55]{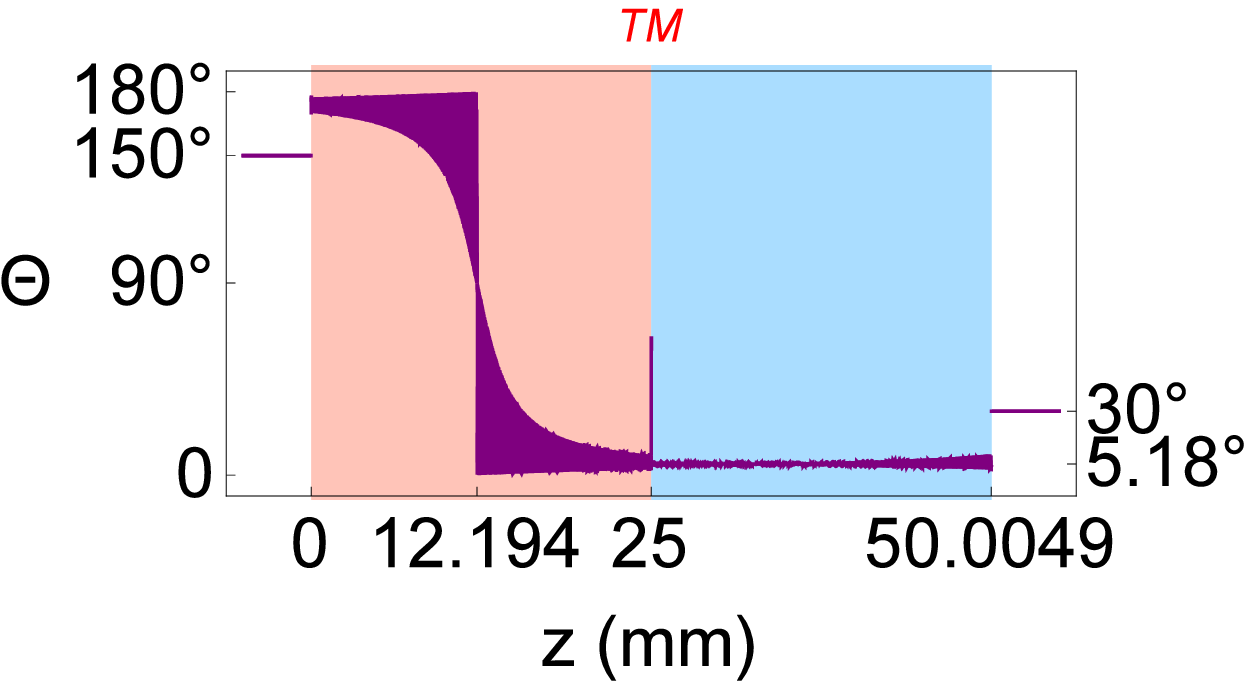}
    \caption{(Color online) Graphs of $\br u\kt$ in units of $\br u_I\kt$ (dashed navy curves) and $|\langle\vec{S}\rangle|$ in units of $|\langle\vec{S}_I\rangle|$ (solid red curves) on the top and the graphs of the angle $\Theta$ between $\langle\vec{S}\rangle$ and the positive $z$-axis in the bottom for the singular TE and TM modes of a destructive configuration of the $\cP\cT$-symmetric Nd:YAG two-slab system with specifications (\ref{sp-gen}) and (\ref{SEM-30=destructivespace}). In contrast to the case $s=0$, $\br u\kt$ and $|\langle\vec{S}\rangle|$ take much smaller values in the interior of the system.}
    \label{fig12a}
    \end{center}
    \end{figure}

\begin{figure}
    \begin{center}
    \includegraphics[scale=.55]{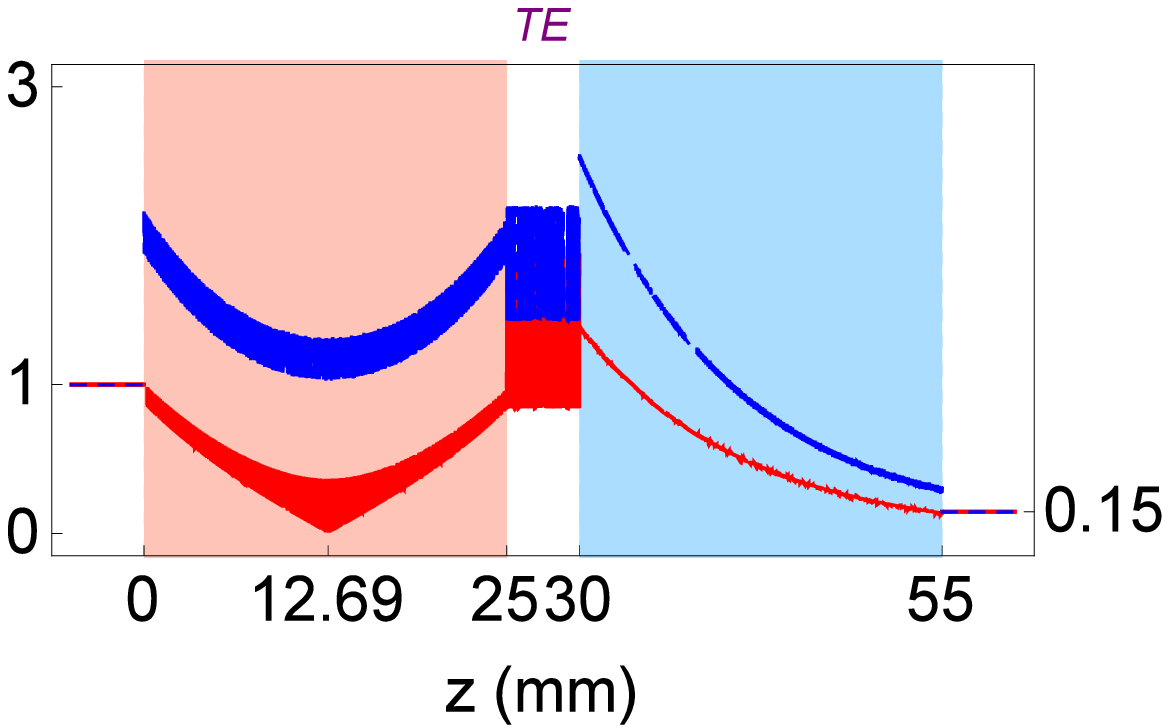}~~~~~
    \includegraphics[scale=.55]{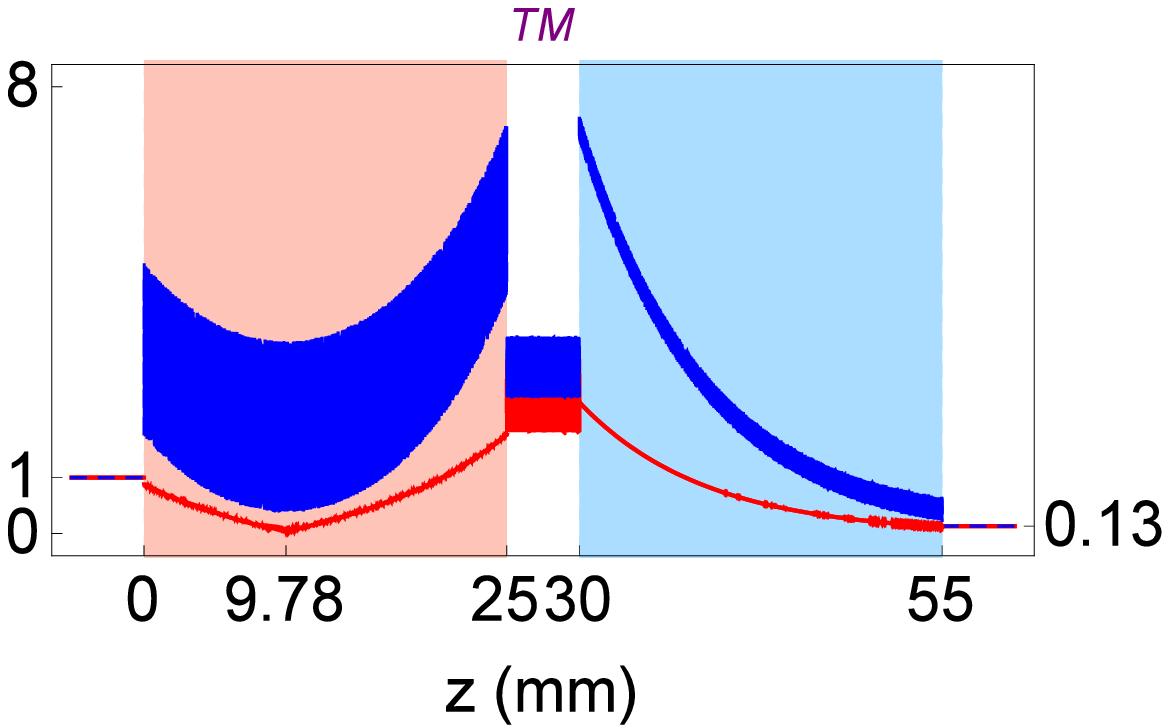}\\
    \includegraphics[scale=.55]{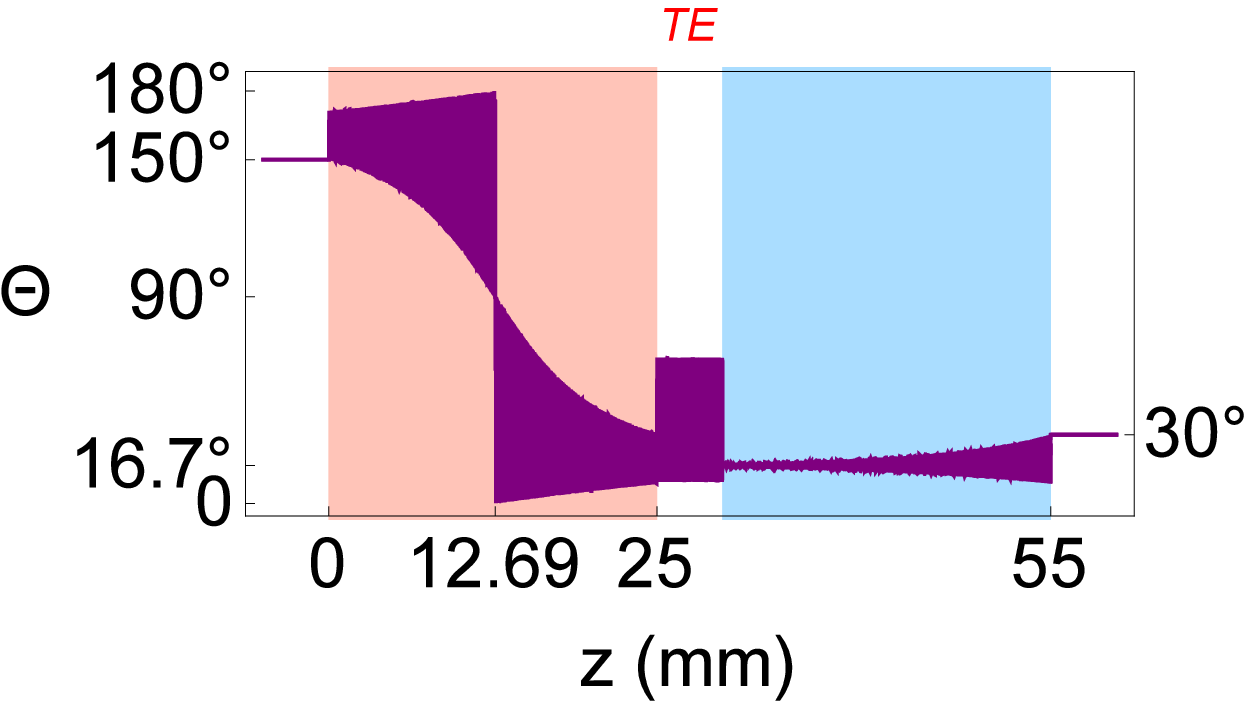}~~~~~~
    \includegraphics[scale=.55]{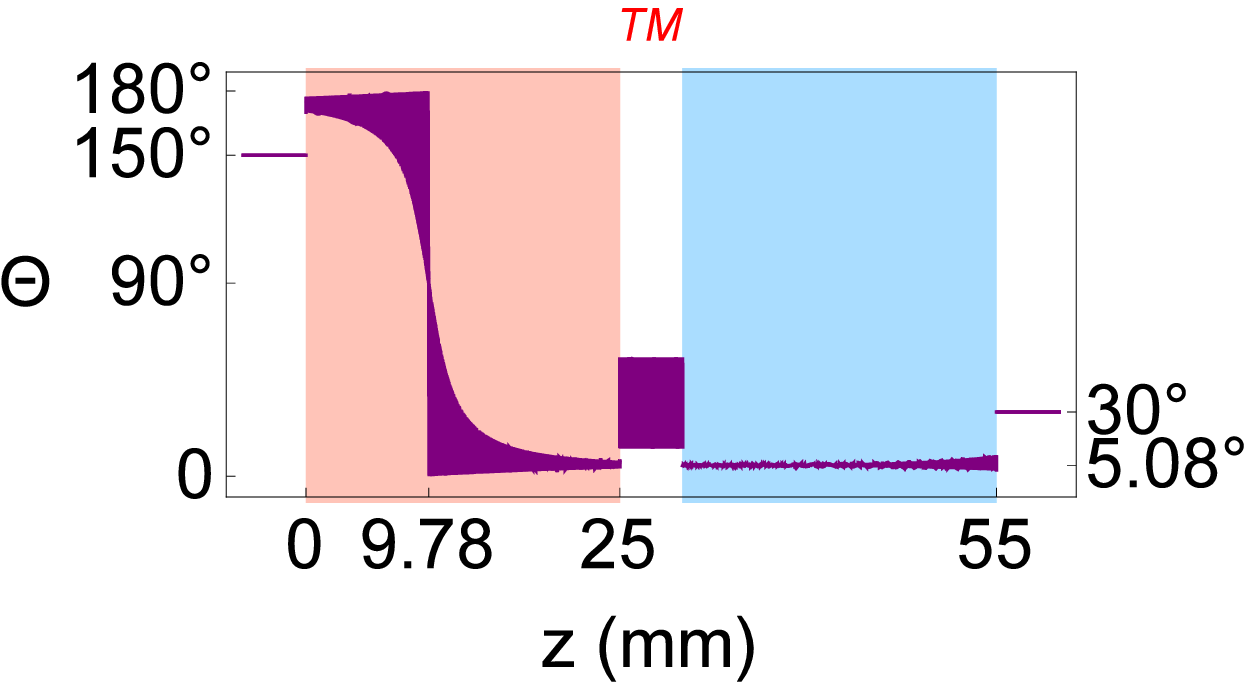}
    \caption{(Color online) Graphs of $\br u\kt$ in units of $\br u_I\kt$ (dashed navy curves) and $|\langle\vec{S}\rangle|$ in units of $|\langle\vec{S}_I\rangle|$ (solid red curves) on the top, and the graphs of the angle $\Theta$ between $\langle\vec{S}\rangle$ and the positive $z$-axis in the bottom for the singular TE and TM modes of a generic configuration of the $\cP\cT$-symmetric Nd:YAG two-slab system with specifications  (\ref{sp-gen}) and (\ref{SEM-30=arbitraryspace}).}
    \label{fig13a}
    \end{center}
    \end{figure}

Figures~\ref{fig11a}-\ref{fig13a} show the plots of $\br u\kt$, $|\br\vec S\kt|$, and the angle $\Theta$ of Eq.~(\ref{big-theta}) for the singular TE and TM modes of a $\cP\cT$-symmetric Nd:YAG two-slab system with the following specifications.
    \begin{align}
    &L=25~{\rm mm},~~~~~~~~~~\eta=1.8217,~~~~~~~~~~~\theta=30^\circ,
    \label{sp-gen}\\[6pt]
    &\left\{
    \begin{aligned}
    &s^{(E/M)}=20 s_0=4.66\,\mu{\rm m}, && \lambda^{(E/M)}=807.996~{\rm nm},\\
    &g^{(E)}=3.8936\,{\rm cm}^{-1}, && g^{(M)}=4.1391\,{\rm cm}^{-1},
    \end{aligned}\right.
    \label{SEM-30=constructivespace}\\[6pt]
    &\left\{
    \begin{aligned}
    &s^{(E/M)}=21s_0=4.90\,\mu{\rm m}, && \lambda^{(E/M)}=807.998~{\rm nm},\\
    &g^{(E)}=0.6366\,{\rm cm}^{-1}, && g^{(M)}=0.8626\,{\rm cm}^{-1},
    \end{aligned}\right.
    \label{SEM-30=destructivespace}\\[6pt]
    &\left\{
    \begin{aligned}
    &s^{(E/M)}=21436.35 s_0=5.00\,{\rm mm}, && \lambda^{(E/M)}=807.997~{\rm nm},\\
    & g^{(E)}=0.8291\,{\rm cm}^{-1}, && g^{(M)}=1.0756\,{\rm cm}^{-1}.
    \end{aligned}\right.
    \label{SEM-30=arbitraryspace}
    \end{align}
Note that (\ref{SEM-30=constructivespace}), (\ref{SEM-30=destructivespace}), and (\ref{SEM-30=arbitraryspace}) correspond to constructive, destructive, and generic configurations of the system, respectively.

\begin{figure}
    \begin{center}
    \includegraphics[scale=.44]{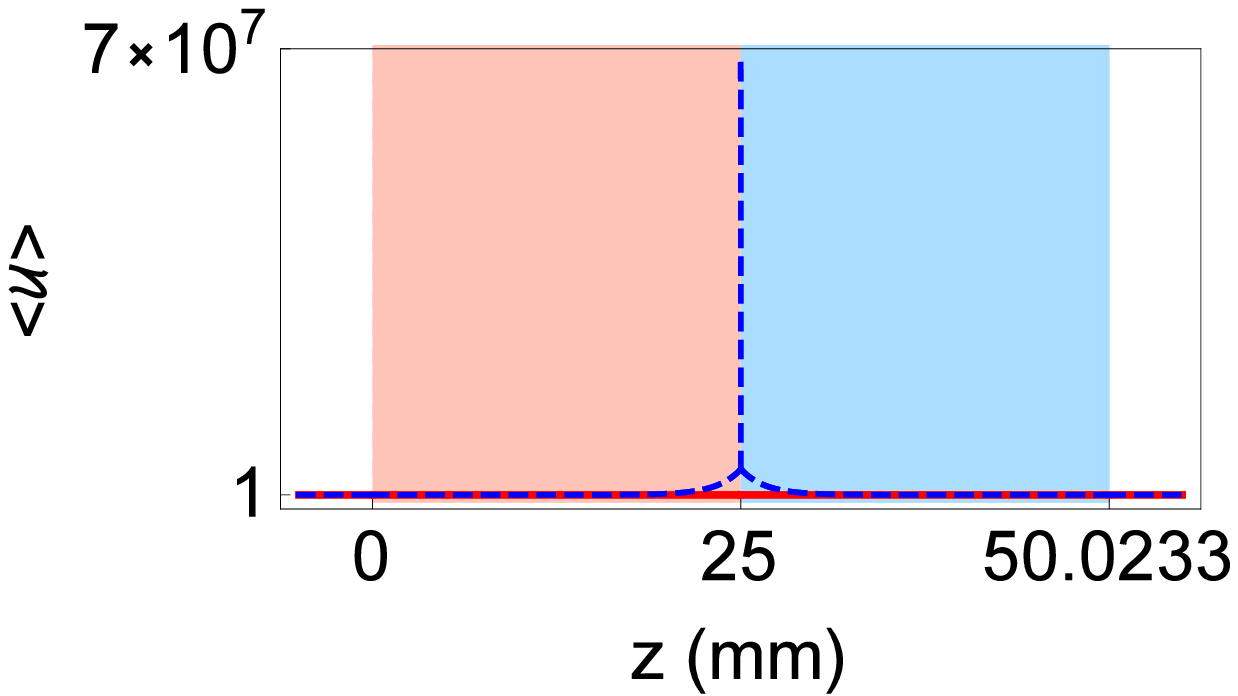}~~~~
    \includegraphics[scale=.44]{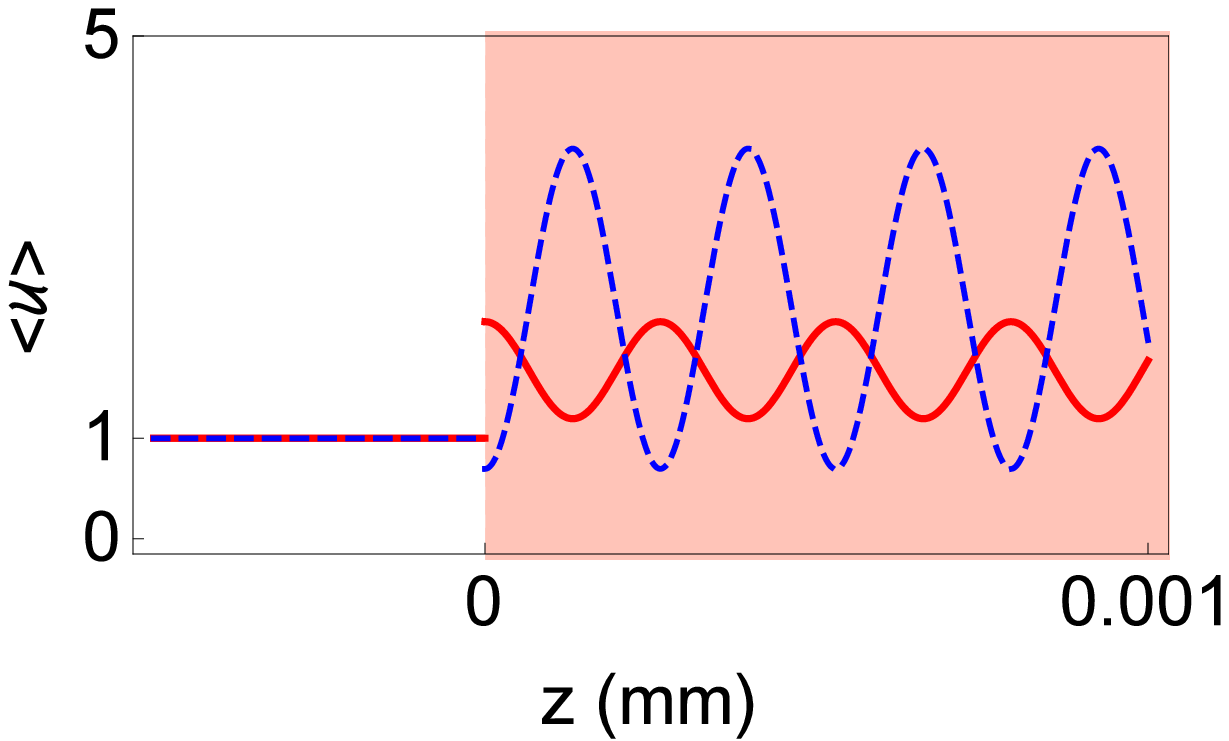}~~~~
    \includegraphics[scale=.44]{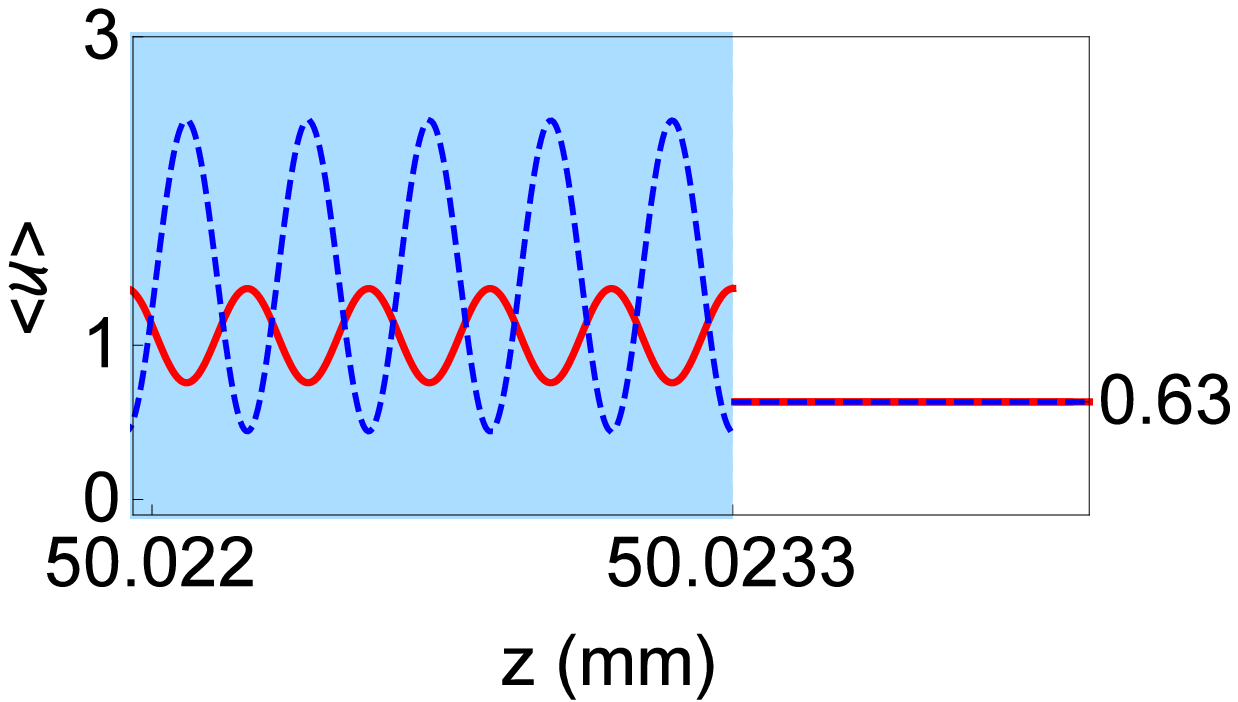}\\
    \includegraphics[scale=.44]{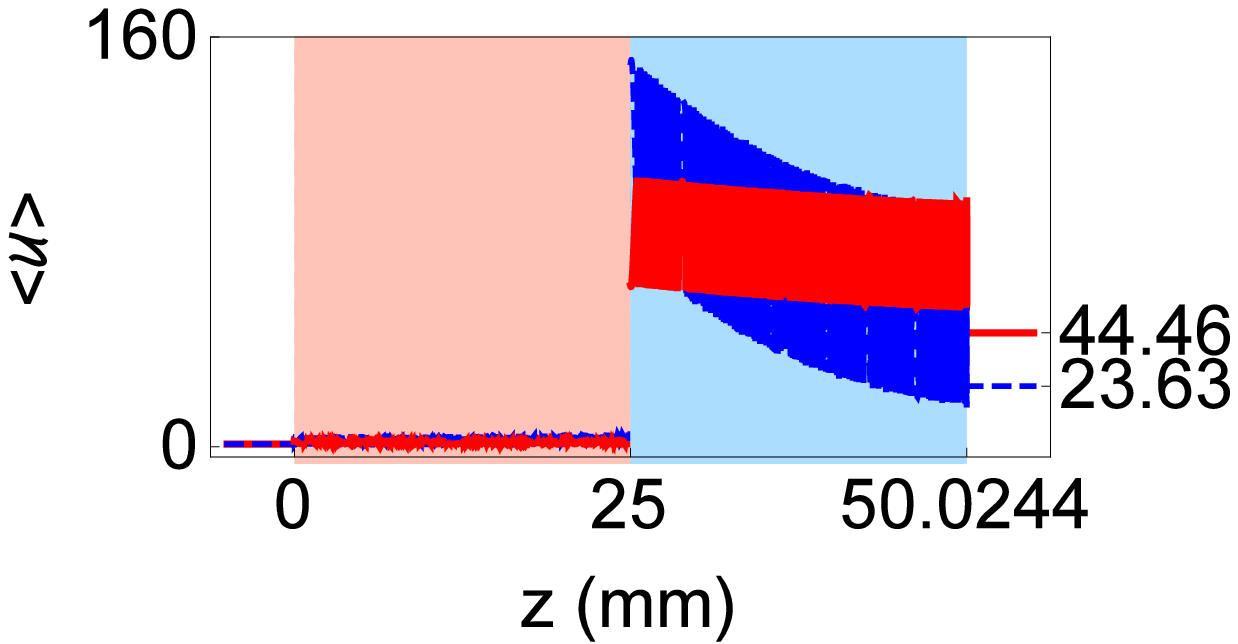}~~~~
    \includegraphics[scale=.44]{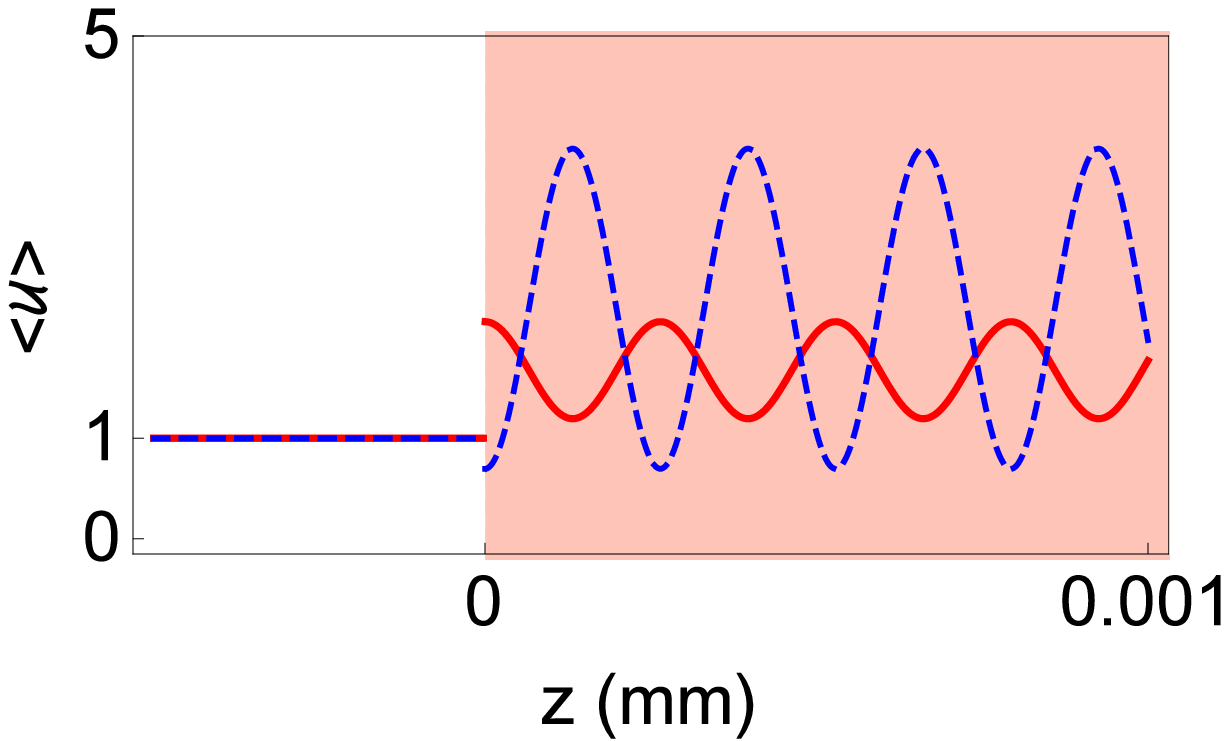}~~~~
    \includegraphics[scale=.44]{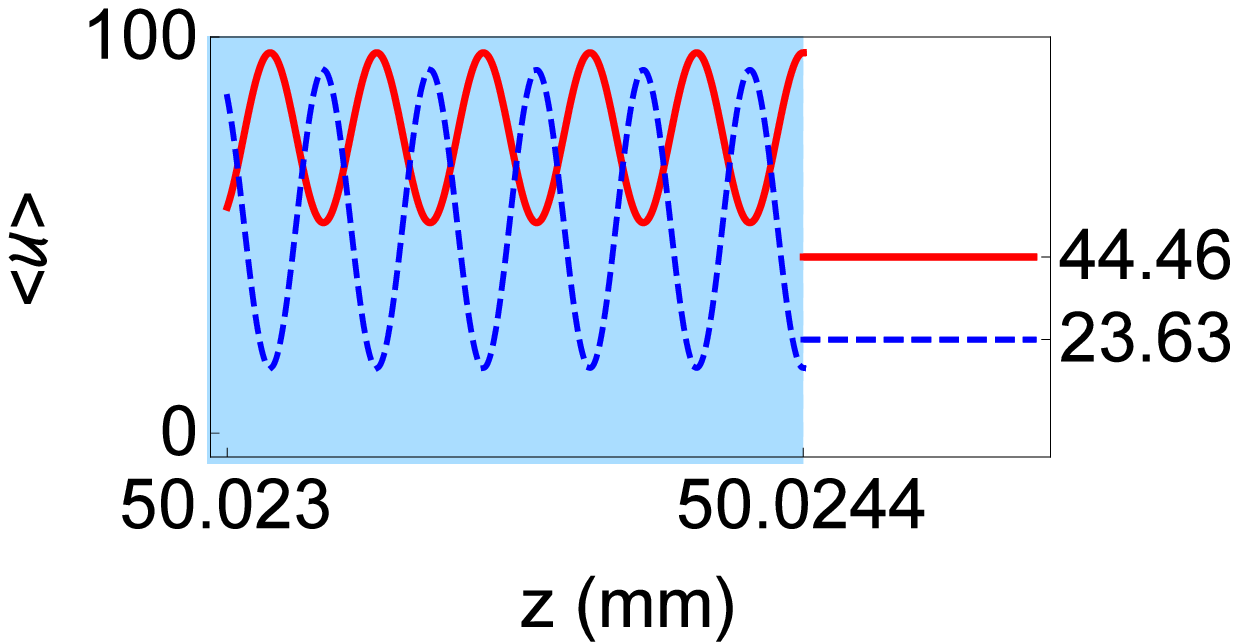}\\
    \includegraphics[scale=.44]{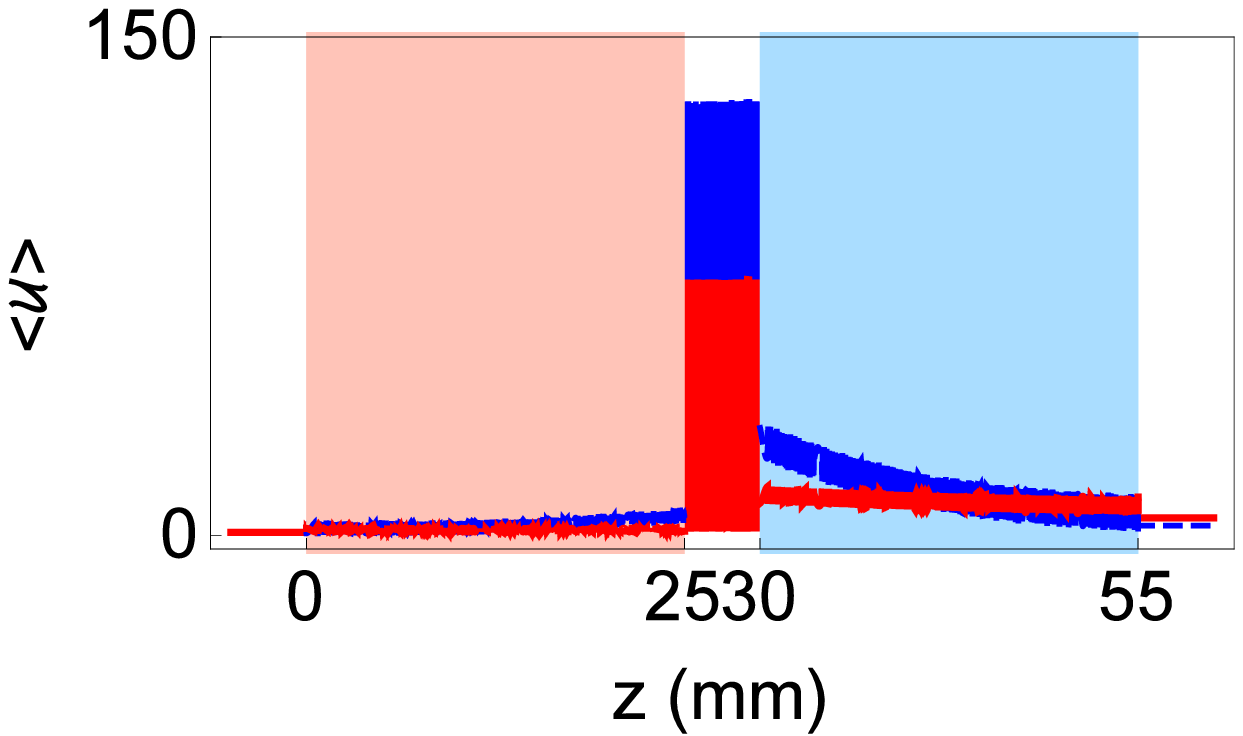}~~~~
    \includegraphics[scale=.44]{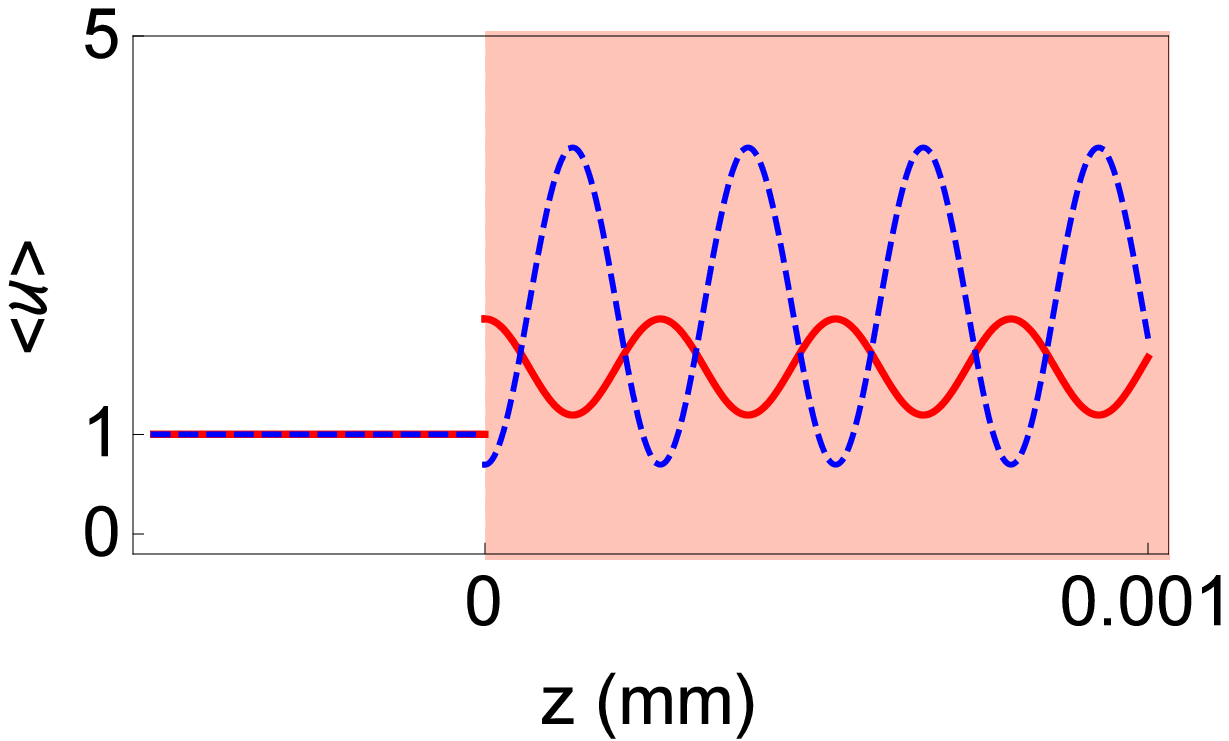}~~~~
    \includegraphics[scale=.44]{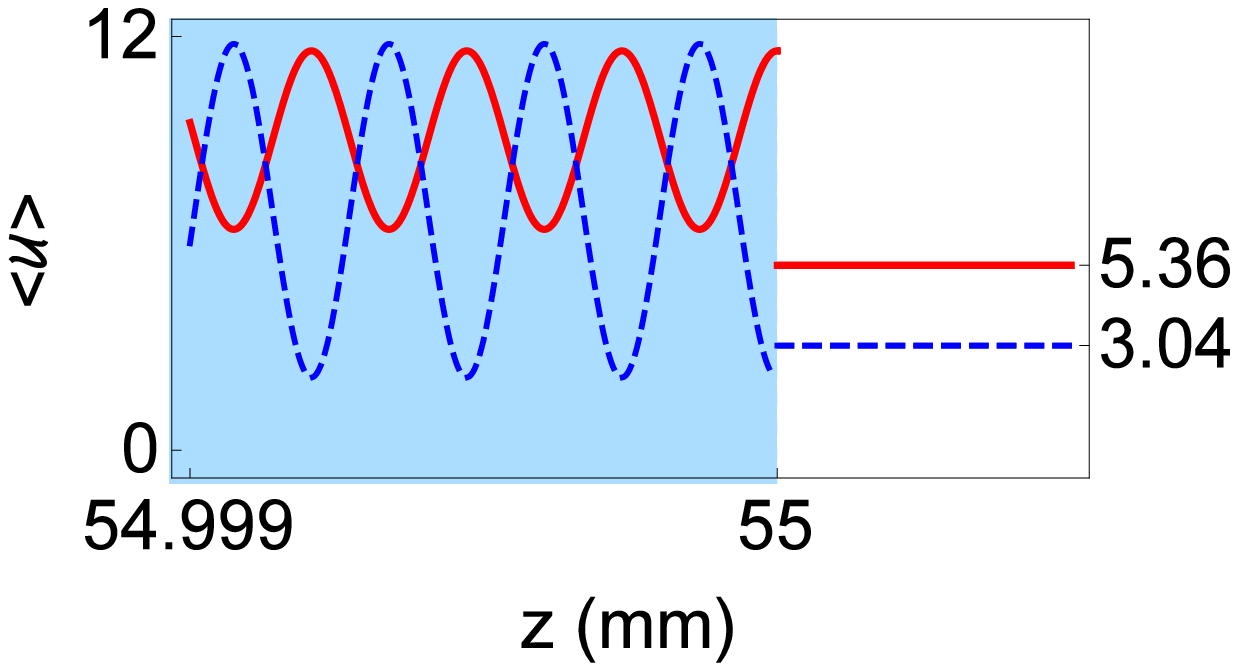}
    \caption{(Color online) Graphs of $\langle u \rangle$ (in units of $\langle u_I\kt$) as a function of $z$ for the singular TE (solid red curves) and TM (blue dashed curves) modes of a $\cP\cT$-symmetric Nd:YAG two-slab system with specifications (\ref{sp-gen-80}) -- (\ref{SEM-80=g}). Plots depicted in the top, middle, and bottom rows correspond to
    the constructive, destructive, and generic configurations given by (\ref{SEM-80=c}), (\ref{SEM-80=d}), and (\ref{SEM-80=g}), respectively.}
    \label{fig14}
    \end{center}
    \end{figure}

We have also made a graphical analysis of the singular TE and TM modes of similar $\cP\cT$-symmetric slab systems that are made of other material and for different values of $\theta$. The results show no qualitative differences from those of the particular cases we have discussed above except that for sufficiently large values of $\theta$ the time-averaged energy density of the emitted waves from the lossy layer, $\br u_V\kt$, can take larger values than that for the emitted waves from the gain layer, $\br u_I\kt$. This surprising effect is present provided that the system is not in one of its constructive configurations, i.e., $s$ is not an even multiple of $s_0$. It is most prevalent for the destructive configurations. Figure~\ref{fig14} provides a graphical demonstration of this behavior. It shows the plots of $\br u\kt$ for the TE and TM modes of a $\cP\cT$-symmetric Nd:YAG two-slab system with the following values for the relevant physical parameters.
    \begin{align}
    &L=25~{\rm mm},~~~~~~~~~~\eta=1.8217,~~~~~~~~~~~\theta=80^\circ,
    \label{sp-gen-80}\\[6pt]
    &\left\{
    \begin{aligned}
    &s^{(E/M)}=20 s_0=23.265\,\mu{\rm m}, \\
    &\lambda^{(E)}=807.996~{\rm nm},&& g^{(E)}=2.656\,{\rm cm}^{-1},\\
    &\lambda^{(M)}=808.000~{\rm nm},&& g^{(M)}=5.140\,{\rm cm}^{-1},
    \end{aligned}\right.
    \label{SEM-80=c}\\[6pt]
    &\left\{
    \begin{aligned}
    &s^{(E/M)}=21 s_0=24.429\,\mu{\rm m}, \\
    &\lambda^{(E)}=808.032~{\rm nm},&& g^{(E)}=0.109\,{\rm cm}^{-1},\\
    &\lambda^{(M)}=807.998~{\rm nm},&& g^{(M)}=0.394\,{\rm cm}^{-1},
    \end{aligned}\right.
    \label{SEM-80=d}\\[6pt]
    &\left\{
    \begin{aligned}
    &s^{(E/M)}=4298.24 s_0=5.000\,{\rm mm}, \\
    &\lambda^{(E)}=807.997~{\rm nm},&& g^{(E)}=0.215\,{\rm cm}^{-1},\\
    &\lambda^{(M)}=807.996~{\rm nm},&& g^{(M)}=0.638\,{\rm cm}^{-1},
    \end{aligned}\right.
    \label{SEM-80=g}
    \end{align}
Clearly (\ref{SEM-80=c}), (\ref{SEM-80=d}), and (\ref{SEM-80=g}) respectively correspond to the constructive, destructive, and generic configurations.
Further examination shows that there is always a critical angle $\theta_c$ such that $\br u_I\kt\geq \br u_V\kt$ for $\theta\leq\theta_c$. For $\theta>\theta_c$, both $\br u_I\kt\geq \br u_V\kt$ and $\br u_I\kt<\br u_V\kt$ can occur. The latter case dominates for larger values of
$\theta-\theta_c$.

Comparing the numerical values of the threshold gain coefficients listed in (\ref{SEM-30=constructivespace}) -- (\ref{SEM-30=arbitraryspace}) and (\ref{SEM-80=c}) -- (\ref{SEM-80=g}), we see that they are smaller for the destructive configurations than the generic and constructive ones. This reveals another advantage of destructive configurations for the purpose of using the system as a laser or a CPA. We have indeed checked that the peaks in the graphs of the threshold gain coefficient that are given in Fig.~\ref{g0theta2n} correspond to the constructive configurations of the system while their minima give the destructive configurations.

\section{Amplitude and Phase Conditions for the CPA Action}
\label{S8}

An optical system functions as a CPA provided that the condition for the realization of a spectral singularity is realized for the time-reversed system. For a $\cP\cT$-symmetric system this coincides with the condition for having a spectral singularity for the system itself. It is however important to notice that this is just a necessary condition. A system fulfilling this condition would absorb incoming waves only if they have appropriate amplitude and phase. For the $\cP\cT$-symmetric slab systems we consider in this article, we have already derived the condition for the emergence of a spectral singularity. In what follows we determine the amplitude and phase conditions for the incoming waves that are absorbed by these systems.

Consider a $\cP\cT$-symmetric slab system of the form depicted in Fig.~\ref{fig1} with $\fn_2=\fn_1^*=\fn^*$. Suppose that the system supports a spectral singularity in a TE or TM mode with wavenumber $k$ and angle $\theta$. Then for $z\notin[0,2L+z]$, i.e., outside the system, the electric field $\vec E(x,y,z)$ for the TE wave and the magnetic field $\vec H(x,y,z)$ for the TM wave have the form
	\be
	\left\{\begin{aligned}
	& b_1 e^{i(k_xx-k_z z)}\hat e_y && {\rm for}~z\in I,\\
	& a_5 e^{i(k_xx+k_z z)}\hat e_y && {\rm for}~z\in V.
	\end{aligned}\right.
    \label{f17a}
	\ee
This in particular shows that, for $z\notin[0,2L+z]$, the corresponding time-reversed waves that are absorbed by the time-reversed system are given by
	\be
	\left\{\begin{aligned}
	& b_1^* e^{-i(k_xx-k_z z)}\hat e_y && {\rm for}~z\in I,\\
	& a_5^* e^{-i(k_xx+k_z z)}\hat e_y && {\rm for}~z\in V.
	\end{aligned}\right.
	\label{time-rev-wave}
	\ee
	
\begin{figure}
    \begin{center}
    \includegraphics[scale=.7]{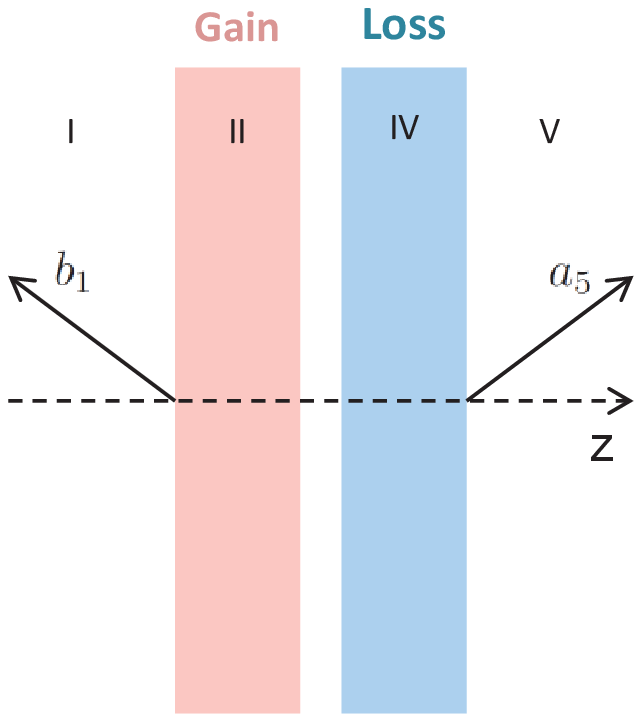}~~~~~
    \includegraphics[scale=.7]{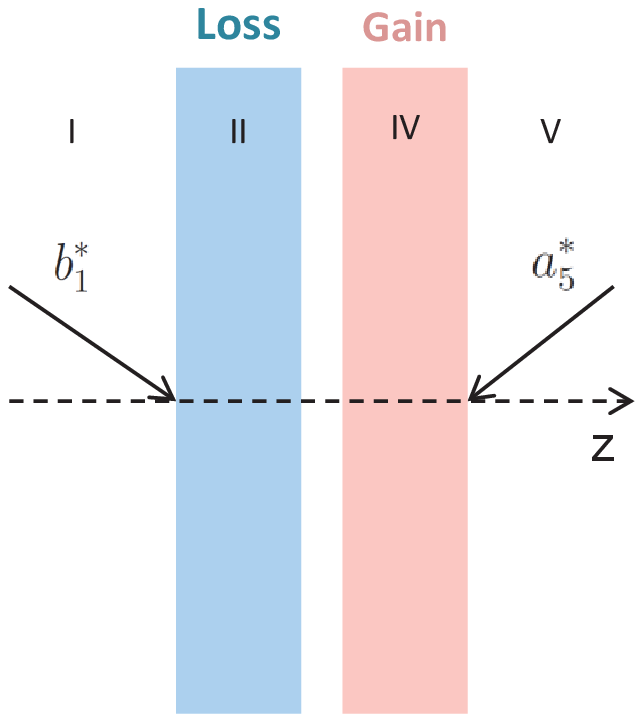}~~~~~
    \includegraphics[scale=.7]{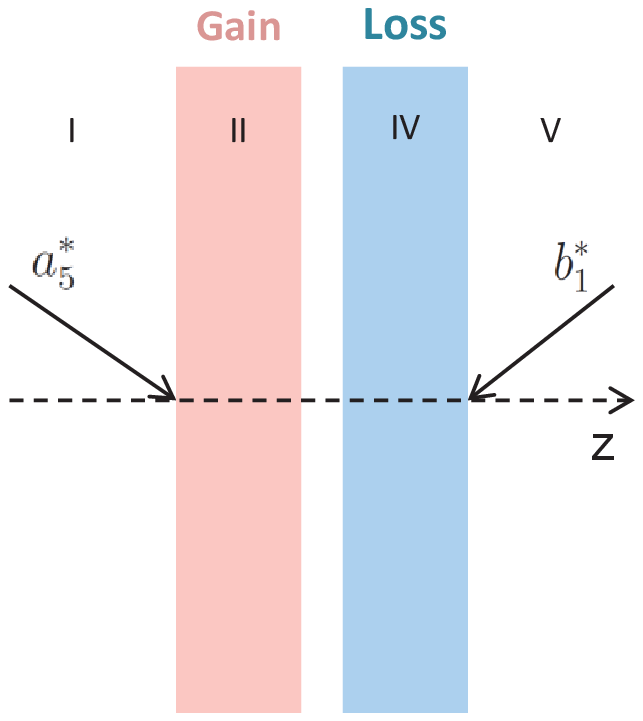}
    \caption{(Color online) The schematic description of the waves emitted by the CPA-laser as given by (\ref{f17a}) on the left, the waves absorbed by the time-reversed system corresponding to (\ref{time-rev-wave}) in the middle, and the waves absorbed by the original CPA-laser as quantified by (\ref{CPA-wave}) on the right.}
    \label{fig17}
    \end{center}
    \end{figure}

The time-reversed system is obtained by swapping the gain and loss layers, i.e., $\fn\to\fn^*$. In light of the equivalence between this operation and the reflection $z\to 2L+s-z$, we infer that the original system (where the gain component is to the left of the lossy one) would absorb waves of the form
	\be
	\left\{\begin{aligned}
	& a_5^* e^{-ik_z(2L+s)}e^{-i(k_xx-k_z z)}\hat e_y && {\rm for}~z\in I,\\
	& b_1^* e^{ik_z(2L+s)} e^{-i(k_xx+k_z z)}\hat e_y && {\rm for}~z\in V.
	\end{aligned}\right.
	\label{CPA-wave}
	\ee
In particular, the incidence angle at $z=0$ is given by $-\theta$, where $\theta$ was the angle determining the singular TE and TM waves in the preceding sections. Figure~\ref{fig17} shows the emitted and absorbed waves given by (\ref{f17a}) -- (\ref{CPA-wave}).
	
Because $b_1$ is a free parameter, it is the ratio $\rho$ of the complex amplitude of the incoming waves for $z\to 0$ and $z\to 2L+s$ that is crucial for their absorption by the system. In light of (\ref{zq2}) and (\ref{CPA-wave}), this is given by
 	\be
	\rho=\frac{e^{-ik_z(2L+s)}a_5^*}{b_1^*}=V_s(\tilde\fn,\tilde\fn^*)^*,
	\ee
where $V_s(\tilde\fn,\tilde\fn^*)$ is introduced in (\ref{Vs-def}). In particular, the ratio of the amplitudes and phase factors for the incident waves from the left to that from the right are respectively given by
	\begin{align}
	&|\rho|=|V_s(\tilde\fn,\tilde\fn^*)|, &&e^{i\delta\phi}=\frac{V_s(\tilde\fn,\tilde \fn^*)^*}{|V_s(\tilde\fn,\tilde\fn^*)|}.
	\end{align}
Tables~\ref{table04}-\ref{table06} give the numerical values of the physical parameters for which  various configurations of our $\cP\cT$-symmetric slab system perform as a CPA for different incident TE and TM waves.
	\begin{table}[!htbp]
	\begin{center}
	\begin{tabular}{|c|c|c|c|c|}
  	\hline
   	& Bilayer & Construtive Conf. & Destructive Conf.\ & Generic Conf.\ \\
   	\hline
  	$s^{TE/TM}$ & 0 & $4.040~\mu{\rm m}$ & $4.242~\mu{\rm m}$ & $5.000~{\rm mm}$ \\
  	\hline
  	$\kappa^{TE}$ & $-7.351\times 10^{-5}$ & $-8.804\times 10^{-6}$ & $-4.927\times 10^{-6}$ & 	 $-5.615\times 10^{-6}$ \\
  	\hline
  	$\kappa^{TM}$ & $-7.351\times 10^{-5}$ & $-2.813\times 10^{-5}$ & $-4.927\times 10^{-6}$ & $-5.615\times 10^{-6}$ \\
  	\hline
  	$g^{TE}$ & $11.433~{\rm cm^{-1}}$ & $1.369~{\rm cm^{-1}}$ & $0.766~{\rm cm^{-1}}$ & $0.873~{\rm cm^{-1}}$ \\
  	\hline
  	$g^{TM}$ & $11.433~{\rm cm^{-1}}$ & $4.375~{\rm cm^{-1}}$ & $0.766~{\rm cm^{-1}}$ & $0.873~{\rm cm^{-1}}$ \\
  	\hline
  	$\lambda^{TE}$ & $808.006~{\rm nm}$ & $808.005~{\rm nm}$ & $807.100~{\rm nm}$ & $807.999~{\rm nm}$ \\
  	\hline
  	$\lambda^{TM}$ & $808.006~{\rm nm}$ & $807.998~{\rm nm}$ & $807.998~{\rm nm}$ & $807.999~{\rm nm}$ \\
  	\hline
  	$|\rho^{TE}|$ & 0.2912 & 0.2912 & 0.2515 & 0.2687 \\
  	\hline
  	$|\rho^{TM}|$ & 0.2912 & 0.2912 & 0.2515 & 0.2687 \\
  	\hline
  	$\delta\phi^{TE}$ & $93.058^{\circ}$ & $135.513^{\circ}$ & $90.022^{\circ}$ & $109.799^{\circ}$ \\
  	\hline
  	$\delta\phi^{TM}$ & $93.058^{\circ}$ & $135.513^{\circ}$ & $90.022^{\circ}$ & $109.799^{\circ}$ \\
  	\hline
	\end{tabular}
	\vspace{6pt}
    	\caption{Physical parameters for the coherent perfect absorption of the TE and TM modes with incidence angle $\theta = 0^{\circ}$. The constructive and destructive configurations correspond to the $s/s_0=10$ and $s/s_0=21$, respectively.}
    	\label{table04}
    	\end{center}
   	 \end{table}%
	\begin{table}[!htbp]
    	\begin{center}
	\begin{tabular}{|c|c|c|c|c|}
  	\hline
   	& Bilayer & Construtive Conf. & Destructive Conf.\ & Generic Conf.\ \\
   	\hline
  	$s^{TE/TM}$ & 0 & $4.665~\mu{\rm m}$ & $4.898~\mu{\rm m}$ & $5.000~{\rm mm}$ \\
  	\hline
  	$\kappa^{TE}$ & $-6.977\times 10^{-5}$ & $-2.503\times 10^{-5}$ & $-4.093\times 10^{-6}$ & $-5.331\times 10^{-6}$ \\
  	\hline
  	$\kappa^{TM}$ & $-7.230\times 10^{-5}$ & $-2.661\times 10^{-5}$ & $-5.546\times 10^{-6}$ & $-6.916\times 10^{-6}$ \\
  	\hline
  	$g^{TE}$ & $10.851~{\rm cm^{-1}}$ & $3.894~{\rm cm^{-1}}$ & $0.637~{\rm cm^{-1}}$ & $0.829~{\rm cm^{-1}}$ \\
  	\hline
  	$g^{TM}$ & $11.244~{\rm cm^{-1}}$ & $4.139~{\rm cm^{-1}}$ & $0.863~{\rm cm^{-1}}$ & $1.076~{\rm cm^{-1}}$ \\
  	\hline
  	$\lambda^{TE/TM}$ & $807.993~{\rm nm}$ & $807.996~{\rm nm}$ & $807.998~{\rm nm}$ & $807.997~{\rm nm}$ \\
  	\hline
  	$|\rho^{TE}|$ & 0.3194 & 0.3194 & 0.5184 & 0.3812 \\
  	\hline
  	$|\rho^{TM}|$ & 0.3194 & 0.3194 & 0.4604 & 0.3622 \\
  	\hline
  	$\delta\phi^{TE}$ & $154.018^{\circ}$ & $131.050^{\circ}$ & $179.346^{\circ}$ & $151.612^{\circ}$ \\
  	\hline
  	$\delta\phi^{TM}$ & $153.903^{\circ}$ & $130.996^{\circ}$ & $179.083^{\circ}$ & $152.690^{\circ}$ \\
  	\hline
	\end{tabular}
	\vspace{6pt}
    	\caption{Physical parameters for the coherent perfect absorption of the TE and TM modes with incidence angle $\theta = -30^{\circ}$. The constructive and destructive configurations correspond to the $s/s_0=10$ and $s/s_0=21$, respectively.}
    	\label{table05}
    	\end{center}
    	\end{table}%
	\begin{table}[!htbp]
    	\begin{center}
	\begin{tabular}{|c|c|c|c|c|}
  	\hline
  	& Bilayer & Construtive Conf. & Destructive Conf.\ & Generic Conf.\ \\
   	\hline
  	$s^{TE/TM}$ & 0 & $23.265~\mu{\rm m}$ & $24.429~\mu{\rm m}$ & $5.000~{\rm mm}$ \\
  	\hline
  	$\kappa^{TE}$ & $-5.803\times 10^{-5}$ & $-1.708\times 10^{-5}$ & $-6.994\times 10^{-7}$ & $-1.383\times 10^{-6}$ \\
  	\hline
  	$\kappa^{TM}$ & $-6.368\times 10^{-5}$ & $-3.305\times 10^{-5}$ & $-2.533\times 10^{-6}$ & $-4.104\times 10^{-6}$ \\
  	\hline
  	$g^{TE}$ & $9.025~{\rm cm^{-1}}$ & $2.656~{\rm cm^{-1}}$ & $0.109~{\rm cm^{-1}}$ & $0.215~{\rm cm^{-1}}$ \\
  	\hline
  	$g^{TM}$ & $9.903~{\rm cm^{-1}}$ & $5.140~{\rm cm^{-1}}$ & $0.394~{\rm cm^{-1}}$ & $0.638~{\rm cm^{-1}}$ \\
  	\hline
  	$\lambda^{TE}$ & $808.009~{\rm nm}$ & $807.996~{\rm nm}$ & $808.032~{\rm nm}$ & $807.997~{\rm nm}$ \\
  	\hline
  	$\lambda^{TM}$ & $807.999~{\rm nm}$ & $808.000~{\rm nm}$ & $807.998~{\rm nm}$ & $807.996~{\rm nm}$ \\
  	\hline
  	$|\rho^{TE}|$ & 0.7959 & 0.7959 & 6.6679 & 2.3158 \\
  	\hline
  	$|\rho^{TM}|$ & 0.7959 & 0.7959 & 4.8607 & 1.7425 \\
  	\hline
  	$\delta\phi^{TE}$ & $89.733^{\circ}$ & $127.382^{\circ}$ & $179.738^{\circ}$ & $16.792^{\circ}$ \\
  	\hline
  	$\delta\phi^{TM}$ & $95.976^{\circ}$ & $38.371^{\circ}$ & $176.942^{\circ}$ & $9.002^{\circ}$ \\
  	\hline
	\end{tabular}
	\vspace{6pt}
    	\caption{Physical parameters for the coherent perfect absorption of the TE and TM modes with incidence angle $\theta = -80^{\circ}$. The constructive and destructive configurations correspond to the $s/s_0=10$ and $s/s_0=21$, respectively.}
    	\label{table06}
    	\end{center}
    	\end{table}

\section{Concluding Remarks}
\label{S9}

The phenomenon of coherent perfect absorption of electromagnetic waves corresponds to the time-reversal of lasing at the threshold gain. The basic mathematical concept describing the latter is the spectral singularity of a complex scattering potential. Spectral singularities are given by the real zeros of the $M_{22}$ entry of the transfer matrix of the potential, while their time-reversal correspond to the real zeros of $M_{11}$. Because for a $\cP\cT$-symmetric potential these coincide, $\cP\cT$-lasers act also as a coherent perfect absorber. Such a CPA-laser would absorb incident waves from the left and right directions only if they have correct amplitude and phase contrasts. In this article we have examined in great detail the conditions for achieving CPA-laser action in a $\cP\cT$-symmetric slab system with adjacent or separated gain and loss layers. This is actually the simplest experimentally realizable model for a CPA-laser that allows for a completely analytic treatment.

Our results show that the presence of the separation between the gain and loss layers can have dramatic effects on the performance of this system as a CPA-laser. In particular if the separation distance $s$ is an even integer multiple of a characteristic length scale, namely $s_0:=\pi/2\lambda\cos\theta$, the time-averaged energy density and the magnitude of the Poynting vector for the singular TE and TM waves take extremely large values inside the system. The opposite is the case when $s$ is an odd integer multiple of $s_0$. These, so-called destructive configurations, provide the optimal situations where we can safely ignore the effects of nonlinearities and operate the system either as a laser or a coherent perfect absorber with smaller values of gain and loss.

Another outcome of our study is the counterintuitive observation that for nonconstructive configurations of the system the intensity of the emitted wave from the lossy layer can be larger than the one from its gain layer. This effect occurs for sufficiently large values of the incidence (emission) angle $\theta$ and is more prevalent for the destructive configurations.

We have also derived an explicit formula for the amplitude and phase contrast for the incoming waves that are absorbed by our CPA-laser, and given the numerical values of the physical quantities for which various configurations of the system function as a CPA. Again for nonconstructive configurations with large values of $\theta$, we find that the intensity of the incoming wave that is absorbed by the gain layer can be larger than that of the wave absorbed by the lossy layer.
\vspace{12pt}

\noindent{\bf Note:} After the completion of this project, we were made aware of Ref.~\cite{baum} where the author explore prospects of an experimental realization of CPA-laser action in a similar
non-$\cP\cT$-symmetric system. 
\vspace{12pt}

\noindent{\bf Acknowledgments:}  We are grateful to Ali Serpeng\"{u}zel for fruitful discussions. This work has been supported by  the Scientific and Technological Research Council of Turkey (T\"UB\.{I}TAK) in the framework of the project no: 112T951, and by the Turkish Academy of Sciences (T\"UBA).

\section*{Appendix: Poynting vector and energy density of the singular waves}

By definition the time-averaged Poynting vector and energy density \cite{jackson} are respectively given by    		
    \begin{align}
    &\langle\vec{S}\rangle = \frac{1}{2}\,\RE \left(\vec{E} \times \vec{H}^*\right),
    \label{S-def}\\
    &\br u\kt := \frac{1}{4}\RE\left(\vec{E}\cdot \vec{D}^* +\vec{B}\cdot \vec{H}^*\right)=
    \frac{1}{4}\left(\epsilon_0\RE[\fz(z)]|\vec{E}|^2 +\mu_0|\vec{H}|^2\right),
    \label{u-def}
    \end{align}
where ``$\RE$'' stands for the real part of its argument. In what follows we compute these quantities for the singular TE and TM modes of our slab system. This requires substituting the formulas given in Table~\ref{table03} in Eqs.~(\ref{S-def}) and (\ref{u-def}). The result is as follows. For $s=0$,
    \begin{align}
    &\langle\vec{S}^{(E/M)}\rangle=
    |\br\vec S_I^{(E/M)}\kt|\times \left\{\begin{array}{ccc}
    \sin\theta\, \hat e_x-\cos\theta\,\hat e_z & {\rm for} & z\in I,\\[3pt]
    \mathcal{Y}_0(\tilde{\fn}_1, z) \sin\theta\,\hat e_x+\mathcal{Z}^{(E/M)}_0(\tilde{\fn}_1, z)\cos\theta\,\hat e_z
     & {\rm for} & z\in I\!I,\\[6pt]
    \begin{aligned}
    &\cX_0(\tilde{\fn}_1, \tilde{\fn}_2, L)\big[\mathcal{Y}_0(\tilde{\fn}_2, 2L-z) \sin\theta\,\hat e_x\\
    &~~~~~~~~-\mathcal{Z}^{(E/M)}_0(\tilde{\fn}_2, 2L-z)\cos\theta\,\hat e_z\big]
    \end{aligned}
     & {\rm for} & z\in I\!V,\\[6pt]
    \cX_0(\tilde{\fn}_1, \tilde{\fn}_2, L)[\sin\theta\,\hat e_x+\cos\theta\, \hat e_z] & {\rm for} & z\in V,
    \end{array}\right.\nn
    \end{align}
    \begin{align}
    &\br u^{(E/M)}\kt = \br u^{(E/M)}_I\kt
    \times\left\{
    \begin{array}{ccc}
    1 & {\rm for} & z\in I,\\[3pt]
    \cU_0^{(E/M)}(\tilde{\fn}_1, z) & {\rm for} & z\in I\!I,\\[3pt]
    \cU_0^{(E/M)}(\tilde{\fn}_2, 2L-z)\cX_0(\tilde{\fn}_1, \tilde{\fn}_2, L)  & {\rm for} & z\in I\!V,\\[3pt]
    \cX_0(\tilde{\fn}_1, \tilde{\fn}_2, L) & {\rm for} & z\in V,\end{array}\right.
    \nn
    \end{align}
    \normalsize
where the superscripts $(E)$ and $(M)$ refer to the TE and TM waves respectively, $|\br\vec S_I^{(E)}\kt|:=|b_1|^2/2Z_0$, $|\br\vec S_I^{(M)}\kt|:=Z_0|b_1|^2/2$, $\cX_0(\tilde{\fn}_1, \tilde{\fn}_2, z):=\left|V_0(\tilde{\fn}_1, \tilde{\fn}_2, z)\right|^2$, $\mathcal{Y}_0(\tilde{\fn}, z):=\RE\left(\frac{\fu}{\tilde{\fn}}\right)\left|U_+(\tilde{\fn}, z)\right|^2$, $\mathcal{Z}^{(E)}_0(\tilde{\fn},z):=\RE\left\{U_+(\tilde{\fn}, z)U^*_-(\tilde{\fn}, z)\, \fu^* \right\}$, $\mathcal{Z}^{(M)}_0(\tilde{\fn}, z):=\RE\left\{U^*_+(\tilde{\fn}, z)\,U_-(\tilde{\fn}, z)\, \fu \right\}$, $\br u^{(E)}_I\kt:= \epsilon_0|b_1|^2/2$, $\br u^{(M)}_I\kt:=\mu_0|b_1|^2/2$,
    \begin{align}
	&\cU^{(E)}_0(\tilde{\fn}, z):=\frac{1}{2}\left\{
    [\sin^2\theta\,+\RE(\fn^2)]\,\mathcal{Y}_0(\tilde{\fn}, z)+
     \cos^2\theta \cW_0(\tilde{\fn}, z)\right\},\nn\\
	&\cU^{(M)}_0(\tilde{\fn}, z):=\frac{1}{2}\left\{
    \cos^2\theta\,\cW_0(\tilde{\fn}, z)+\left[\left(\RE\Big[\frac{\fu}{\tilde{\fn}}\Big]\right)^{-1}+ \sin^2\theta \right]\mathcal{Y}_0(\tilde{\fn},z)\right\},\nn\\
    &\cW_0(\tilde{\fn}, z):=\RE\left(\frac{\tilde{\fn}}{\fu}\right)\left|U_-(\tilde{\fn}, z)\right|^2\,\left|\fu\right|^2,\nn
	\end{align}
and $V_0$ and $U_\pm$ are given by (\ref{Vs-def}) and (\ref{Upm-def}). For $s>0$,
    \bea
    \langle\vec{S}^{(E/M)}\rangle&=&
    |\br\vec S_I^{(E/M)}\kt|\times \left\{\begin{array}{ccc}
    \sin\theta\, \hat e_x-\cos\theta\,\hat e_z & {\rm for} & z\in I,\\[3pt]
    \mathcal{Y}_0(\tilde{\fn}_1, z) \sin\theta\,\hat e_x+\mathcal{Z}^{(E/M)}_0 (\tilde{\fn}_1, z)\cos\theta\,\hat e_z
     & {\rm for} & z\in I\!I,\\[3pt]
     \mathcal{Y}_1(\tilde{\fn}_1, z-L) \sin\theta\,\hat e_x+\mathcal{Z}^{(E/M)}_1 (\tilde{\fn}_1, z-L)\cos\theta\,\hat e_z
     & {\rm for} & z\in I\!I\!I,\\[6pt]
     \begin{aligned}
     &\cX_s(\tilde{\fn}_1, \tilde{\fn}_2, L)\Big[\mathcal{Y}_0(\tilde{\fn}_2, 2L+s-z) \sin\theta\,\hat e_x\\
     &~~~~~~~~~-\mathcal{Z}^{(E/M)}_0 (\tilde{\fn}_2, 2L+s-z)\cos\theta\,\hat e_z\Big]
     \end{aligned}
     & {\rm for} & z\in I\!V,\\[6pt]
    \cX_s(\tilde{\fn}_1, \tilde{\fn}_2, L)[\sin\theta\,\hat e_x+\cos\theta\, \hat e_z] & {\rm for} & z\in V,
    \end{array}\right.
    \nn
    \eea
    \bea
    \br u^{(E/M)}\kt &=& \br u^{(E/M)}_I\kt
    \times\left\{
    \begin{array}{ccc}
    1 & {\rm for} & z\in I,\\[3pt]
    \cU^{(E/M)}_0(\tilde{\fn}_1, z) & {\rm for} & z\in I\!I,\\[3pt]
    \cU_1(\tilde{\fn}_1, z-L)  & {\rm for} & z\in I\!I\!I,\\[3pt]
     \cX_s(\tilde{\fn}_1, \tilde{\fn}_2, L)\,\cU^{(E/M)}_0(\tilde{\fn}_2, 2L+s-z)  & {\rm for} & z\in I\!V,\\[3pt]
    \cX_s(\tilde{\fn}_1, \tilde{\fn}_2, L) & {\rm for} & z\in V,\end{array}\right.
    \nn
    \eea
where $\cX_s(\tilde{\fn}_1, \tilde{\fn}_2, z):=\left|V_s(\tilde{\fn}_1, \tilde{\fn}_2, z)\right|^2$,
$\mathcal{Y}_1(\tilde{\fn}, z):=\left|V_{+}(\tilde{\fn}, z)\right|^2$, $\mathcal{Z}^{(E)}_1(\tilde{\fn},z):=\RE\left\{V_{+}(\tilde{\fn}, z)V^{*}_{-}(\tilde{\fn}, z) \right\}$, $\mathcal{Z}^{(M)}_1(\tilde{\fn},z):=\RE\left\{V_{-}(\tilde{\fn}, z)V^{*}_{+}(\tilde{\fn}, z) \right\}$, and
	\begin{align}
	&\cU_1(\tilde{\fn}, z):=\frac{1}{2}\left[\cos^2\theta \left|V_{-} (\tilde{\fn}, z)\right|^2 + (1+\sin^2\theta) \left|V_{+} (\tilde{\fn}, z)\right|^2\right].\nn
	\end{align}

\end{document}